\documentclass[format=acmsmall, review=false]{acmart}

\RequirePackage{etoolbox}
\makeatletter

\newcommand{\ACMbetweenAuthorsAndAbstract}{%
  \noindent\emph{Authors affiliated with the Harvard  School of Engineering and Applied Sciences}\par
  \smallskip
  \noindent{\small Corresponding author: \ \texttt{Mark York: markyork@g.harvard.edu | markjamesyork@gmail.com}}\par
  \medskip
}

\@ifundefined{@mkabstract}{%
  \AtBeginEnvironment{abstract}{\ACMbetweenAuthorsAndAbstract}
}{%
  \pretocmd{\@mkabstract}{\ACMbetweenAuthorsAndAbstract}{}{}
}

\makeatother

\usepackage{xr-hyper}
\usepackage{hyperref}
\usepackage{acm-ec-24}
\usepackage{booktabs} 
\usepackage[ruled]{algorithm2e} 

\SetAlFnt{\small}
\SetAlCapFnt{\small}
\SetAlCapNameFnt{\small}
\SetAlCapHSkip{0pt}
\IncMargin{-\parindent}

\setcitestyle{authoryear}

\usepackage{graphicx} 
\usepackage{booktabs} 
\usepackage[ruled]{algorithm2e} 
\usepackage{amsmath}
\usepackage{amsthm}
\usepackage{caption}
\usepackage{subcaption}
\usepackage{xcolor}
\usepackage[normalem]{ulem}
\usepackage{thmtools}
\usepackage{thm-restate}
\usepackage{tikz}
\usepackage{comment}
\usepackage{lipsum}
\usepackage{cancel}
\usepackage{calc}
\usepackage{mathtools}
\usepackage{empheq}

\usepackage{amsfonts}
\usepackage{amsmath}
\usepackage{amsthm}
\usepackage{thmtools}
\usepackage{stmaryrd}

\usepackage{xcolor}
\usepackage{nameref}

\definecolor{blueish}{RGB}{103, 135, 176}
\definecolor{reddish}{RGB}{ 205, 102, 7}
\hypersetup{
	colorlinks,
	linkcolor={reddish!70!black},
	citecolor={blueish!70!black},
	urlcolor=black,
}

\usepackage[nameinlink,noabbrev]{cleveref}
\Crefname{equation}{}{}
\Crefname{equation}{}{}
\Crefname{defn}{Definition}{Definitions}
\Crefname{lem}{Lemma}{Lemmas}
\Crefname{lem}{Lemma}{Lemmas}
\Crefname{cor}{Corollary}{Corollaries}
\Crefname{cor}{Corollary}{Corollaries}
\Crefname{thm}{Theorem}{Theorems}
\Crefname{thm}{Theorem}{Theorems}
\Crefname{exmp}{Example}{Examples}
\Crefname{exmp}{Example}{Examples}
\usepackage{thm-restate}
\newtheorem{defn}{Definition}\declaretheorem[name=Theorem]{thm}
\newtheorem{lem}{Lemma}
\newtheorem{example}{\protect\examplename}

\newtheoremstyle{informalstyle}
  {3pt}                 
  {3pt}                 
  {\itshape}            
  {}                    
  {\bfseries}           
  {.}                   
  {.5em}                
  {\thmname{#1}\thmnote{ #3}}

\theoremstyle{informalstyle}
\newtheorem*{informalthmbase}{Theorem}

\newenvironment{informalthm}[1]
  {\begin{informalthmbase}[\ref{#1} (informal)]}
  {\end{informalthmbase}}

\theoremstyle{plain}

\usepackage{comment}
\colorlet{proofcolor}{black}

\providecommand{\revisionmode}{draft}
\usepackage[\revisionmode,authormarkup=none,commentmarkup=footnote,commandnameprefix=always]{changes}
\definecolor{english}{rgb}{0.0, 0.5, 0.0}
\definecolor{airforceblue}{rgb}{0.36, 0.54, 0.66}
\definecolor{royalblue}{RGB}{65,105,225}
\definecolor{manuelpurple}{RGB}{128,40,185}
\makeatletter
\ifbool{Changes@optiondraft}{
  \newcommand{\revisioncomment}[3]{}
  \newcommand{\revisionadd}[2]{#2}
  \newcommand{\revisiondelete}[2]{}
  \newcommand{\revisionreplace}[3]{#2}
  \newcommand{\revisionmathawaredelete}[2]{}
  \newcommand{\revisionmathawarereplace}[4]{#3}
  \newcommand{\revisiondeleteddisplay}[2]{}
}{
  \newcommand{\revisioncomment}[3]{}
  \newcommand{\revisionadd}[2]{#2}
  \newcommand{\revisiondelete}[2]{}
  \newcommand{\revisionreplace}[3]{#2}
  \newcommand{\revisionmathawaredelete}[2]{}
  \newcommand{\revisionmathawarereplace}[4]{#3}
  \newcommand{\revisiondeleteddisplay}[2]{}
}
\newcommand{\dcp}[1]{\revisioncomment{cyan}{DCP}{#1}}
\newcommand{\dcpadd}[1]{\revisionadd{english}{#1}}

\newcommand{\mjyadd}[1]{\revisionadd{royalblue}{#1}}
\newcommand{\mjydelete}[1]{\revisiondelete{royalblue}{#1}}
\newcommand{\mjyreplace}[2]{\revisionreplace{royalblue}{#1}{#2}}

\newcommand{\manueladd}[1]{\revisionadd{manuelpurple}{#1}}
\newcommand{\manueldelete}[1]{\revisionmathawaredelete{manuelpurple!45}{#1}}
\newcommand{\manuelreplace}[2]{\revisionmathawarereplace{manuelpurple!45}{manuelpurple}{#1}{#2}}
\newcommand{\manueldeleteddisplay}[1]{\revisiondeleteddisplay{manuelpurple!45}{#1}}

\newcommand{\manuelmathreplace}[2]{\manuelreplace{#1}{#2}}
\newcommand{\mathauditadd}[1]{\manueladd{#1}}
\newcommand{\mathauditdelete}[1]{\manueldelete{#1}}
\newcommand{\mathauditreplace}[2]{\manuelreplace{#1}{#2}}

\newcommand{\mathauditdeleteddisplay}[1]{\manueldeleteddisplay{#1}}

\newcommand{\revisionredadd}[1]{\manueladd{#1}}
\newcommand{\revisionredreplace}[2]{\manuelreplace{#1}{#2}}
\makeatother

\title[Belief Elicitation from Agents with Competing Incentives]{Belief Elicitation from Agents with Competing Incentives}

 \author{Manuel W\"uthrich, Mark York, David C.~Parkes}

\providecommand{\examplename}{Example}

\makeatother

\providecommand{\examplename}{Example}

\begin{document}

\begin{titlepage}
    \begin{abstract}
	We study the problem of eliciting beliefs about an uncertain future event, where the reported
    beliefs are used to inform a decision. 
    Unlike the standard model of belief elicitation, in our model the belief-reporting agents (or {\em recommenders}) have a vested interest in the decision and may attempt to change the decision by reporting untruthfully.
    For instance, a climate scientist with ties to construction companies may overstate the risk of an extreme weather event to encourage investment into flood barriers. A mechanism takes reported beliefs, and consists of a decision rule and a payment rule. 
    We quantify the lowest achievable manipulation under any mechanism in terms of the (i) sensitivity of the target decision function, (ii) the size of competing incentives, (iii) the mechanism's budget, and (iv) the number of recommenders. We design a mechanism that achieves these bounds up to log factors for a family of target decision functions. When the competing incentives \dcpadd{are not intrinsic but} correspond to a bribe from a rational third party, we characterize the size of the mechanism's budget required for there to be no bribe in equilibrium.
    \dcpadd{We illustrate the effectiveness
    of the adaptive-payment mechanism in a simple loan setting.}
    \end{abstract}  

	\maketitle

	\setcounter{tocdepth}{2} 
\end{titlepage}	

\section{Introduction}

There are a wide range of settings where it is helpful, in making a decision, to elicit beliefs about the probability of an uncertain event. For example, suppose a city government must decide how much money to invest into flood defenses. To make an informed decision, the government may elicit predictions or beliefs from experts (who we call {\em recommenders}), such as climate scientists, on whether an extreme weather event such as a hurricane will occur within the next five years. 
The government will aggregate these beliefs and decide how much to invest into flood defenses. However, these beliefs are privately held, and the government must  design a payment scheme to elicit accurate belief reports and then make its investment decision based on these reports.

In the standard problem of {\em information elicitation with verification}~\citep{brier1950verification}, one elicits a belief from each participant and scores this report against the realization of a future event. This problem has been studied extensively and many payment schemes ({\em scoring rules}) that ensure truthful reporting of beliefs are known. However, these approaches do not address a fundamental problem: recommenders may have a vested interest in the decision taken on the basis of the reported beliefs. This interest may outweigh the incentive provided by the payment scheme and cause recommenders to seek to manipulate the decision by reporting untruthfully. In the above example, a climate scientist may own real estate that would lose value due to construction of nearby flood defense barriers, which may lead them to understate the risk. Others may be bribed by a construction company interested in building a flood barrier, which may lead them to overstate the risk.

In this paper, we study how such vested interests (which we call {\em competing incentives}) influence the reporting of recommenders, given different decision and payment rules that comprise the design of the  belief elicitation mechanism.
We consider different methods for aggregating the  beliefs of recommenders into a decision (we call this the {\em target decision function}). For a given target decision function, we seek to design a {\em mechanism} that is able to realize (or {\em implement}) the target decision function, even though the decision function is defined on private beliefs while the mechanism must rely on reports produced by recommenders with competing incentives. A mechanism consists of a {\em decision rule}, which maps reports to a decision, and a {\em payment rule}, which maps reports and the realized outcome to a payment for each recommender. 

The main results are: 
\begin{enumerate}
\item We show that for any nontrivial target decision function, no mechanism can exactly implement the  decision function when recommenders have intrinsic competing incentives. As a corollary, there are incentives to deviate from truthful reports in any mechanism that simply adopts the target decision function as its decision rule.
\item We give a lower bound on the extent to which the decision taken based on  reports will deviate, for any mechanism, from the target decision function. This bound depends on the {\em sensitivity} of the target decision function, which measures how responsive the target  function is to changes in recommenders' beliefs and
captures the susceptibility to manipulation.
%
\item We introduce  the {\em adaptive-payment mechanism} which, given a target decision function, uses \dcpadd{as its decision rule the target decision function and adopts} a payment rule that concentrates incentives where the target  function is most sensitive to beliefs. We show that for any  target  decision function within a piecewise-linear family, this mechanism achieves the lower bound on manipulation up to a log factor.
\item When the competing incentives  stem from a bribe made by a \emph{rational briber} who gains value from  influencing the decision in a particular direction, rather than being intrinsic to a recommender,  we characterize
the size of the mechanism's budget  for which the adaptive-payment mechanism precludes bribes 
in equilibrium, the budget depending on the sensitivity of the target decision function and the 
potential gain to the briber.
\end{enumerate}

\dcpadd{We also illustrate the effectiveness of the adaptive-payment mechanism in a loan setting, 
with the set-up of the simulation calibrated to data collected from a study with loans and 
recommenders in Uganda. The results show that the mechanism can be effective in boosting 
profit and return-on-investment of a lender in a realistic setting.}

\subsection{Overview of the Results}

To illustrate our results, we return to the example of constructing a flood barrier. 
\begin{example}
\label{ex:dam}
A city government  must decide how much money ($A\geq0$) to invest into flood barriers. \dcpadd{This is the decision to make, or action to take by the government.} 
For this, it wants to assess the probability of an extreme weather event within the next five years ($O=1$) or not ($O=0$) by consulting $n\geq1$ experts (recommenders). Each recommender has a private belief $Q_{i}\in[0,1]$ of the probability of \dcpadd{outcome} $O=1$, and this belief may depend on a signal $Y_i$ which the recommender observes. The target decision function $f:[0,1]^{n}\to\mathbb{R}_{+}$ specifies how much the government would invest into flood barriers given knowledge of  beliefs $Q=(Q_{1},..,Q_{n})$. For example, 
\begin{equation}
f(Q)=\ell\cdot\bar{Q},\label{eq:example}
\end{equation}
where $\bar{Q}=\frac{1}{n}\sum_{i=1}^{n}Q_{i}$, and
$\ell\ge0$ is the slope of the target decision function.

However, the government 
must rely on reports $R=(R_{1},\ldots,R_{n})$, 
and  designs a mechanism with  decision rule, $F(R)\geq0$, \dcpadd{based on reports,} and a payment rule, $\textsc{pay}_{i}(R,O)$, \dcpadd{based on reports and the realized outcome}, that makes a payment to each recommender $i$ after observing whether an extreme weather event happened ($O=1$) or not ($O=0$).
The government has total budget $\beta>0$ to pay recommenders.

In choosing a report to make, a  recommender may have competing incentives \dcpadd{for the action $A$ taken by the city}.  
For example,  recommender $i$ with  ties to a construction company 
may earn value $C_{i}\cdot A$ \dcpadd{($C_i$ is the competing incentive)}, for
$C_{i}>0$ and   have an incentive to over-report, with $R_{i}>Q_{i}$. 
Another recommender $j$ may own real estate  and face loss $C_{j}\cdot A$ for $C_{j}<0$
depending on the extent of  flood barriers that are constructed, 
and \dcpadd{have an incentive to under-report, with} $R_{j}<Q_{j}$.   

Can we design a
mechanism such that $f(Q)=F(R)$, for all beliefs $Q$ and rational choices made \dcpadd{in equilibrium} by
 recommenders about reports $R$,  despite these competing incentives? 
\end{example}

For a mechanism, $\textsc{mech}=(\textsc{F},\textsc{pay})$, we define  a {\em payment rule}, $\textsc{pay}=(\textsc{pay}_{i})_{i\in[n]}$, for each recommender, and a {\em decision rule} $\, \textsc{F}:[0,1]^{n}\to\mathbb{R}_{+}$, that maps any report profile $R$ to a decision, \dcpadd{or action} $A\geq0$. The total of all payments to recommenders must not exceed the budget $\beta$.
To implement a {\em target decision function} $f$, it would be natural to consider $\textsc{F}:=f$. However, we shall see that 
no payment scheme can ensure truthful reporting $R=Q$
for nontrivial target \dcpadd{decision} functions.  Hence, we allow  $\textsc{F}\neq f$, seeking to reduce the amount by which recommenders misreport or to reduce the effect of misreports.\footnote{The {\em revelation principle}~\citep{Hurwicz60} suggests that w.l.o.g.~we can focus on truthful mechanisms where agents report their true private types.
However, as we will make formal, the full type of a recommender in this setting may include, for example, any private data they  have, their beliefs about what data others may have, how they think  this data relates to the outcome $O$, 
and their competing incentives. Hence, direct revelations are not practical 
and we must work with   indirect mechanisms. This is why we study
mechanisms with    decision rules $\textsc{F}\neq f$ for simpler reports $R_i\in[0,1]$ \dcpadd{about a recommender's posterior belief regarding the probability that the outcome takes on value $O=1$}, 
and why we study \dcpadd{accuracy through} $\left|\textsc{F}(R)-f(Q)\right|$.}

\subsubsection{Intrinsic Competing Incentives}


We show that for any nontrivial target decision function $f$, there is no mechanism $\textsc{mech}$ that can \dcpadd{always} ensure $f(Q)=\textsc{F}(R)$, given \dcpadd{equilibrium} reports $R$ from rational  recommenders in the presence of competing incentives  and give a lower bound on the extent to which the realized decision \dcpadd{must} deviate from the decision of the target decision function. From~\Cref{ex:dam}, \dcpadd{we can for now give a simple 
idea of the} competing incentive as $C_i \in \mathbb{R}$, 
which provides recommender $i$ with  utility $C_i \cdot A$, 
linear in the \dcpadd{action} $A$ \dcpadd{and that the recommender
cares about along with} any payment from the mechanism. 
\dcpadd{This allows us to state an} informal version of our main negative result: 


\begin{informalthm}{thm:lower}
\label{thm:informalLower} For any target decision function $f$\mjyreplace{,}{ and} any mechanism $\textsc{mech}=(\textsc{F},\textsc{pay})$ with budget $\beta>0$, \mjyadd{and any bound on competing incentives $\gamma > 0$ such that $\sum_{i=1}^{n}|C_{i}|\le\gamma$, }there exist recommender \dcpadd{subjective posterior} beliefs $Q=(Q_{1},\ldots,Q_{n})$ \mjyreplace{}{and competing incentives $C=(C_{1},\ldots,C_{n})$ (bounded by $\sum_{i=1}^{n}|C_{i}|\le\gamma$, for some $\gamma>0$)}
such that \mjyreplace{in every}{recommenders'} Bayes-Nash equilibrium, \dcpadd{recommender} reports $R$ 
\dcpadd{have the effect}
\[
\left|\textsc{F}(R)-f(Q)\right|\gtrapprox\frac{\gamma}{\beta}\cdot\textsc{sensitivity}_{f},
\]
where $\textsc{sensitivity}_{f}$ is a measure of the sensitivity of $f$.
\end{informalthm}

We  formally define sensitivity in \dcpadd{\Cref{sec:results}}. It captures how fast the target \dcpadd{decision} function, $f$, varies as a function of beliefs, and  equals $0$ if and only if $f$ completely ignores beliefs. 
Hence, if the sensitivity of $f$ is nonzero and there are competing incentives, then no finite budget $\beta>0$ can guarantee  there will be no manipulation. Further, we see that the
\dcpadd{effect of} manipulation scales \dcpadd{at least} linearly with the ratio between the size of competing incentives, $\gamma$, and the budget of the mechanism, $\beta$. 

Applying \mjyadd{\Cref{thm:lower}}  
to the target decision function in~\Cref{ex:dam}, and ignoring some constants and log factors, we obtain
that \dcpadd{for budget, $\beta>0$, and bound on competing incentives, $\gamma>0$}, 
that 
for any mechanism $\textsc{mech}=(\textsc{F},\textsc{pay})$ there exists 
recommender beliefs $Q$ 
such that \dcpadd{in every Bayes-Nash equilibrium, the recommender reports $R$} have the effect 
\[
\left|\textsc{F}(R)-f(Q)\right|\gtrapprox\frac{\gamma}{\beta}\cdot\frac{\ell^{2}}{n},
\]
where $\ell\geq0$ is the slope of the target decision function.\footnote{\dcpadd{This follows from~\Cref{lem:dcptight}, with $\textsc{sensitivity}_{f}=\Delta_f$, and $\Delta_f \gtrapprox\frac{\alpha \ell^{2}}{m}$, substituting $\alpha=1,m=n$ to reflect the target decision function in~\Cref{ex:dam}.}}
We see that the effect of the manipulation scales with the ratio between the size of the square
of the slope and the number of recommenders. 

This \dcpadd{dependence on $\gamma, \beta, \ell$ and $n$ is not an} artifact of the \dcpadd{analysis}. To show this, we 
introduce the {\em adaptive-payment mechanism}, which achieves this \dcpadd{manipulation} bound up to factors that are logarithmic in $n$ for a family of {\em piecewise-linear target decision functions} \dcpadd{(\Cref{thm:upper}, \Cref{thm:bldfTight})}. The novelty of 
the adaptive-payment  mechanism, \dcpadd{which adopts the target decision function as its decision rule,} is that its payment rule concentrates incentives where the target  function is most sensitive to reports. 

 We adopt {\em rationalizability}~\citep{Bernheim1984-nc,Pearce1984-ld} as a solution concept
for our positive results. \dcpadd{Rationalizability is}  
 a general class of behaviors which includes  all strategies a rational agent could possibly follow \dcpadd{and is broader than a particular equilibrium notion such as Bayes-Nash equilibrium}. 
 We  define rationalizability formally in~\Cref{sec:problem-definition}.
 We obtain the following result: 
\begin{informalthm}{thm:bldfTight} \dcpadd{(Second part of theorem)}
For any target decision function $f$ in \dcpadd{a suitable} family of piecewise-linear functions, the adaptive-payment mechanism with budget $\beta>0$ achieves the lower bound from \mjyadd{\Cref{thm:lower}} up to log factors. In particular, for the target decision function in~\Cref{ex:dam}, we obtain, for any \dcpadd{recommender subjective posterior beliefs} $Q$, and for any rationalizable \dcpadd{recommender reports} $R$, that 
\[
\left|\textsc{F}(R)-f(Q)\right|\lessapprox\frac{\gamma}{\beta}\cdot\frac{\ell^{2}}{n},
\]
for competing incentives bounded by $\sum_{i=1}^{n}|C_{i}|\le\gamma$, a target decision function with slope $\ell\geq0$, and $n$ recommenders. 
\dcpadd{Here, $F$ is the decision rule  in the adaptive-payment mechanism, which is $F:=f$.}
\end{informalthm}

We see from Theorem~\ref{thm:bldfTight} that there are three ways of decreasing the effect of manipulation in the context of these piecewise-linear target decision functions: (i) make the target decision function $f$ less sensitive, by decreasing slope $\ell$, which is however often not desirable as it becomes less responsive to \dcpadd{recommender}  beliefs, (ii) increase the mechanism's budget $\beta$; or, (iii) increase the number of recommenders $n$. 

The third option \mjyreplace{may appeal to practitioners}{is interesting in practice}, as it allows to maintain the sensitivity of the target decision function without having to increase the mechanism's budget. \mjyadd{However, increasing the number of recommenders with a fixed budget also reduces the incentive for each recommender to put in effort to get a good prediction. There may also be costs to recruiting additional recommenders. Each of these factors is outside of our formal model but could apply in practice.}

\subsubsection{Competing Incentives due to a Briber}

In a second setting, we study competing incentives that are not intrinsic to  recommenders, but  stem from bribes offered to recommenders by some third party. In~\Cref{ex:dam}, a recommender may not have direct ties to a construction company but may  be bribed by the  company. Suppose the construction company gains a value $\upsilon\cdot A$, that is linear in the investment $A\geq0$ into the flood barrier, where $\upsilon\in\mathbb{R_{+}}$ 
\mjyreplace{controls}{is controls} the amount by which the briber stands to gain. The briber may share part of their upside  with recommenders; i.e., they may promise a recommender $i$ a bribe $C_{i}\cdot A$, which the briber
\dcpadd{pays to the recommender} in \dcpadd{the event of action $A$ by the mechanism}.

What is new in this setting
is that we can, \dcpadd{in effect, remove} competing incentives by making it too costly for a rational briber 
to pay bribes. \dcpadd{In this way, we overcome the impossibility of~\Cref{thm:lower} and identify conditions
under which a target decision function can be exactly implemented.} 

We first give a necessary condition on the \dcpadd{mechanism's budget, the briber's value, and the sensitivity of the
target decision function}
for the existence of mechanism that implements a target decision function $f$. \dcpadd{In the presence of a} briber, the mechanism rules induce a sequential game, and we use {\em  perfect Bayesian equilibrium} as the solution concept,  which is a 
suitable refinement of the Bayes-Nash equilibrium \dcpadd{to sequential games}. 
\begin{informalthm}{thm:briberNecessary}
\mjyreplace{Given}{For any} target decision function $f$ \mjyadd{with sensitivity to reports, $\textsc{sensitivity}_{f}$, and a briber with value $\upsilon>0$, then for} any mechanism $\textsc{mech}$ with budget $\beta>0$, \mjyreplace{}{and any briber with value $\upsilon>0$, }such that 
\[
\frac{\beta}{\upsilon}\lessapprox\textsc{sensitivity}_{f},
\]
\mjyreplace{}{where $\textsc{sensitivity}_{f}$ is a measure of the sensitivity of $f$,} there exist recommender 
\dcpadd{subjective posterior} beliefs $Q$ such that all perfect Bayesian equilibrium 
\dcpadd{recommender} reports $R$ satisfy $\textsc{F}(R)\neq f(Q)$. 
\end{informalthm}


\dcpadd{In other words, it is necessary that $\beta/\upsilon>\textsc{sensitivity}_{f}$ for it to be possible to implement the target function in the presence of a briber.}
For the target decision function in Example~\ref{ex:dam},  \dcpadd{and as long as the maximum amount of money the city will invest in barriers is bounded}\dcpadd{\footnote{This requirement is needed for the $\max_q f(q)$ component of the formal statement of~\Cref{thm:briberNecessary} to be well defined.}}
this  
\dcpadd{translates to it being necessary for success that}
\[
\frac{\beta}{\upsilon}\dcpadd{\gtrapprox} \frac{\ell^{2}}{n},
\]
where $\ell>0$ is the slope of the target decision function.\footnote{\dcpadd{This follows from~\Cref{lem:dcptight}, with $\textsc{sensitivity}_{f}=\Delta_f$, and $\Delta_f \gtrapprox\frac{\alpha \ell^{2}}{m}$, substituting $\alpha=1,m=n$ to reflect the target decision function in~\Cref{ex:dam}.}}

%
Hence, if the budget is too low or the \dcpadd{slope} is to high, compared to the value a briber may gain,
there will be bribes and \dcpadd{any mechanism must sometimes fail to implement the target decision function}. 

We also show, \dcpadd{in this  setting with bribes, that this condition on 
the mechanism's budget, the briber's value, and the sensitivity of the
target decision function is in effect sufficient for success, again by making use of the adaptive-payment mechanism.} 
\begin{informalthm}{thm:briberSufficient}
\mjyreplace{Given}{For any} target decision function $f$ in some family of piecewise-linear functions, \dcpadd{and where this target function is strictly positive}, 
and in the briber setting with the adaptive-payment mechanism, 
then if the mechanism budget $\beta\ge0$ and briber value $\upsilon>0$ satisfy
\[
\frac{\beta}{\upsilon}\gtrapprox \dcpadd{\textsc{sensitivity}'_{f}},
\]
where \dcpadd{$\textsc{sensitivity}'_{f}$} is a suitable measure of the sensitivity of $f$, 
then for any \dcpadd{recommender subjective posterior} beliefs $Q$, and for \dcpadd{any rationalizable recommender reports $R$}, we have $\textsc{F}(R)=f(Q)$. \dcpadd{Here, $F$ is the decision rule used in the adaptive-payment mechanism, which is  $F:=f$.}
\end{informalthm}

The  notion of sensitivity, denoted here as $\textsc{sensitivity}'_{f}$, is slightly different than that
used so far, and is also defined in~\Cref{sec:results}.
For the target decision function in~\Cref{ex:dam}, \dcpadd{and as long as the  city will invest a non-zero amount in barriers,}\dcpadd{\footnote{This requirement is needed for the $\min_q f(q)$ component of the formal statement of~\Cref{thm:briberSufficient} to be well defined.}} the condition in~\Cref{thm:briberSufficient} translates \dcpadd{to
it being sufficient for success that}
\[
\frac{\beta}{\upsilon}\gtrapprox\frac{\ell^{2}}{n},
\]
where $\ell>0$ is the slope of the target decision function.\footnote{\dcpadd{This follows from~\Cref{lem:dcptight}, with $\textsc{sensitivity}'_{f}=\nabla_f$, and $\nabla_f \lessapprox\frac{\alpha \ell^{2}}{m}$, substituting $\alpha=1,m=n$ to reflect the target decision function in~\Cref{ex:dam}.}}


\subsection{Related work}

\label{subsec:related-work}

The literature on elicitation with verification uses \emph{proper scoring rules}, which elicit subjective information from a single agent about an uncertain future event and align incentives with truthful reports~\citep{brier1950verification,winkler1994evaluating,GneitingTilmann2007SPSR}. A (strictly) proper scoring rule pays the agent a score based on their information and the event outcome; the score is (strictly) maximized when they truthfully-report their subjective information. 
\emph{Wagering mechanisms}~\citep{freeman2018axiomatic} and \emph{prediction markets}~\citep{wolfers2004prediction} extend mechanisms for belief elicitation to multiple agents and allow for belief aggregation, but without modeling a decision to be made (and consequently, without handling competing incentives); see also~\citet{zhang2011task}. An empirical study demonstrates that
prediction markets can be manipulated, with  manipulated prices persisting for 60 days, especially for markets with few players and low volume \citep{rasooly2025manipulable}. This suggests manipulation resistance is an important open question.

In \textit{decision scoring rules}, a principal uses agent reports to make a decision ~\citep{oesterheld2020decision, ChenYilingDMwG,york2021eliciting}. These papers consider the problem that arises when whether an uncertain event is observed depends on the decision that is made. For example, if collecting information about the market success of a new electric vehicle (EV) design, with reports used to decide whether the EV is built and goes to market, then whether the outcome (market success or not) is observed depends on reports. 
In contrast to our work, \dcpadd{none of these papers} consider agents with outside incentives in regard to the decision that is made. 
To our knowledge, ours is the first paper in the information elicitation literature to formalize the problem of belief elicitation for the purpose of making a decision in the presence of participants with
these kinds of competing incentives. 


\mjyadd{A family of papers considers} \dcpadd{eliciting beliefs} \mjyadd{from agents about an uncertain
event where the agents can also take actions to influence the likelihood of the event.} \dcpadd{For example, 
a software company wants to predict whether a product
will be released on time and an employee who predicts a late release
may then work to sabotage  an on-time release.
\citet{shi2009prediction} consider the distinction between desirable and undesirable actions; e.g., choosing to work to expedite a software release is desirable. They say that an elicitation mechanism is {\em principal-aligned} if it
removes incentives for undesirable actions.} 
\mjyadd{These authors show the principal may have
to pay up to $n$ times as much, for $n$ recommenders, as in
the non-influence case to prevent undesirable actions.} \dcpadd{\citet{chakraborty2014manipulation} and \citet{freeman2017crowdsourced} add more structure, modeling  agents who can vote or otherwise
make reports in a second  stage that determines the realized event.}
%
\mjyadd{\citet{chakraborty2014manipulation} consider the two-player case} 
\dcpadd{and adopt a simple voting game to determine the event outcome.}
\mjyadd{\citet{freeman2017crowdsourced} consider multiple agents and show that with 
a limit to the amount of budget any agent can spend in the first, belief elicitation stage,
truthful event outcomes can be obtained}
\dcpadd{through suitable mechanism design of elicitation in the second
stage.}
\dcpadd{\citet{bacon12} also consider a specific setting of a worker sharing beliefs about
task completion time with a manager, the worker then exerting effort on a task. These papers do not model a principal who needs
to make a decision, taking up instead the orthogonal problem 
of agents who can influence the  event outcome.}

The \emph{peer prediction} literature~\citep{miller2005eliciting,36128,prelec2004bayesian,witkowski2012robust,olckers2024manipulation} studies \emph{information elicitation without verification}---eliciting information from multiple agents in the absence of an uncertain future event against which to score. This absence of verifiable information is fundamental to the problem of peer prediction and solutions rely on  correlation  between agent signals. 
\mjyadd{Unlike our setting, the outcome of events about which beliefs
are elicited in peer prediction mechanisms are not later observed}. We are not aware of any
 peer prediction model that handle the challenges arising from competing incentives from outside the mechanism. \dcpadd{A hybrid model between peer prediction and observable outcomes
 has also been studied, where bribes  seek to influence reports, claims can be made
 about inaccurate aggregation, and  the mechanism
 can fall back on costly access to ground truth about the 
 event outcome,  but where
 in equilibrium the mechanism operates in the style of peer prediction~\citep{moroz26}.}



A body of work proposes strategy-proof mechanisms for cases when the mechanism designer and all agents know the distribution of valuations, and valuations are interdependent (e.g. \citet{cremermclean}, \citet{mezzetti2004mechanism}). While these mechanisms are theoretically-interesting they are less relevant to practice because the mechanism designer 
needs knowledge of the distribution on values including correlation between values. In our case, agents' private information---beliefs and competing incentives---are independent,  rendering the inference in these papers not possible. 
Moreover, whereas these papers have single dimensional types, our work differs in that agents' beliefs  and their external, \dcpadd{competing} incentives 
are separate.

There are models of group decision problem with competing incentives, specifically selecting a committee from a set of voters where each voter would prefer to be selected~\citep{alon2011sum}. A  mechanism is introduced to align incentives through the use of randomization to introduce suitable decoupling between an agent's own report and whether or not it is selected; see also~\citet{kurokawa2015impartial} and \citet{niemeyer2024optimal}. Another setting of competing incentives arises in the \emph{transitive trust} literature, where reports on the trust one agent has for another are made, and a mechanism is used to aggregate these reports. Competing incentives arise because a participant may care about their own trust ranking (e.g., being ranked above others, or above some threshold). \citet{hopcroft2007manipulation} develop a variation on PageRank~\citep{page1999pagerank} that is robust to misreports; see also~\citet{parkes_seuken}. Recent work from \citet{milionis2025incentive} also characterizes mechanisms that incentivize nodes in communication networks to honestly report their locations. These settings are different from ours in that they are not about eliciting beliefs about an uncertain event, do not make use of payments, and in that the competing incentives stem from competition across participants. 
 %

Bribery has  been studied in voting settings~\citep{keller2018approximating,elkind2009swap,faliszewski2016control,ParkesDavidC.2017TVBT}, for example in regard to how many votes a briber must flip to change the outcome of an election or poll. In contrast, our setting treats the aggregation of belief reports and choosing an outcome in a continuous range, which is quite different from a discrete social choice problem; e.g., a participant with competing incentives may always usefully be able to affect the outcome  by a small amount in our setting. The issue is that incentives to influence the decision is linear in the decision probability, while the derivative of proper score reduction asymptotically approaches zero at the true belief, meaning that there will be a small zone where the slope of outside incentive is greater than the slope of score incentive. 

In summary, the components that make our problem setting unique include:
\begin{enumerate}
    \item the social choice functions of interest depend on \dcpadd{recommenders' subjective posterior} beliefs, which are
    payoff-independent signals,  and
    not the \dcpadd{competing} incentive $C_i$,
    \item direct-revelation mechanisms are precluded because the mechanism designer does not know the signal space (which may correspond to data), and
    \dcpadd{the subjective belief about the outcome as a function of signals is similarly complex}, 
    \item a \dcpadd{recommender's prior  on signals,  realized signal, 
    prior  on beliefs about the outcome conditioned on signals,  posterior
    belief, and competing incentives are unknown to the mechanism designer}, and
    \item we have a contractible outcome \dcpadd{of an event}, meaning payments to agents can be conditioned on a verifiable outcome.
\end{enumerate}

\section{Problem Definition}

\label{sec:problem-definition}

In this section, we formally define the model and introduce the solution concepts used in stating the main results.

\subsection{Notation}



We  use $[n]$ to denote the set $\{1,..,n\}$. If there is a variable $x_{i}$ associated with each recommender $i\in[n]$, we use $x$ to denote $(x_{1},..,x_{n})$ and if there is a set $\mathcal{X}_{i}$ associated with each recommender, we use $\mathcal{X}$ to denote the Cartesian product $\mathcal{X}_{1}\times..\times\mathcal{X}_{n}$. Further, we use \mjydelete{t }$\textsc{distr}(\mathcal{X}|\mathcal{Z})$ to denote the set of all conditional probability distributions over a random variable in $\mathcal{X}$ conditioned on a random variable in $\mathcal{Z}$. We use $\textsc{supp}(D)$ to denote the support of  probability distribution $D$.

\subsection{Basic Set-up}

There is a {\em principal}, who seeks to elicit recommenders' beliefs about an unknown \emph{outcome}, $O\in\mathcal{O}:=\{0,1\}$, in order to take an \emph{action}, $A\in\mathcal{A}:=\mathbb{R}_{\ge0}$. 
There are $n$ \emph{recommenders}, and each recommender $i\in[n]$ observes a {\em private signal} $Y_{i}\in\mathcal{Y}_{i}$ related to the outcome $O$ (this signal, $Y_{i}$, could for example be some data available to $i$). Each recommender $i$ has a {\em subjective belief}, $P_{i}\in\mathcal{P}_{i}:=\textsc{distr}(\mathcal{O}|\mathcal{Y})$, \dcpadd{on} the outcome \dcpadd{$O$}. While $i$ will typically be uncertain about the signals of others, 
 the conditional distribution $P_{i}$ describes $i$'s belief regarding $O$ as if $i$ had access to the entire signal profile. This allows $i$ to reason about the reports that others may make and how informed their reports may be. It also allows recommenders to hold different beliefs regarding outcome $O$ even if they have access to the same signal profile $Y$; i.e., 
it allows subjective beliefs, with different recommenders arriving at different conclusions even when 
accessing the same data. 

Each recommender has a  {\em subjective posterior belief, $Q_{i}\in\mathcal{Q}_{i}:=[0,1]$, regarding the probability of $O=1$}, given their private signal. As we shall see in the following sections, the belief $Q_{i}$ follows from $i$'s subjective belief $P_{i}(O|Y)$ \dcpadd{on outcome $O$}, a common prior regarding the private signals $D(Y)$, and the \dcpadd{recommender's} observed signal $Y_{i}$. Recommender $i$ makes a {\em report} $R_{i}\in\mathcal{R}_{i}:=[0,1]$ to a mechanism. If truthful, \dcpadd{a recommender's report} $R_i = Q_{i}$, \dcpadd{and is equals to subjective posterior belief regarding the probability of $O=1$.} 

The goal of the principal is modeled through a {\em target decision function}, $f:\mathcal{Q}\to\mathcal{A}$, that expresses the action $A$ that should  be taken based on recommenders'  beliefs $Q$. Since the principal does not have direct access to beliefs $Q$, the question is how to build a mechanism such that the actual action, $A$, is as close as possible to the target action, $f(Q)$.

 A recommender may  have a {\em competing incentive}, $C_{i}\in\mathcal{C}_{i}:=\mathbb{R}$, which may be positive or negative 
and yields value $C_{i}\cdot A$ to the recommender. This competing incentive is private, and distinct from any utility the recommender will gain from payments in the mechanism. We study two settings: 
\begin{enumerate}
\item Each recommender $i$ may have an \emph{intrinsic competing incentive} regarding the outcome, which we model as \mjyreplace{$C_{i}\in \mathbb{R}$.}{$C_{i}\geq0$.} 

\item Each recommender $i$ \mjyreplace{may}{mat} have a  competing incentive $C_{i}$ that represents a bribe offered by a  rational briber. The briber gains a value $V\cdot A$ that is proportional to the decision (e.g. the investment) where $V$ is the type of the briber. The briber  may offer a payment $C_i\cdot A$ to recommender $i$, trying to get them to misreport.
The briber seeks to maximize the net value, $(V-\sum_{i\in[n]}|C_{i}|)\cdot A$.
\end{enumerate}

\subsection{An Elicitation Mechanism\label{sec:mech}}

To determine payments to recommenders and the action to be taken, the principal makes use of a \emph{mechanism}, which is known to all agents and defined as follows: 
\begin{defn}[Mechanism]
\label{def:mechanism} A {\em mechanism} is defined by the tuple $\textsc{mech}=(\textsc{F},\textsc{pay})$, where 
\begin{enumerate}
\item $\textsc{F}:\mathcal{R}\to\mathcal{A}$ is the   {\em decision rule}, and produces an action based on the reports, and 
\item $\textsc{pay}:=(\textsc{pay}_{i})_{i\in[n]}$ is the   {\em payment rule}, with $\textsc{pay}_{i}:\mathcal{R}\times\mathcal{O}\to\mathbb{R}_{+}$  a strictly proper scoring rule on the report of $i$,  determining recommender $i$'s payment as a function of the report profile and  observation. 
\end{enumerate}
\end{defn}

 \mathauditadd{The budget constraint is $\sum_{i\in[n]}\textsc{pay}_{i}(r,o)\le\beta$ for all $r,o$.} The strictly proper property requires $\mathbb{E}_{O\sim q_{i}}\left[\textsc{pay}_{i}\left((q_{i},r_{-i}),O\right)\right]>\mathbb{E}_{O\sim q_{i}}\left[\textsc{pay}_{i}\left(r,O\right)\right]$ for all $i\in[n],r\in\mathcal{R},q_{i}\in\mathcal{Q}\mathauditadd{_{i}}$ \mathauditadd{with $r_i\ne q_i$}. This  ensures that
the mechanism will align incentives with truthful reporting
 for recommenders without  competing incentives. 
 Many strictly proper scoring rules are known~\citep{GneitingTilmann2007SPSR}.

In the simplest case, the decision rule  is simply the target decision function, i.e., $\textsc{F}=f$ and hence $A=f(R)$. However,   this may not yield a decision $A$ closest to the target decision $f(Q)$ due to possible misreports $R\neq Q$. 
An $\textsc{F}$ that differs slightly from $f$ may (i) be less susceptible to misreports $R$, or (ii) may  lead to better equilibria, with reports $R$  closer to $Q$. 
Therefore, the mechanism design problem allows for  $\textsc{F}\neq f$.

As discussed in \Cref{subsec:related-work}, using a proper scoring rule to determine recommenders' payments  
is necessary and sufficient
to ensure that the mechanism is truthful (dominant-strategy incentive compatible) in the absence of competing incentives
and  for this reason we restrict attention to mechanisms with  strictly proper payment rules. 
This imposes some restrictions on the form of $\textsc{pay}_{i}(r,o)$ in $r_{i}$ and $o$ but allows for any dependence on
$r_{-i}$, including discontinuities. 

For simplicity of exposition, we do not allow for randomization,
but our results straightforwardly extend to this case because the analysis only relies on the expected action \dcpadd{(the action is on the non-negative real line)} and expected payments. We can  alternatively  think of $\textsc{F}(R)$ as being the expected action, and $\textsc{pay}_{i}(R,O)$ as the expected payment. For instance, we could interpret $A$ as the probability of some binary decision (e.g., whether or not to build a particular flood barrier design). 

\subsection{The Elicitation Game with Intrinsic Competing Incentives\label{sec:game}}

In the first setting, {\em recommender $i$'s type} is $(P_{i},Y_{i},C_{i})$, and consists of $i$'s subjective belief, $P_{i}\in\mathcal{P}_{i}:=\textsc{distr}(\mathcal{O}|\mathcal{Y})$, \dcpadd{on outcome $O$}, private signal $Y_{i}\in\mathcal{Y}_{i}$, and competing incentive $C_{i}\in\mathcal{C}_{i}$. Agent types are private, and we model them as being drawn from a common prior, $D\in\textsc{distr}(\mathcal{P}\times\mathcal{Y}\times\mathcal{C})$, that is independent across recommenders and subtypes, i.e., the prior $D$ can be written as a product, 
\[
D(p,y,c)=\prod_{i\in[n]}D_{i}^{p}(p_{i})\cdot\prod_{i\in[n]}D_{i}^{y}(y_{i})\cdot\prod_{i\in[n]}D_{i}^{c}(c_{i}),
\]
for some $D_{i}^{p}\in\textsc{distr}(\mathcal{P}_{i})$, $D_{i}^{y}\in\textsc{distr}(\mathcal{Y}_{i})$, $D_{i}^{c}\in\textsc{distr}(\mathcal{C}_{i})$. 

\mjyadd{For recommender $i$,  its subjective belief $P_i(O|Y)$ on outcome $O$
is sampled from $D_{i}^{p}(p_{i})$, independently of its realized signal $Y_i$, which is sampled from $D_{i}^{y}(y_{i})$. However, its subjective posterior belief, $Q_i$, is formed
after observing $Y_i$, and thus depends on both $P_i$ and $Y_i$.}

The game played by recommenders in this \dcpadd{first} setting \dcpadd{with intrinsic competing incentives} is defined by the tuple $(\textsc{mech},D)$, and proceeds as follows: 
\begin{enumerate}
\item The recommender types are drawn from the prior $P,Y,C\sim D$.
\item Each recommender $i$ observes their type $(P_{i},Y_{i},C_{i})$.
\item Each recommender \dcpadd{forms its subjective posterior belief $Q_i$ regarding the probability of $O=1$,} and makes a report $R_{i}\in\mathcal{R}_{i}$.
\item The mechanism takes action $A=\textsc{F}(R)\in \mathbb{R}_{\ge0}$.
\item The outcome $O\in\mathcal{O}$ is observed. 
\item Each recommender $i$ receives utility $\textsc{pay}_{i}(R,O)+C_{i}\cdot A$. 
\end{enumerate}

We model recommenders with utility linear in the payment received from the mechanism and their value $C_{i}\cdot A$ for the outcome.  
Each recommender $i$ holds a different subjective belief $P_{i}$ regarding \dcpadd{outcome} $O$.

To formulate the game in the Bayesian setting, we express each recommender's subjective expected utility given the type profile $p,y,c$ and report profile $r$.  The subjective expected utility does not depend on the full type profile, but only on $p_{i},y,c_{i}$, and we  define:
\begin{align}
\textsc{util}_{i}\left(p_{i},y,c_{i},r\right): & =\mathbb{E}_{O\sim p_{i}(\cdot|y)}\left[\textsc{pay}_{i}(r,O)+c_{i}\cdot\textsc{F}(r)\right].\label{eq:recUtil}
\end{align}

This utility absorbs the random variable $O$ by taking its expectation with respect to $i$'s subjective belief \dcpadd{on the outcome}. Hence, we will not need to reason about $O$ explicitly in the sequel. Rather, we can  think of an equivalent version of the game, where steps (4) and (5) are replaced by each recommender $i$ receiving utility in the amount
$\textsc{util}_{i}\left(P_{i},Y,C_{i},R\right)$.

\subsubsection{Common Knowledge}

We assume that the prior $D$ and the mechanism $\textsc{mech}$ are common knowledge
to all recommenders. However, we do not assume the principal knows the prior $D$, and thus  the principal cannot tailor the mechanism to  $D$ but needs a mechanism that works well for any prior.
The principal may not even know what kind of signal is available to the recommenders  (the space of observations $\mathcal{Y}$).
In \Cref{ex:dam}, it is not plausible that one could model \mjyreplace{all the data that different experts}{all the data different that experts} could  have regarding an extreme weather event. 

\subsubsection{Solution Concepts\label{sec:solutions}}

Each recommender $i\in[n]$ will select their report $R_{i}$ based on their type, $(p_{i},y_{i},c_{i})$, hence $i$'s {\em strategy} has the form, 
\[
s_{i}\in\mathcal{S}_{i}:=\textsc{distr}\left(\mathcal{R}_{i}^{\mathcal{P}_{i}\times\mathauditreplace{\mathcal{Y}_{i}\times\mathcal{C}_{i}}{\mathcal{C}_{i}\times\mathcal{Y}_{i}}} \right),
\]
where we allow for mixed strategies. Recommender $i$'s expected utility, given their type $p_{i},y_{i},c_{i}$ and strategy profile $s$, is 
\begin{align*}
\textsc{eutil}_{i}\left(s,p_{i},y_{i},c_{i}\right) & :=\mathbb{E}\left[\textsc{util}_{i}\left(P_{i},Y,C_{i},R\right)\big|s,p_{i},y_{i},c_{i}\right]\\
 & =\mathbb{E}_{P_{-i},Y_{-i},C_{-i}\sim D,\hat{S}\sim s}\left[\textsc{util}_{i}\left(p_{i},\left(y_{i},Y_{-i}\right),c_{i}, \mathauditadd{(}\hat{S}_{i}(p_{i},y_{i},c_{i}),\hat{S}_{-i}(P_{-i},Y_{-i},C_{-i})\mathauditadd{)} \right) \right].
\end{align*}

We model recommenders who play a strategy that is a best-response to some belief about other recommenders' strategies. 
\begin{defn}
Strategy $s_{i}\in\mathcal{S}_{i}$ is a {\em best-response} to a belief over others' strategies $\phi_{i}\in\textsc{distr}(\mathcal{S}_{-i})$ if 
\[
\mathbb{E}_{S_{-i}\sim\phi_{i}}\left[\textsc{eutil}_{i}\left(\left(s_{i},S_{-i}\right),p_{i},y_{i},c_{i}\right)\right]\ge\mathbb{E}_{S_{-i}\sim\phi_{i}}\left[\textsc{eutil}_{i}\left(\left(s_{i}',S_{-i}\right),p_{i},y_{i},c_{i}\right)\right]\quad\forall s_{i}'\in\mathcal{S}_{i},p_{i},y_{i},c_{i}.
\]
\end{defn}

A natural solution concept in our setting is a Bayes-Nash equilibrium:
\begin{defn}[Bayes-Nash Equilibrium (BNE):]
Strategy profile $s\in\mathcal{S}$ is a {\em BNE} if for all $i\in[n]$, $s_{i}$ is a best response to the point belief $\phi_{i}\in\textsc{distr}(\mathcal{S}_{-i})$ that assigns probability $1$ to $s_{-i}$. 
\end{defn}

To strengthen our positive results, we also work with a solution concept that captures a larger set of strategies that rational agents might play,  based on {\em rationalizability}~\citep{Pearce1984-ld,Bernheim1984-nc}. 
Informally, a strategy $s_{i}\in\mathcal{S}_{i}$ is rationalizable if it is a best-response to some belief $\phi_{i}\in\textsc{distr}(\mathcal{S}_{-i})$ that $i$ may hold about other agents' strategies. Here, it is important that $\phi_{i}$ is not arbitrary;
it must be consistent with the knowledge that other agents are best-responding to their beliefs about other agents' strategies and that these beliefs are themselves also consistent with that knowledge, etc.

The set of rationalizable strategies is typically defined recursively:  start with the set of all strategy profiles, then, for each $i$, remove the strategies $s_{i}$ that are not a best response to some belief over others' remaining strategies $s_{-i}$ etc. until convergence; see e.g.,~\citet{Shoham2008-vp}. For our purpose, it is sufficient to apply this recursion twice, which substantially simplifies the formal definition and yields a superset of all rationalizable strategies, since each recursion only removes strategies:
\begin{defn}[2-Rationalizability (\dcpadd{$2\textsc{rat}$}):]\label{def:rat}
Let $\Phi:=\textsc{distr}(\mathcal{S}_{-1})\times..\times\textsc{distr}(\mathcal{S}_{-n})$ be the set of belief profiles describing each agent's belief about other agents' strategies. A strategy $s_{i}\in\mathcal{S}_{i}$ is {\em \mjyreplace{2-Rationalizable (\dcpadd{$2\textsc{rat}$})}{2-Rationalizability (2R)}} if there exists a belief $\pi_{i}\in\textsc{distr}(\mathcal{S}_{-i}\times\Phi_{-i})$ about other agents' strategies $S_{-i}$ and beliefs $\phi_{-i}$ (recall, $\phi_{j}\in\textsc{distr}(\mathcal{S}_{-j})$ is $j$'s belief regarding other agents' strategies), such that: (i) $s_{i}$ is a best response to the marginal belief over strategies $\pi_{i}(S_{-i})$ and (ii) $\pi_{i}$ is consistent with the knowledge that other agents will also play a best response strategy, i.e., for all $j\neq i$ and $\phi_{j}\in\Phi_{j}$, the conditional belief $\pi_{i}(S_{j}|\phi_{j})$ only assigns nonzero probability to strategies $S_{j}$ that are a best response to $\phi_{j}$. We say that a strategy profile $s\in\mathcal{S}$ is \dcpadd{$2\textsc{rat}$} if all individual strategies $s_{1},..,s_{n}$ are \dcpadd{$2\textsc{rat}$}. 
\end{defn}

Hence, the set of \dcpadd{$2\textsc{rat}$} solutions contains the set of rationalizable solutions, which in turn contains the set of BNE solutions (see, e.g., \citet{Shoham2008-vp}). Hence, our negative results, which hold for some BNE, also hold for a worst-case \dcpadd{$2\textsc{rat}$} profile. 
We  state our positive results for all \dcpadd{$2\textsc{rat}$} strategies, and hence they also hold for all BNE.

\subsection{The Elicitation Game with a Briber}

In the second setting, we introduce a briber who may choose to provide competing incentives to recommenders
when doing so is in the self interest of the briber.
The elicitation game in the presence of a briber is similar, except that the competing incentives
do not come from prior $D^{c}$ but are generated instead by a rational agent (the briber). 
The solution concepts are  similar, and we relegate a detailed discussion 
to \Cref{sec:briber}.  

\section{Manipulation due to Intrinsic Competing Incentives}

\label{sec:results} In this section, we quantify to what extent  decision $A$ may deviate from the target decision $f(Q)$ under different mechanisms. More precisely, we want to bound the difference $\left|A-f(Q)\right|=\left|\textsc{F}(R)-f(Q)\right|$ between the action that is prescribed by the target decision function $f$ for true beliefs $Q$ and the  action $A=\textsc{F}(R)$ taken by the mechanism based on the reported beliefs $R$ under an equilibrium strategy.

The design space $\textsc{mech}=(\textsc{F},\text{\textsc{pay}})$ is  large, as we allow for any arbitrary, even discontinuous $\textsc{F}$ and $\textsc{pay}$ (as long as each $\textsc{pay}_{i}$ is strictly proper in $r_{i},o$). 
For instance, we may hope to reduce or eliminate manipulation by designing $\textsc{F}$ and $\textsc{pay}$ such that $\textsc{pay}$ incentivizes recommenders to report truthfully whenever they may have a large influence on $\textsc{F}$. 
Concretely, one could design $\textsc{F}$ such that around any given report profile $r$ only a small set of recommenders are influential, with $\textsc{pay}$ that concentrates the entire budget on these recommenders.

We first derive a lower bound on the manipulation, relative to a given target decision function, that can be achieved by
any  mechanism.
We show that for any nontrivial target decision function and any mechanism with limited budget, there are, in the presence of  competing incentives, equilibria with  manipulation. 
The main task is to quantify the size of this manipulation, depending on the target decision function $f$, the mechanism's budget $\beta$, the total bound on competing incentive $\gamma$, and the number of recommenders $n$. We also present a  mechanism, \dcpadd{the adaptive-payment mechanism}, which achieves this lower bound up to log factors
for a family of piecewise-linear \dcpadd{target} decision functions.

\subsection{Lower Bounds on Manipulation}

\label{sec:lower}

Intuitively, the more sensitive the target decision function $f$, the more attractive is manipulation for recommenders as they can influence the decision with even small misreports (and thus with a  smaller loss in the payment due to the scoring rule $\textsc{pay}$). 

In the following, we define the {\em sensitivity}, $\Delta_{f}$, which describes the sensitivity of target  \dcpadd{decision} function 
$f$. We   give a lower bound on the manipulation as a function of sensitivity.
In the next section, we  see that this lower bound is tight for a family of target decision functions, which indicates that $\Delta_{f}$ is a fundamental property in capturing  the manipulability
of target  \dcpadd{decision} function $f$.

\subsubsection{Sensitivity of Target Decision Functions}

The notion of sensitivity relies on a particular type of random walk. The idea is that this random walk moves through the domain of target decision function $f$, recommender by recommender\mjyreplace{.}{:} \mjyadd{ We use a random walk rather than a simpler derivative-based constraint to cover a  wide range of target decision functions, including non-differentiable and discontinuous ones. It is  natural to consider discontinuous target decision functions; e.g., the city will build a bridge with probability 1 if the probability of staying under budget is at least 0.8 and will not build the bridge otherwise.}

\begin{defn}[Random walk]
\label{def:walk} The random walk is defined by the tuple $\mathcal{W}=(\mathcal{K},\mathcal{D}_{0},\mathcal{D}_{z})$ where $\mathcal{K}\subseteq[n]$ is a subset of recommenders,  \mathauditadd{with $k:=|\mathcal{K}|$, } \manuelreplace{$\mathcal{D}_0$ is a distribution with support in $[0,1]^n$ and $\mathcal{D}_z$ is a distribution with support in $[-1,1]^n$}{and $\mathcal{D}_{0},\mathcal{D}_{z}$ are distributions with support in $[0,1]^{n}$} such that $X^{0}\sim\mathcal{D}_{0}$ and $Z\sim\mathcal{D}_{z}$ satisfy $X^{0}+Z\in[0,1]^{n}$ with probability $1$. We define a random walk to be the following process: First, $\Theta^{1},..,\Theta^{k}\in\mathcal{K}$, are drawn uniformly without replacement from $\mathcal{K}$. Then, $X^{1},..,X^{k}$ are generated as follows
\begin{align*}
X^{0} & \sim\mathcal{D}_{0};\quad Z\sim\mathcal{D}_{z};\quad X_{i}^{j}=\begin{cases}
X_{i}^{j-1}+Z_{i} & \text{if }i=\Theta^{j}\\
X_{i}^{j-1} & \text{otherwise}.
\end{cases}
\end{align*}
Let $\textsc{walks}(k,\delta)$ be the set of all such random walks $\mathcal{W}=(\mathcal{K},\mathcal{D}_{0},\mathcal{D}_{z})$ with $|\mathcal{K}|=k$ and $\mathcal{D}_{z}$ having support in $[-\delta,\delta]^{n}$.
\end{defn}

Given this definition of a random walk, we   define sensitivity. 
\begin{defn}[Sensitivity $\Delta_{f}$]
\label{def:sensitivityLower}For a given target decision function $f$, we define
\[
\Lambda_{f}(k,\delta):=\max_{\mathcal{W}\in\textsc{walks}(k,\delta)}\mathbb{E}_{X\sim\mathcal{W}}\left[\sum_{i=1}^{k}\left|f(X^{i})-f(X^{i-1})\right|\right],
\]
and 
\[
\Delta_{f}:=\max_{k\in[n],\delta\in\mathauditreplace{(}{[}0,1]} \frac{\Lambda_{f}(k,\delta)^{2}}{\delta\cdot k\cdot\log(3\cdot k)}.
\]
We will call $\Lambda_{f}$ the {\em sensitivity function} and $\Delta_{f}$ the {\em sensitivity}.
\end{defn}

The sensitivity function $\Lambda_{f}(k,\delta)$ corresponds to the sum of variations of the target decision function $f$ along a random walk through its domain, involving $k$ recommenders, each of whom vary their report/belief by at most \mjyreplace{$\delta$}{$\gamma$}. We will see how to construct a random walk and obtain $\Lambda_{f}(k,\delta)$ for some families of target decision functions in the next section. But first, we show how this notion of sensitivity translates into a lower bound on manipulation.

\subsubsection{General Lower Bound on Manipulation}

We provide a lower bound on the extent to which the  decision $A$ may deviate from the target decision $f(Q)$ under any mechanism $\textsc{mech}$ (and hence also under the optimal mechanism for $f$). This lower bound depends on the sensitivity $\Delta_{f}$ of the target  \dcpadd{decision} function, the number of recommenders $n$, the mechanism's budget $\beta$, and the total competing incentive $\gamma$, which we define as $\gamma\revisionredreplace{\ge}{:=}\max_{c\in\textsc{supp}(D^{c})}\left\Vert c\right\Vert _{1}$. This is the same $\gamma$ 
that bounded the sum of absolute values of outside incentives in the 
 informal theorem statement. \dcpadd{We say that prior $D$ {\em satisfies} competing incentive $\gamma$
 when $\gamma \ge \max_{c\in\textsc{supp}(D^{c})}\left\Vert c\right\Vert _{1}$.}
\begin{thm}[Lower-bound on manipulability]
\label{thm:lower} Suppose we want to implement some target decision function $f$ with sensitivity function $\Lambda_{f}$. 
Then, for  any mechanism \dcpadd{$\textsc{mech}=(\textsc{F},\textsc{pay})$} with budget $\beta>0$, any bound on competing incentive $\gamma>0$, 
there \dcpadd{exists} a prior $D$ \dcpadd{satisfying $\gamma$} \dcpadd{and recommender subjective posterior beliefs $Q=(Q_1,\ldots,Q_n)$} 
such \dcpadd{in every} BNE,  the recommender reports $R$ have the effect
\begin{align*}
\left|\textsc{F}(R)-f(Q)\right|\ge & \frac{1}{16}\cdot\max_{k\in[n],\delta\in\mathauditreplace{(}{[}0,1]} \frac{\Lambda_{f}(k,\delta)^{2}}{k\cdot\delta\cdot\frac{\beta}{\gamma}\cdot\log(3\cdot k)+k\cdot\Lambda_{f}(k,\delta)},
\end{align*}
with probability $1$.
For  \mathauditadd{sensitivity, $\Delta_f$, with $\Delta_f$ finite, and}  competing incentives $\gamma$, 
this implies
\[
\left|\textsc{F}(R)-f(Q)\right|\ge\frac{1}{16}\cdot\frac{\gamma}{\beta}\cdot\Delta_{f}-\mathcal{O}(\gamma^{2})\quad\text{as }\gamma\to0.
\]
\end{thm}

Even with an arbitrarily small competing incentive $\gamma>0$, no mechanism can prevent manipulation, unless it ignores the recommenders' reports, i.e., sensitivity function $\Lambda_{f}(k,\delta)=0$. Even if we co-design the decision $\textsc{F}$ and payment $\textsc{pay}$ function in some intricate way, also allowing  for discontinuities, 
 it is impossible to guarantee that there will be no manipulation. 
 
 Further, this result implies that any target decision function $f$ with a discontinuity of size $d$ (which implies \revisionredreplace{$\liminf_{\delta\downarrow0}\Lambda_f(1,\delta)\ge d$}{$\Lambda_{f}(1,0)=d$}) has manipulability of at least $\frac{d}{16}$ for any $\gamma>0$ and any budget $\beta$. By a discontinuity of size $d$, we mean that there is a point $a$ where $\left| \lim_{x \to a^-} f(x) - \lim_{x \to a^+} f(x) \right| = d.$
 %

 Finally,   for small competing incentives, $\gamma$, we see the manipulation scales linearly with $\gamma/\beta$ (the ratio between the competing incentive and the budget of the mechanism) and with the sensitivity $\Delta_{f}$ of $f$. We see that the quantity $\Delta_{f}$ captures the vulnerability of the target decision function, $f$, to manipulation. This sensitivity will play a central role in what follows. We now show how to express this lower bound for particular classes of target decision functions.

\subsubsection{Lower Bounds on Manipulation for Target Decision Functions with Bounds on the Derivative}

It is instructive to specialize \Cref{thm:lower} to the case where, in some part of its domain, the target decision function $f$ is known to have a lower bound on its derivative.\footnote{This does not impose any restrictions on the design space of the mechanism, i.e., the decision rule is still allowed to be arbitrary, even discontinuous.} The proof of the following result also illustrates how a random walk can be constructed in order to find the sensitivity $\Delta_{f}$ of a given $f$.
\begin{lem}
\label{lem:deltaLinearCube}Consider a target decision function $f$ with the following property: There is a  \mathauditadd{nonempty } set of recommenders $\mathcal{M}\subseteq[n]$, a subset of the domain $\mathcal{L}\subseteq\mathcal{Q}$, and an $L\in\mathbb{R}_{+}$, such that for each recommender $i\in\mathcal{M}$, the \dcpadd{target} decision function is monotonic and has slope at least $L>0$ \dcpadd{for reports} in $\mathcal{L}$, i.e. 
\[
\frac{\partial f}{\partial x_{i}}(x)\ge L\quad x\in\mathcal{L},\quad\text{or}\quad\frac{\partial f}{\partial x_{i}}(x)\le-L\quad x\in\mathcal{L}
\]
where $\mathcal{L}$ is a hypercube, $\mathcal{L}=\left\{ x\in\mathbb{R}^{n}:x_{-\mathcal{M}}=\mu_{-\mathcal{M}},\left\Vert x_{\mathcal{M}}-\mu_{\mathcal{M}}\right\Vert _{\infty}\le\sigma\right\} ,$ parametrized by $\sigma\in[0,1]$, and $\mu\in[\sigma,1-\sigma]^{n}$. For any such target decision function $f$, we have (using $m:=|\mathcal{M}|$) that \dcpadd{the sensitivity satisfies}
\begin{align*}
\Delta_{f} & \ge\frac{m}{\log(3\cdot m)}\cdot\sigma\cdot L^{2}.
\end{align*}
\end{lem}

\begin{proof}
 \mathauditreplace{If $\sigma=0$, this is immediate; otherwise, for }{For } a given $k\le m,\delta\le\sigma$, consider the random walk with $\mathcal{K}\subseteq\mathcal{M}$ some set of size $k$, $X^{0}=\mu$, and $Z_{i}=\delta$ for $i\in\mathcal{K}$ and $Z_{i}=0$ else. 
Then, for any such path, we have 
\begin{align*}
\sum_{j=1}^{k}\left|f(X^{j})-f(X^{j-1})\right| & \ge\delta\cdot L\cdot k.
\end{align*}

Hence, for any $k\le m$ and $\delta\le\sigma$, we have 
\dcpadd{sensitivity function} $\Lambda_{f}(k,\delta)\ge\delta\cdot L\cdot k.$ It then follows that \dcpadd{sensitivity}
\begin{align*}
\Delta_{f} & =\max_{k\in[n],\delta\in\mathauditreplace{(}{[}0,1]} \frac{\Lambda_{f}(k,\delta)^{2}}{k\cdot\delta\cdot\log(3\cdot k)}\\
 & \ge\max_{k\in[m],\mathauditadd{0<}\delta\le\sigma} \frac{\Lambda_{f}(k,\delta)^{2}}{k\cdot\delta\cdot\log(3\cdot k)}\\
 & \ge\max_{k\in[m],\mathauditadd{0<}\delta\le\sigma} \frac{\delta\cdot L^{2}\cdot k}{\log(3\cdot k)}\\
 & \ge\frac{\sigma\cdot L^{2}\cdot m}{\log(3\cdot m)}\quad(\text{with }\delta=\sigma,k=m).
\end{align*}
\end{proof}
Combining this result with \Cref{thm:lower}, we immediately obtain that, under the conditions of \Cref{thm:lower}, we have 
\[
\left|\textsc{F}(R)-f(Q)\right|\ge\frac{1}{16}\cdot\frac{\gamma}{\beta}\cdot\frac{m}{\log(3\cdot m)}\cdot\sigma\cdot L^{2}-O(\gamma^{2}).
\]

The larger the set $\mathcal{M}$ of recommenders \dcpadd{for whom the} slope of the target decision function is large, the larger the manipulation. Further, the manipulation also grows with the size $\sigma$ of the domain within which these recommenders have a large influence. The manipulation also grows quadratically with the \dcpadd{minimum} slope $L>0$ of \dcpadd{target decision function}$f$. \manuelreplace{Using a more sophisticated random walk, we can weaken the assumptions of \Cref{lem:deltaLinearCube}. Specifically, the following result requires the derivative bounds $L$ to hold only on the intersection of the hypercube in \Cref{lem:deltaLinearCube} with a central slab that restricts the average displacement of the recommenders' reports from $\mu$, rather than on the entire hypercube. It therefore applies to a larger class of target decision functions:}{Using a more sophisticated random walk, we can obtain the following more general result:}
\begin{lem}
\label{lem:deltaLinear}Consider a decision function $f$ with the following property: There is a  \mathauditadd{nonempty } set of recommenders $\mathcal{M}\subseteq[n]$, a subset of the domain $\mathcal{L}\subseteq\mathcal{Q}$, and an $L\in\mathbb{R}_{+}$, such that for each recommender $i\in\mathcal{M}$, the decision function is monotonic and has slope at least $L$ in $\mathcal{L}$, i.e. 
\[
\frac{\partial f}{\partial x_{i}}(x)\ge L\quad x\in\mathcal{L},\quad\text{or}\quad\frac{\partial f}{\partial x_{i}}(x)\le-L\quad x\in\mathcal{L},
\]
where $\mathcal{L}$ is the set 
\[
\mathcal{L}:=\left\{ x\in\mathbb{R}^{n}:x_{-\mathcal{M}}=\mu_{-\mathcal{M}},\left\Vert x_{\mathcal{M}}-\mu_{\mathcal{M}}\right\Vert _{\infty}\le\sigma,\left|\frac{1}{|\mathcal{M}|}\sum_{i\in\mathcal{M}}\left(x_{i}-\mu_{i}\right)\right|\le\epsilon\right\} ,
\]
and parametrized by $\sigma\in[0,1],\mu\in[\sigma,1-\sigma]^{n}$, and $\epsilon\in[0,1]$. For any target decision function $f$ with this property, we have, using $m:=|\mathcal{M}|$, \dcpadd{that sensitivity satisfies}
\[
\Delta_{f}\ge\manuelmathreplace{\frac{1}{4}}{\frac{1}{3}}\cdot\frac{m}{\log(3\cdot m)}\cdot\sigma\cdot L^{2}\cdot\min\left\{ \frac{\sqrt{m}\cdot\epsilon}{\sigma},1\right\} .
\]
\end{lem}
\manueladd{When $\epsilon\ge\sigma$, the final constraint in the definition of $\mathcal{L}$ is redundant, because $\left\Vert x_{\mathcal{M}}-\mu_{\mathcal{M}}\right\Vert_{\infty}\le\sigma$ implies $\left|\frac{1}{m}\sum_{i\in\mathcal{M}}(x_i-\mu_i)\right|\le\sigma$. Thus, in this case the region $\mathcal{L}$ coincides with the hypercube in \Cref{lem:deltaLinearCube}, and the lower bound in \Cref{lem:deltaLinear} recovers the bound in \Cref{lem:deltaLinearCube} up to a factor of $1/4$. For $\epsilon<\sigma$, \Cref{lem:deltaLinear} requires the derivative bounds to hold only within a central slab of the hypercube, at the cost of the additional factor $\min\{\sqrt{m}\epsilon/\sigma,1\}$.}

We  study the implications of this result for specific decision functions in \Cref{sec:bldf}. But first, we give a novel mechanism and prove an upper bound on its degree of manipulation. The negative results from the present section are helpful for designing mechanisms, since they serve to give a bound on what is possible.

\subsection{The Adaptive-Payment Mechanism}

We start by defining a second, more conservative notion of sensitivity that will be helpful for our upper bounds.
\begin{defn}[\mjyadd{Differentiable }sensitivity $\nabla_{f}$]
\label{def:nabla}For a target decision function $f$ that is \revisionredreplace{Lipschitz}{differentiable almost everywhere}, we define {\em differentiable sensitivity},
\[
\nabla_{f}:=\left(m\cdot\manuelmathreplace{\max_{i\in[n]}\int_{0}^{1}\max_{q_{-i}\in\mathcal{Q}_{-i}}\left|\partial_{x}f(x,q_{-i})\right|dx}{\int_{0}^{1}\max_{i\in[n],q_{-i}\in\mathcal{Q}_{-i}}\left|\partial_{q_{i}}f(q)\right|dq_{i}}\right)\cdot\left(\max_{i\in[n],q\in\mathcal{Q}}\left|\partial_{q_{i}}f(q)\right|\right),
\]
where $m$ is the number of recommenders that $f$ has nonzero dependence on, and $\partial$ denotes \manuelreplace{the generalized derivative with the largest absolute value}{the largest generalized derivative} (identical to the derivative wherever $f$ is differentiable).
\end{defn}

\dcpadd{The first term in differentiable sensitivity, $\nabla_f$, is $m$ times what is 
in effect an upper bound on the total variation on target decision function $f$ due to 
any single recommender. The second term is the maximum absolute partial derivative of 
the target decision function $f$ with respect to the report of any recommender.}


We  see in \Cref{sec:bldf} that \dcpadd{differentiable sensitivity} $\nabla_{f}$ is equivalent to \dcpadd{sensitivity} $\Delta_{f}$ (up to log factors) for a family of piecewise-linear \dcpadd{target decision} functions. We now give a novel mechanism, 
\dcpadd{the adaptive-payment mechanism,}
and show it achieves the lower bounds from \Cref{sec:lower} for a particular class of target \dcpadd{decision} functions. The key idea is to define the payment rule to concentrate the incentives where the target  
\dcpadd{decision} function $f$ has a large slope. 
\begin{defn}[Adaptive-payment mechanism]
\label{def:adaPay}Given a budget $\beta$ and a target decision function $f$ that is \revisionredreplace{Lipschitz}{differentiable almost everywhere}, we define the
\dcpadd{{\em adaptive-payment mechanism}, $\textsc{mech}=(\textsc{F}=f,\textsc{pay}=(f,\beta)\textsc{-ada-pay})$, 
with  decision rule, $F$, set  to the target decision function $f$, and payment rule, $\textsc{pay}$,
set to the \emph{adaptive-payment rule} $(f,\beta)\textsc{-ada-pay}$.} 
Let $\mathcal{M}\subseteq[n]$ be the set of recommenders on which $f$ has nonzero dependence. 
Define
\[
\dcpadd{(f,\beta)}\textsc{-ada-pay}_{i}(r_{i},o)=\begin{cases}
\frac{\beta}{3\cdot|\mathcal{M}|}\cdot\left(\frac{\int_{0}^{r_{i}}\delta_{i}(x)(o-x)\cdot dx}{\int_{0}^{1}\delta_{i}(x)\cdot dx}+1\right) & \text{if }i\in\mathcal{M}\\
\frac{\beta}{6\cdot n}\cdot\left(\int_{0}^{r_{i}}(o-x)\cdot dx+1\right) & \text{otherwise,}
\end{cases}
\]
with $\delta_{i}:[0,1]\to\mathbb{R}_{>0}$ being a  \mathauditdelete{continuous } function such that 
\begin{align*}
\delta_{i}(r_{i}) & =\max_{r_{-i}}\left|\partial_{r_{i}}f(r)\right|+\manuelmathreplace{\int_{0}^{1}\max_{r_{-i}}\left|\partial_{x}f(x,r_{-i})\right|dx}{\int_{0}^{1}\max_{r_{-i}}\left|\partial_{r_{i}}f(r)\right|dr_{i}},
\end{align*}
where $\partial$ denotes \manuelreplace{the generalized derivative with the largest absolute value}{the largest generalized derivative} (just the derivative wherever $f$ is differentiable).
\end{defn}

%

The adaptive-payment mechanism  pays 
recommenders who are outside of $\mathcal{M}$ and even with no influence on the decision 
to preserve strict properness for all recommenders, as required by \Cref{def:mechanism}.

To gain an intuition for this payment rule, consider a single recommender,
with true belief $q$. \dcpadd{In the following, we write $(f,\beta)\textsc{-ada-pay}$ as $\textsc{ada-pay}$.}
The derivative of the
expected payment, under the adaptive-payment rule, has the form 
\begin{align}
\partial_{r}\mathbb{E}_{o\sim q}\left[\textsc{ada-pay}(r,o)\right] & =\partial_{r}\left(\frac{\beta}{3}\cdot\frac{\int_{0}^{r}\delta(x)(q-x)\cdot dx}{\int_{0}^{1}\delta(x)\cdot dx}\dcpadd{+1}\right) \label{eq:penultstep} \\
 & =\frac{\beta}{3}\cdot\frac{\delta(r)(q-r)}{\int_{0}^{1}\delta(x)\cdot dx}\notag \\
 & =\frac{\beta}{6}\cdot\left(1+\frac{\left|\partial_{r}f(r)\right|}{\int_{0}^{1}\left|\partial_{x}f(x)\right|dx}\right)(q-r), \label{eq:laststep}
\end{align}
\dcpadd{where~\Cref{eq:penultstep} follows by the Fundamental Theorem of Calculus 
and~\Cref{eq:laststep} follows
because $\delta(r)=|\partial_rf(r)|+\int_0^1|\partial_xf(x)|dx$ and $\int_{0}^{1}\delta(x)\cdot dx=\int_0^1|\partial_xf(x)|dx+\int_{y=0}^1\left[\int_{x=0}^1|\partial_xf(x)|dx\right]dy=2\int_0^1|\partial_xf(x)|dx$.} 

We see that the derivative is zero only if $q=r$,
and hence the rule is proper. The novelty of the payment rule is that
\dcpadd{the derivative of the expected payment} scales with $\left|\partial_{r}f(r)\right|$, i.e.,
it concentrates \dcpadd{payment} incentives where \dcpadd{reports}; have a large influence
on the decision.
\begin{example}
    \label{ex:dam-payment-rule}
 \dcpadd{Building from \Cref{ex:dam}, suppose now that the city government 
 plans to invest between 30 and 130 million USD
into flood barriers, proportional to the estimated probability $\bar{Q}=\frac{1}{n}\sum_{i=1}^nQ_{i}$ of an extreme weather event; i.e.,
\begin{align}
\label{eq:simplef}
f(Q) &=30+100\cdot\bar{Q},
\end{align}
in millions of dollars. Here, the city wants to consider the average
of the beliefs of $n$ experts.}
 \if 9
 
 with probability equal to the average report of the recommenders. That is, the target decision function is
    \begin{equation}
        f(r) = \frac{\sum_{i \in \mathcal{M}} r_i}{|\mathcal{M}|}
    \end{equation}
\fi
\dcpadd{Suppose the city has a budget of 5 million USD
to spend on elicitation  from these experts $(\beta=5)$, and  
$n=20$.}
%
\dcpadd{Following the adaptive-payment mechanism, the
decision rule will be $F:= f$,  and the same as the target decision
function.} 
\mjyadd{For the payment rule, 
we have
\begin{align*}
\delta_{i}(r_{i}) & =\max_{r_{-i}}\left|\partial_{r_{i}}f(r)\right|+\int_{0}^{1}\max_{r_{-i}}\left|\partial_{x}f(x,r_{-i})\right|dx =
\frac{100}{n}+\int_0^1\frac{100}{n}dx=\frac{200}{n}.
\end{align*}}

Since every expert influences the decision,  $\mathcal{M}=[n]$ in
the definition of the payment rule, and we 
have that the payment to  recommender $i$ is
\mjyadd{\begin{align}
(f,\beta)\textsc{-ada-pay}_{i}(r_{i},o) &=
\frac{\beta}{3n}\cdot\left(\frac{\int_{0}^{r_{i}}\delta_{i}(x)(o-x) dx}{\int_{0}^{1}\delta_{i}(x) dx}+1\right)\notag \\
&= \frac{\beta}{3n} \cdot \left( \int_0^{r_i}(o - x) dx +1 \right)\label{eq:mid} \\
&= \frac{\beta}{3n}\cdot\left(o \cdot r_i - \frac{r_i^2}{2} + 1 \right),\label{eq:mid2}
\end{align}}
\dcpadd{where $o$ is the outcome ($o=1$ if a severe flood event), 
and~\Cref{eq:mid} follows because
$\int_{0}^{r_{i}}\delta_{i}(x)(o-x) dx=(200/n)\int_0^{r_i}(o-x)dx$
and $\int_{0}^{1}\delta_{i}(x) dx=200/n$.

In this simple example, we recover the form of the
strictly proper, {\em quadratic scoring rule}~\citep{GneitingTilmann2007SPSR}, which would make payment $s_q(r_i,o)=-(1/2)(r_i-o)^2=o\cdot r_i-(1/2)r_i^2-(1/2)o^2$ and neither $(1/2)o^2$ in the quadratic scoring rule or the ``$+1$" term in~\Cref{eq:mid2} affect the incentives of the recommender since they
do not depend on report $r_i$.

With 20 recommenders, budget $\beta=5$, and a report from
recommender $i$ of $r_i = 0.9$ for the probability
that a severe flood occurs within five years,
then in the event of such a flood, the
payment to recommender $i$ is  $\frac{5}{3\cdot 20}\left(1\cdot 0.9-\frac{0.9^2}{2}+1\right)=\frac{1}{12}(1.495)=0.1246$
million USD. 
If a severe flood does not occur, the payment to the recommender
is $\frac{5}{3\cdot 20}\left(0\cdot 0.9-\frac{0.9^2}{2}+1\right)=\frac{1}{12}(0.595)=0.0496$
million USD.}
\end{example}

Using the adaptive-payment rule, we obtain a mechanism, \dcpadd{for a suitably differentiable target decision function}, that achieves the following upper bound on \dcpadd{the effect of} manipulation: 
\begin{thm}
\label{thm:upper}Suppose that we are given target decision function $f$ that is \revisionredreplace{Lipschitz}{differentiable almost everywhere} and \dcpadd{has differentiable sensitivity $\nabla_f$}.
Using the adaptive-payment mechanism with budget $\beta\mathauditadd{>0}$, 
then for any  \dcpadd{bound on competing incentive} $\gamma\ge0$,
any prior $D$ \dcpadd{that satisfies $\gamma$, any recommender subjective posterior beliefs $Q$}, 
and any \dcpadd{$2\textsc{rat}$} recommender reports, $\dcpadd{R}$,
we have 
\begin{align*}
\left|\textsc{F}(R)-f(Q)\right| & \le6\cdot\frac{\gamma}{\beta}\cdot\nabla_{f},
\end{align*}
with probability 1.
\dcpadd{Here, $F$ is the decision rule used in the adaptive-payment mechanism, which is  $F:=f$.}
\end{thm}

This dependence on \mjyreplace{$\frac{\gamma}{\beta}$ times the differentiable sensitivity $\nabla_f$ is similar to }{$\gamma$ and $\beta$ matches }the way the lower bound \mjyadd{depends on $\frac{\gamma}{\beta}$ times sensitivity $\Delta_f$ in} (\Cref{thm:lower}). To gain further understanding, we  focus on a particular class of decision functions. 
\begin{lem}
\label{lem:nablaLinearCube}Consider a target decision function $f$ with the following properties: It is \revisionredreplace{Lipschitz}{differentiable almost everywhere}, and there is some set of recommenders $\mathcal{M}\subseteq[n]$ of size $m$ and some parameters $\mu\in\mathcal{R},\sigma\in[0,1]$ (with $\mu\pm\sigma\in\mathcal{R}$), \dcpadd{and $L\geq 0$}
such that the derivative of $f$ is upper bounded by
\begin{align*}
\left|\frac{\partial f}{\partial x_{i}}(x)\right| & \le\begin{cases}
L & \text{if }i\in\mathcal{M}\text{ and }x\in\mathcal{L},\\
0 & \text{otherwise}
\end{cases}
\end{align*}
where $\mathcal{L}:=\left\{ x\in\mathbb{R}^{n}:\left\Vert x-\mu\right\Vert _{\infty}\le\sigma\right\} $ is a hypercube with center $\mu$ and side length $\mathauditadd{2}\sigma$.

For any such decision function, we have \dcpadd{differentiable sensitivity,}
\begin{align*}
\nabla_{f} & \le2\cdot m\cdot\sigma\cdot L^{2}.
\end{align*}
\end{lem}
\begin{proof}
Such a target decision function $f$ satisfies for all $i$
\begin{align*}
\int_{0}^{1}\max_{r_{-i}}\left|\partial_{x}f(x,r_{-i})\right|dx & \le2\cdot\sigma\cdot L\quad\text{and}\quad\max_{r}\left|\partial_{r_{i}}f(r)\right|\le L.
\end{align*}
Further,   $f$ only depends on $m$ recommenders. Hence, it follows immediately from~\Cref{def:nabla} that 
\begin{align*}
\nabla_{f} & \le\left(m\cdot2\cdot\sigma\cdot L\right)\cdot\left(L\right)=2\cdot\sigma\cdot m\cdot L^{2}.
\end{align*}
\end{proof}

It follows immediately from  \Cref{thm:upper} that
 \dcpadd{the adaptive-payment mechanism 
with budget $\beta>0$ and bound on competing incentive $\gamma$ satisfies, and
with a target decision function $f$ in the class of~\Cref{lem:nablaLinearCube},
has the property that
for 
any subjective posterior beliefs $Q$ and any $2\textsc{rat}$ reports $R$,}
\begin{align*}
\left|\textsc{F}(R)-f(Q)\right| & \le12\cdot\frac{\gamma}{\beta}\cdot m\cdot\sigma\cdot L^{2}.
\end{align*}

\dcpadd{Here, $m$ is the cardinality of the set of recommenders in~\Cref{lem:nablaLinearCube} on which 
the target decision function depends, but only in a limited way as prescribed by max slope $L\geq 0$
and over an interval that depends on $\sigma\geq 0$.}
This bound matches the lower bound below \Cref{lem:deltaLinearCube} up to log factors. Hence, for decision functions that are linear on some hypercube $\mathcal{L}$ and constant elsewhere, the mechanism is essentially optimal and the lower and upper bounds are tight.

\subsection{Bounded-Linear Target Decision Functions}

\label{sec:bldf} We  now define a family of piecewise-linear target decision functions 
that allow us to  illustrate and compare the general results we obtained in the previous sections. 
For convenience, we will use the notation $\left\llbracket x\right\rrbracket _{y}^{z}$ (for some $x,z\ge y\in\mathbb{R}$) to denote the value of $x$ clipped between $y$ and $z$, i.e. $\left\llbracket x\right\rrbracket _{y}^{z}:=\min\left\{ \max\left\{ x,y\right\} ,z\right\} $.
\begin{defn}[Bounded-linear target decision function]
\label{def:bldf} 
A target decision function $f$ is {\em bounded-linear} if it can be written as
\[
f(q)=\bar{f}+\left\llbracket \ell\cdot\frac{1}{m}\sum_{i\in\mathcal{M}}\left\llbracket q_{i}-t\right\rrbracket _{-\alpha}^{\alpha}\right\rrbracket _{-\varepsilon}^{\varepsilon},
\]
with parameters $\mathauditadd{\varnothing\ne}\mathcal{M}\subseteq[n]$ (we use $m:=|\mathcal{M}|$), $\bar{f}\in\mathbb{R}$, $t\in\left[\frac{1}{10},\frac{9}{10}\right]$, $\alpha\in\mathauditreplace{(}{[}0,1]$, $|\ell|\le\frac{\sqrt{m}}{\alpha}$, $\varepsilon\ge\frac{1}{10}\cdot\alpha\cdot\mathauditadd{|}\ell\mathauditadd{|}$. Hence, target decision function $f$ is parametrized by $(\mathcal{M},\bar{f},\varepsilon,\ell,\alpha,t)$.
\end{defn}

\dcpadd{This is a family of piecewise-linear functions and it is Lipschitz differentiable
almost everywhere. Thus, differentiable sensitivity is well-defined for any bounded-linear target decision function.} 
In effect, recommenders' reports are  constrained to $q_i \in [t - \alpha, t + \alpha].$ 
 For a specific example, with $\mathcal{M}=[n]$ and $\alpha=1$, $\varepsilon=\infty$, we obtain a target decision function that is linear in the average report $\bar{q}=\mathauditadd{\frac{1}{n}}\sum_{i\in[n]}q_{i}$,
\[
f(q)=\bar{f}+\ell\cdot(\bar{q}-t).
\]

We will see other examples of  bounded-linear \dcpadd{target} decision functions below, but first we derive some general results. Recall that we introduced two notions of sensitivity, $\Delta_{f}$ and $\nabla_{f}$. We used $\Delta_{f}$ in our lower bounds on manipulation and \mjyadd{differentiable sensitivity}, $\nabla_{f}$, in our upper bounds. For bounded-linear \dcpadd{target} decision functions, the two notions \dcpadd{of sensitivity} are equivalent up to log factors and hence our bounds are tight. \manueladd{To see this, note that both notions are finite and the conditions of \Cref{thm:lower,thm:upper} hold. Applying \Cref{thm:lower} to the adaptive-payment mechanism and comparing it with \Cref{thm:upper} gives $\Delta_f\le6\cdot 16 = 96\cdot\nabla_f$. The following lemma establishes the reverse comparison up to a logarithmic factor.}
\begin{lem}
\label{lem:dcptight}
For any bounded-linear target decision function, $f$, as defined in \Cref{def:bldf}, \dcpadd{and with parameters $\alpha, \ell$, and $m$, we have the following relationship between differentiable sensitivity and sensitivity}:
\begin{align*}
\nabla_{f} & \le2\cdot\frac{\alpha\cdot\ell^{2}}{m}\le\manuelmathreplace{80}{60}\cdot\log(3\cdot m)\cdot\Delta_{f}.
\end{align*}
\end{lem}
\begin{proof}
We first prove an upper bound on \mjyadd{differentiable sensitivity}, $\nabla_{f}$, and then a lower bound on \dcpadd{sensitivity,} $\Delta_{f}$, and then we relate them. It follows straightforwardly from \revisionredreplace{\mbox{\Cref{def:nabla}}}{\mbox{\Cref{lem:nablaLinearCube}}}\hskip0pt\mathauditdelete{, with $L=\frac{\ell}{m}$ and $\sigma=\alpha$,}\ that
\begin{align}
\nabla_{f} & \le2\cdot\frac{\alpha\cdot\ell^{2}}{m}.\label{eq:nablalower}
\end{align}
\mathauditadd{If $\ell=0$, the claim is immediate; by symmetry, assume $\ell>0$.}

We now proceed to prove a lower bound on \dcpadd{sensitivity,} $\Delta_{f}$. 
To apply \Cref{lem:deltaLinear}, note that any $q\in\mathcal{Q}$ such that
\begin{align*}
\left\Vert q_{\mathcal{M}}-t\right\Vert _{\infty} & \le\min\{t,1-t,\alpha\}\quad\text{ and }\quad\left|\frac{1}{m}\sum_{i\in\mathcal{M}}q_{i}-t\right|\le\varepsilon/\ell
\end{align*}
we have $\frac{\partial f}{\partial q_{i}}(q)=\frac{\ell}{m}.$ Further, \manuelreplace{since}{it follows from} \Cref{def:bldf} \manuelreplace{specifies that $t,1-t\ge1/10\ge\alpha/10$ and $\alpha\ge\alpha/10$, we have}{that} $\min\{t,1-t,\alpha\}\manuelmathreplace{\ge}{\le}\frac{\alpha}{10}$, and we can hence apply \Cref{lem:deltaLinear}, with $\sigma\to\frac{\alpha}{10},L\to\frac{\ell}{m},\epsilon\to\varepsilon/\ell$, and obtain \dcpadd{sensitivity}
\begin{align*}
\Delta_{f} & \ge\manuelmathreplace{\frac{1}{4}}{\frac{1}{3}}\cdot\frac{m}{\log(3\cdot m)}\cdot\sigma\cdot L^{2}\cdot\min\left\{ \frac{\sqrt{m}\cdot\epsilon}{\sigma},1\right\} =\manuelmathreplace{\frac{1}{40}}{\frac{1}{30}}\cdot\frac{1}{\log(3\cdot m)}\cdot\frac{\alpha\cdot\ell^{2}}{m}\cdot\min\left\{ 10\cdot\frac{\varepsilon\cdot\sqrt{m}}{\alpha\cdot\ell},1\right\} ,
\end{align*}
and since \Cref{def:bldf} specifies that $\text{\ensuremath{\varepsilon}}\ge\frac{1}{10}\cdot\alpha\cdot\ell$, we have 
\[
\Delta_{f}\ge\manuelmathreplace{\frac{1}{40}}{\frac{1}{30}}\cdot\frac{1}{\log(3\cdot m)}\cdot\frac{\alpha\cdot\ell^{2}}{m}.
\]
The statement in the lemma follows straightforwardly by comparing this inequality with \Cref{eq:nablalower}. 
\end{proof}

We  now state both the lower bound on manipulation (\Cref{thm:lower}) and the upper bound (\Cref{thm:upper}) in terms of \dcpadd{sensitivity}, $\Delta_{f}$. 
\begin{thm}
\label{thm:bldfTight} 
Consider a bounded-linear target decision function $f$ (\Cref{def:bldf}) \dcpadd{with sensitivity $\Delta_f$}. For any mechanism $\textsc{mech}=(\text{\textsc{F}},\textsc{pay})$ with budget $\beta\mathauditadd{>0}$ and any total competing incentive $\gamma\mathauditadd{>0}$, there exists a prior $D$ \dcpadd{satisfying $\gamma$ and recommender subjective posterior beliefs $Q$} such that \dcpadd{in every} BNE, \dcpadd{recommender} reports $R$ have the effect, with probability 1,
\begin{align*}
\left|\textsc{F}(R)-f(Q)\right| & \ge\frac{1}{16}\cdot\frac{\gamma}{\beta}\cdot\Delta_{f}-\mathcal{O}(\gamma^{2})\\
\text{(with }m,\alpha,\ell\text{ being params of }f\text{, see \Cref{def:bldf})} & \ge\manuelmathreplace{\frac{1}{640}}{\frac{2}{10^{3}}}\cdot\frac{1}{\log(3\cdot m)}\cdot\frac{\gamma}{\beta}\cdot\frac{\alpha\cdot\ell^{2}}{m}-\mathcal{O}(\gamma^{2})\\
 & \ge\manuelmathreplace{\frac{1}{1280}}{\frac{1}{10^{3}}}\cdot\frac{1}{\log(3\cdot m)}\cdot\frac{\gamma}{\beta}\cdot\nabla_{f}-\mathcal{O}(\gamma^{2})\quad\text{as }\gamma\to0.
\end{align*}

Using the adaptive-payment mechanism with budget $\beta\mathauditadd{>0}$, we have for any \dcpadd{total competing incentive} $\gamma\ge0$, any prior $D$ \dcpadd{satisfying}   $\gamma$, \dcpadd{any recommender subjective posterior beliefs $Q$},
and any \dcpadd{$2\textsc{rat}$} \dcpadd{recommender reports $R$},
\begin{align*}
\left|\textsc{F}(R)-f(Q)\right| & \le6\cdot\frac{\gamma}{\beta}\cdot\nabla_{f}\le12\cdot\frac{\gamma}{\beta}\cdot\frac{\alpha\cdot\ell^{2}}{m}\le\manuelmathreplace{480}{360}\cdot\log(3\cdot m)\cdot\frac{\gamma}{\beta}\cdot\Delta_{f},
\end{align*}
with probability 1.
\end{thm}

Hence, the two bounds are equivalent, up to a log factor, and for the family of bounded-linear target decision functions the quantities $\Delta_{f}$ and $\nabla_{f}$ are essentially equivalent, and each one  fully captures how  \dcpadd{function $f$ depends on reports}. Further, Theorem~\ref{thm:bldfTight} implies that the adaptive-payment mechanism   is optimal (up to log factors) for bounded-linear \dcpadd{target} decision functions. 

The dependence of the \dcpadd{effect of the} manipulation on the parameters of the decision function is $\frac{\alpha\cdot\ell^{2}}{m}$. With $\alpha=1$, we obtain a linear target decision function in the mean estimator, $\bar{q}$, i.e., 
\[
f(q)=\bar{f}+\left\llbracket \ell\cdot \mathauditadd{(} \bar{q}-t \mathauditadd{)} \right\rrbracket _{-\varepsilon}^{\varepsilon}
\]
with manipulation $\frac{\ell^{2}}{m}$. We can interpret an $\alpha<1$ as a more robust estimator, where we limit the influence of each recommender to $\alpha$, and obtain a better manipulation bound $\frac{\alpha\cdot\ell^{2}}{m}$.

\section{Manipulation due to Bribery\label{sec:briber}}

In this section, we study the second setting, where the competing incentives are not intrinsic to the recommenders but stem from a rational third party,  the briber. In this setting, 
we  give necessary and sufficient conditions for avoiding manipulation entirely, in terms of the sensitivity of the decision function $f$, the strength of the incentive of the briber, and the budget of the mechanism.

\subsection{Setting}

We  adapt the setting for intrinsic competing incentives defined in \Cref{sec:game} to incentives stemming from a rational briber. The {\em briber's type} $(V,W)$ consists of the value $V\in\mathcal{V}:=\mathbb{R}_{+}$ the briber gains from an increase in the decision $A$ (in our example, the construction company gaining value from investments into flood barriers), where $W\in\mathcal{W}:=2^{[n]}$ denotes the set of recommenders the briber can access.

The competing incentive $C$ is  determined by the briber and not part of the recommenders' types. \dcpadd{As a result,} each {\em recommender's type} \dcpadd{in this second setting is} $(P_{i},Y_{i})$ \dcpadd{and} consists of the subjective belief $P_{i}\in\mathcal{P}_{i}:=\textsc{distr}(\mathcal{O}|\mathcal{Y})$ \dcpadd{on outcome $O$}
and private signal $Y_{i}\in\mathcal{Y}_{i}$. 
Agent types are private, \dcpadd{now including the recommenders' types and the briber's type}, 
and   drawn from a common prior, $D\in\textsc{distr}(\mathcal{P}\times\mathcal{Y}\times\mathcal{V}\times W)$,  independent across agents and subtypes; i.e., prior $D$ can be written as a product, 
\[
D(p,y,v,w)=\prod_{i\in[n]}D_{i}^{p}(p_{i})\cdot\prod_{i\in[n]}D_{i}^{y}(y_{i})\cdot D^{v}(v)\cdot D^{w}(w),
\]
for some $D_{i}^{p}\in\textsc{distr}(\mathcal{P}_{i})$, $D_{i}^{y}\in\textsc{distr}(\mathcal{Y}_{i})$, $D^{v}\in\textsc{distr}(\mathcal{V})$, $D^{w}\in\textsc{distr}(\mathcal{W})$. 
\dcpadd{As in the first setting, while a recommender's subjective belief $P_i(O|Y)$ on outcome $O$
is sampled independently from its  realized signal $Y_i$, 
its subjective posterior belief, $Q_i$, is formed
after observing $Y_i$, and thus depends on both $P_i$ and $Y_i$.} 
In \dcpadd{this second setting with the} presence of a briber, the  game \dcpadd{played by the recommenders and the briber} now proceeds as follows:
\begin{enumerate}
\item The \dcpadd{recommender} types \dcpadd{$P,Y$ and the briber's type $V,W$} are drawn from the prior $P,Y,V,W\sim D$.
\item Each recommender $i$ observes their type $(P_{i},Y_{i})$, and the briber observes \dcpadd{its} type $V,W$.
\item \emph{The briber determines the bribe profile $C\in\mathcal{C}$ (note that it must satisfy $C_{i}=0$ for all $i\notin W$).}
\item Each recommender $i$ observes their bribe $C_{i}$.
\item Each recommender \dcpadd{forms its subjective posterior belief $Q_i$ regarding the probability of $O=1$ and} makes a report $R_{i}\in\mathcal{R}_{i}$.
\item The mechanism takes action $A=\textsc{F}(R)\in \mathbb{R}_{\ge0}$. 
\item The outcome $O\in\mathcal{O}$ is observed. 
\item Each recommender $i$ receives utility $U_{i}=\textsc{pay}_{i}(R,O)+C_{i}\cdot A$. 
\item \emph{The briber receives utility $U^{\textsc{bri}}:=\left(V-\sum_{i\in[n]}|C_{i}|\right)\cdot A$.}
\end{enumerate}

We again formulate this game in the Bayesian setting, and express each agent's subjective expected utility given the type profile $p,y,v,w$ and action profile $r,c$. Recommender $i$'s subjective expected utility  depends on $p_{i},y,c_{i},r$:
\begin{align}
\textsc{util}_{i}\left(p_{i},y,c_{i},r\right): & =\mathbb{E}_{O\sim p_{i}(\cdot|y)}\left[\textsc{pay}_{i}(r,O)+c_{i}\cdot\textsc{F}(r)\right],\label{eq:recUtil-1}
\end{align}
which is identical to the recommenders' utility  with intrinsic competing incentives~\Cref{eq:recUtil}, 
except that $c_{i}$ now represents the briber's action rather than $i$'s own type.

The briber's utility only depends on briber's own type, $v,w$, the bribe profile $c$, and the report profile $r$, with
\[
\textsc{util}^{\textsc{bri}}\left(v,w,c,r\right):=\left(v-\sum_{i\in[n]}|c_{i}|\right)\cdot\textsc{F}(r).
\]

\subsubsection{Common Knowledge}
The prior $D$ and the mechanism $\textsc{mech}$ are common knowledge to all recommenders and the briber.
The principal does not know the prior $D$.  

\subsubsection{Solution Concepts}
For recommenders, the set of possible strategies $\mathcal{S}$ is the same as in the game without briber, i.e. 
\[
s_{i}\in\mathcal{S}_{i}:=\textsc{distr}\left(\mathcal{R}_{i}^{\mathcal{P}_{i}\times\mathauditreplace{\mathcal{Y}_{i}\times\mathcal{C}_{i}}{\mathcal{C}_{i}\times\mathcal{Y}_{i}}} \right).
\]
For the briber, the strategy has the form
\[
s^{\textsc{bri}}\in\mathcal{S}^{\textsc{bri}}:=\textsc{distr}\left(\mathcal{C}^{\mathcal{V}\times\mathcal{W}}\right),
\]
and we will use $s_{i}^{\textsc{bri}}$ to denote the bribe paid to recommender $i$. The briber's expected utility, given the briber type and a strategy profile, is 
\begin{align*}
\textsc{eutil}^{\textsc{bri}}\left(s,s^{\textsc{bri}},v,w\right): & =\mathbb{E}\left[\textsc{util}^{\textsc{bri}}\left(V,W,C,R\right)\big|s,s^{\textsc{bri}},v,w\right]\\
 & =\mathbb{E}_{P,Y,V,W\sim D,\hat{S}\sim s,\hat{S}^{\textsc{bri}}\sim s^{\textsc{bri}}}\left[\textsc{util}^{\textsc{bri}}\left(V,W,\hat{S}^{\textsc{bri}}(V,W),\hat{S}\left(P,Y,\hat{S}^{\textsc{bri}}(V,W)\right)\right)\right].
\end{align*}

Recommender $i$'s expected utility given their type $p_{i},y_{i}$, bribe $c_{i}$ (which they know at the time of reporting), and the strategy profile, is 
\begin{align*}
\textsc{eutil}_{i}\left(s,s^{\textsc{bri}},p_{i},y_{i},c_{i}\right) & :=\mathbb{E}\left[\textsc{util}_{i}\left(P_{i},Y,C_{i},R\right)\big|s,s^{\textsc{bri}},p_{i},y_{i},c_{i}\right].
\end{align*}

We express this expectation in full in \Cref{ap:perfect}. 
 As a solution concept in this sequential move game between the briber and the recommenders, we adopt the notion of a {\em perfect Bayesian equilibrium}, where a belief is said to be consistent if it follows from Bayes rule whenever it is well-defined.
 For   ill-defined cases (i.e., when the event we condition on has zero probability, as happens when $c_{i}$ is not in the support of the mixed strategy of the briber $s^{\textsc{bri}}$), 
any belief is said to be consistent.
We discuss the technical details of what consistency implies regarding recommenders' beliefs in \Cref{ap:perfect}. 

\begin{defn}[Perfect Bayesian equilibrium (PBE)]
\label{def:pbe} A strategy profile $s,s^{\textsc{bri}}$ is a {\em perfect Bayesian equilibrium} if 
\[
\textsc{eutil}^{\textsc{bri}}\left(s,s^{\textsc{bri}},v,w\right)\ge\textsc{eutil}^{\textsc{bri}}\left(s,s'^{\textsc{bri}},v,w\right)\quad\forall s'^{\textsc{bri}},v,w
\]
and for each recommender $i$, there is a consistent belief such that the implied expected utility satisfies 
\[
\textsc{eutil}_{i}\left(s,s^{\textsc{bri}},p_{i},y_{i},c_{i}\right)\ge\textsc{eutil}_{i}\left(s_{i}',s_{-i},s^{\textsc{bri}},p_{i},y_{i},c_{i}\right)\quad\forall s_{i}',p_{i},y_{i},c_{i}.
\]
\end{defn}
For our rationalizability notion, \dcpadd{$2\textsc{rat}$}, we do not need to introduce such a refinement, as we only use it for positive results
and these are not weakened by including non-credible strategy profiles.

\subsection{Results: Elicitation in the Presence of a Briber}

We know that whenever the competing incentive of the recommender is non-zero and the decision function is nontrivial, then no mechanism is truthful. But what if the competing incentive comes from a rational briber? In this case we recover a positive result, by using a mechanism in which it is not in the interest of the briber to make a bribe. 

We characterize the conditions under which a target decision function $f$ can be implemented, by providing a necessary and then a sufficient condition on the sensitivity of $f$, the mechanism's budget $\beta$, and the briber incentive $\upsilon$, which we define as $\upsilon:=\max_{v\in\textsc{supp}(D^{v})}v$.
\begin{thm}
\label{thm:briberNecessary} Given target decision function $f$ with sensitivity $\Delta_{f}$ (\Cref{def:sensitivityLower}), and the briber setting with briber incentive $\upsilon$, then for any mechanism $\textsc{mech}$ with budget $\beta$ such that
\begin{align*}
\frac{\beta}{\upsilon}\le & \frac{1}{\revisionredreplace{16}{8}\cdot\max_{q\in\mathcal{Q}}f\left(q\right)}\cdot\Delta_{f},
\end{align*}
there exists a \dcpadd{prior  $D$ and recommender subjective posterior beliefs $Q$}
such that \dcpadd{in every} PBE (\Cref{def:pbe}), \dcpadd{recommender reports $R$} satisfy
$F(R)\neq f(Q)$ with probability $1$.
\end{thm}

\dcpadd{Ignoring the complicating terms, this says that it is necessary for $\frac{\beta}{\upsilon}> \Delta_{f}$ for it to be possible to 
implement a target decision function, $f$.}
We also provide the following sufficient condition. 
\begin{thm}
\label{thm:briberSufficient}
Given target decision function $f$ \dcpadd{that is Lipschitz differentiable almost everywhere},
with \mjyadd{differentiable }sensitivity $\nabla_{f}$ (\Cref{def:nabla}),
and in the briber setting  with 
the adaptive-payment mechanism, then for any prior $D$ \dcpadd{and any recommender subjective posterior beliefs $Q$},
if the mechanism budget $\beta$
and  briber incentive $\upsilon$ satisfy
\[
\frac{\beta}{\upsilon} 
  \ge \frac{6}{\min_{q\in\mathcal{Q}}f\left(q\right)}\cdot\nabla_{f},
\]
then all \dcpadd{$2\textsc{rat}$} \dcpadd{recommender reports $R$} \mathauditadd{in the presence of a best-responding briber} provide $F(R)=f(Q)$ with probability $1$.
\end{thm}

\dcpadd{Ignoring the complicating terms, this says that it is sufficient for $\frac{\beta}{\upsilon}\geq \nabla{f}$ for it to be possible to 
implement a target decision function, $f$.
This  reverses~\Cref{thm:lower} and shows that is is possible, in the presence of a briber and  depending on the sensitivity of $f$, the mechanism's budget $\beta$, and the briber incentive $\upsilon$, to  implement a suitably differentiable target decision function with the adaptive-payment mechanism. 
 In particular, this is achieved where all rationalizable strategies are truthful in the
adaptive-payment mechanism.}
%
%
%

To gain  intuition into the dependence on the sensitivity of the target decision function $f$, we  apply these results to the family of bounded-linear target decision functions. Considering~\Cref{def:bldf}, \dcpadd{the necessary condition for successful implementation
is that 
\[
\frac{\beta}{\upsilon}\le\frac{\revisionredreplace{1}{2}}{\revisionredreplace{640}{10^{3}}\cdot\log(3\cdot m)\cdot\max_{q\in\mathcal{Q}}f\left(q\right)}\cdot\frac{\alpha\cdot\ell^{2}}{m},
\]
where we replace the RHS in the necessity (\Cref{thm:briberNecessary}) of 
$\frac{\beta}{\upsilon}> \frac{1}{\revisionredreplace{16}{8}\cdot\max_{q\in\mathcal{Q}}f\left(q\right)}\cdot\Delta_{f}$
with a smaller quantity, recognizing from~\Cref{lem:dcptight} that 
$\Delta_{f}\ge \frac{1}{40}\cdot\frac{1}{\log(3\cdot m)}\cdot\frac{\alpha\cdot\ell^{2}}{m}$.}
\dcpadd{Considering~\Cref{def:bldf}, the sufficient condition for successful implementation
is  
\begin{align}
\frac{\beta}{\upsilon} 
  \ge \frac{12}{\min_{q\in\mathcal{Q}}f\left(q\right)}\cdot\frac{\alpha\cdot\ell^{2}}{m},
\end{align}
where we replace the RHS in the sufficiency (\Cref{thm:briberSufficient}) of 
$\frac{\beta}{\upsilon} 
  \ge \frac{6}{\min_{q\in\mathcal{Q}}f\left(q\right)}\cdot\nabla_{f}$
with a larger quantity, recognizing from~\Cref{lem:dcptight} that 
$\nabla_{f}  \le2\cdot\frac{\alpha\cdot\ell^{2}}{m}$.}

These two conditions match up to log factors and the factor ${\max_{q\in\mathcal{Q}}f\left(q\right)}/{\min_{q\in\mathcal{Q}}f\left(q\right)}$. Assuming that this ratio is bounded, the key dependence of $\frac{\beta}{\upsilon}\propto\frac{\alpha\cdot\ell^{2}}{m}$ appears in both bounds, and \dcpadd{establishes $\frac{\beta}{\upsilon}\approx \frac{\alpha\cdot\ell^{2}}{m}$} as a necessary and sufficient condition for the existence of a mechanism
that will exactly implement the target decision function.

 It is noteworthy that the factor $\frac{\alpha\cdot\ell^{2}}{m}$ also controls the effect of manipulation in the setting of intrinsic competing incentives (\Cref{thm:bldfTight}). Hence, this factor  captures, in a very general sense, the sensitivity to competing incentives of bounded-linear \dcpadd{target} decision functions.


\dcpadd{
\begin{example}
\label{ex:dam1}
To gain intuition into the size of a sufficient budget for the adaptive-payment mechanism to avoid manipulation,
we continue from~\Cref{ex:dam-payment-rule},  where
the city  plans to invest between 30 and 130 million USD
into flood barriers, with target decision function
\begin{align}
f(Q) &=30+100\cdot\bar{Q},
\end{align}

in millions of dollars, and   $\bar{Q}=\frac{1}{n}\sum_{i=1}^nQ_{i}$. 
Suppose a construction company gains a profit of 5\% of the investment,
so $\upsilon=0.05$, and is willing
to bribe recommenders. 
By~\Cref{thm:briberSufficient}, it is sufficient for truthful reporting
to the mechanism that the mechanism's budget $\beta$ satisfies 
\begin{align}
\beta&\geq \frac{6 \cdot \upsilon}{\min_{q\in Q} f(q)}\cdot \nabla_f=\frac{6}{30\cdot 20}\cdot \nabla_f=\frac{1}{100}\cdot\nabla_f.
\end{align}

 We can now appeal to~\Cref{lem:nablaLinearCube}, and obtain a bound on
 differentiable sensitivity, as 
 \begin{align}
     \nabla_f & \leq 2m\cdot \sigma\cdot L^2 = 2n(1/2)(\ell/n)^2=\ell^2/n=100^2/n, 
     \end{align}
where $\ell=100$ from~\Cref{eq:simplef},
and $m, \sigma$, and $L$ are as defined in~\Cref{lem:nablaLinearCube}. 
Considering~\Cref{eq:simplef},
all $n$ agents influence $f$ (and thus $m=n$), and 
$\left|\frac{\partial f}{\partial x_{i}}(x)\right| = \frac{\ell}{n}$ for reports $x$ on a unit hypercube
with sidelength $2\sigma=1$ centered at $\mu=(0.5,\ldots,0.5)$, and thus we  adopt $L=\ell/n$ and $\sigma=1/2$.
Since $100^2/n$ is a larger quantity than $\nabla_f$, it is sufficient 
for truthful reports that the mechanism's
budget $\beta\geq (1/100)(100^2/n)=100/n$, in millions of dollars. 

With  a single recommender this is a budget of 100 million USD, 
which is unlikely to be feasible. However, with 20 recommenders, a  5 million USD budget would be sufficient,
and with 100 recommenders  a 1 million USD budget would be sufficient.
As the number of recommenders increases, 
the cost of elicitation becomes modest relative to the fiscal decision facing the city.
\end{example}}

\section{Conclusion}
\label{sec:conclusion} 

We have fully characterized the extent to which recommenders will manipulate a decision that a principal wants to make according to a target decision function $f$. The key quantities influencing the size of the manipulation are (i) the strength of recommenders'  competing incentives, $\gamma$, (ii) the mechanism's budget, $\beta$, and (iii) the sensitivity $\Delta_{f}$ of the target decision function $f$. The quantity that has appeared repeatedly throughout
our results, and characterizes to what extent the decision is manipulated, is $\frac{\gamma}{\beta}\cdot\Delta_{f}$. For the class of bounded-linear target decision functions, this quantity determines the upper and the lower bound for manipulation, indicating that the notion of sensitivity captures a fundamental property of the decision function. 
For this class, the sensitivity takes the form $\frac{\alpha\cdot\ell^{2}}{n}$, where $\ell$ is the slope \dcpadd{of the target decision function}, $\alpha$ determines how much influence a single recommender has, and $n$ is the number of recommenders. 
In this case, manipulation can be reduced, or even avoided in the briber setting, by scaling up the budget $\beta$ proportionally to the competing incentive $\gamma$, reducing the influence $\alpha$ each recommender has, reducing the slope of the decision function $\ell$, or increasing the number of recommenders. 
This last point is particularly interesting, as it suggests that one may reduce manipulation without increasing the total budget, \dcpadd{at least while the incentive payments remain high enough to motivate participation by recommenders}. 

\mjyadd{Our most important positive result shows that we can eliminate manipulation when the external incentive comes from a rational briber, where our target decision function $f$ only needs to be Lipschitz differentiable almost everywhere.} \dcpadd{In effect, this yields a family of mechanisms since $f$ can be configured to suit the domain at hand, and used together with the adaptive payment mechanism.} Importantly, we do not assume  the principal has knowledge of the kind of information  available to recommenders or their prior beliefs.  \dcpadd{It is our belief that this aspect} makes our positive results \dcpadd{in regard to the adaptive-payment mechanism} widely applicable, and we hope they can guide the design of practical mechanisms for eliciting beliefs from experts with incentives that do not align with those of the principal.

\newpage{}  \bibliographystyle{ACM-Reference-Format}
\bibliography{refs}


\newpage{}

\appendix

\section{Consistent Beliefs\label{ap:perfect}}

In the briber setting, recommender $i$'s subjective expected utility is (recall that $s$ is a mixed strategy and we use $\hat{S}$ to denote the pure strategy samples from the mixed strategy):

\begin{align*}
&\textsc{eutil}_{i}\left(s,s^{\textsc{bri}},p_{i},y_{i},c_{i}\right)\\ 
& =\mathbb{E}\left[\textsc{util}_{i}\left(P_{i},Y,C_{i},R\right)\big|s,s^{\textsc{bri}},p_{i},y_{i},c_{i}\right]\\
 & =\mathbb{E}\left[\textsc{util}_{i}\left(p_{i},\left(y_{i},Y_{-i}\right),c_{i},R\right)\big|s,s^{\textsc{bri}},p_{i},y_{i},c_{i}\right]\\
 & =\mathbb{E}\left[\textsc{util}_{i}\left(p_{i},\left(y_{i},Y_{-i}\right),c_{i},\left(\hat{S}_{i}\left(p_{i},y_{i},c_{i}\right),\hat{S}_{-i}\left(P_{-i},Y_{-i},\hat{S}_{-i}^{\textsc{bri}}(V,W)\right)\right)\right)\bigg|s,s^{\textsc{bri}},p_{i},y_{i},c_{i}\right]\\
 & =\mathbb{E}\left[\textsc{util}_{i}\left(p_{i},\left(y_{i},Y_{-i}\right),c_{i},\left[\hat{S}_{i}\left(p_{i},y_{i},c_{i}\right),\hat{S}_{-i}\left(P_{-i},Y_{-i},\hat{S}_{-i}^{\textsc{bri}}(V,W)\right)\right]\right)\bigg|s,s^{\textsc{bri}},c_{i}\right],
\end{align*}
where the expectation is taken with respect to the conditional distribution that follows from Bayes' rule, i.e.
\begin{align*}
 & \mathbb{P}\left[p_{-i},y_{-i},\hat{s},\hat{s}_{-i}^{\textsc{bri}},v,w\bigg|s,s^{\textsc{bri}},c_{i}\right]\\
 & =\mathbb{P}\left[p_{-i},y_{-i},\hat{s}\mathauditdelete{,\hat{s}_{-i}^{\textsc{bri}}}\bigg|v,w,s,s^{\textsc{bri}},c_{i}\right]\mathbb{P}\left[\hat{s}_{-i}^{\textsc{bri}},v,w\bigg|s,s^{\textsc{bri}},c_{i}\right]\\
\text{(independences)} & =\mathbb{P}\left[p_{-i},y_{-i},\hat{s}\mathauditdelete{,\hat{s}_{-i}^{\textsc{bri}}}\bigg|s,s^{\textsc{bri}}\right]\mathbb{P}\left[\mathauditadd{\hat{s}_{-i}^{\textsc{bri}},} v,w\bigg|s^{\textsc{bri}},c_{i}\right].
\end{align*}
The only term that is potentially problematic is
\[
\mathbb{P}\left[\mathauditadd{\hat{s}_{-i}^{\textsc{bri}},} v,w\bigg|s,s^{\textsc{bri}},c_{i}\right].
\]
This distribution is not well-defined if there is no $\hat{s}^{\textsc{bri}}\in\textsc{supp}(s^{\textsc{bri}}),v\in\textsc{supp}(D^{v}),w\in\textsc{supp}(D^{w})$ such that $c_{i}=\hat{s}_{i}^{\textsc{bri}}(v,w)$, i.e. if we conditioned on a zero-probability event. In that case, any belief regarding the briber type, $V,W$, is said to be consistent and can hence be part of a perfect Bayesian equilibrium.


\section{Simulations}


\subsection{Simulation Setup}

\mjyadd{We model a lender deciding how much loan to make to each borrower $j$ in a population, with $n$ recommenders for a borrower, each with a belief about the probability  the borrower will repay the loan $(o=1)$.
The principal (lender) uses these reports to decide the loan size (we may equivalently think of this as deciding the probability with which the lender makes a loan of a fixed size). Through this simulation, we seek to understand the adaptive-payment mechanism, the budget required to avoid manipulation under different numbers of recommenders and minimum loan size levels, and the profit to the principal  from using the mechanism.}

\subsubsection{Borrower model}
\begin{figure}[t]
\centering
\includegraphics[width=0.8\linewidth]{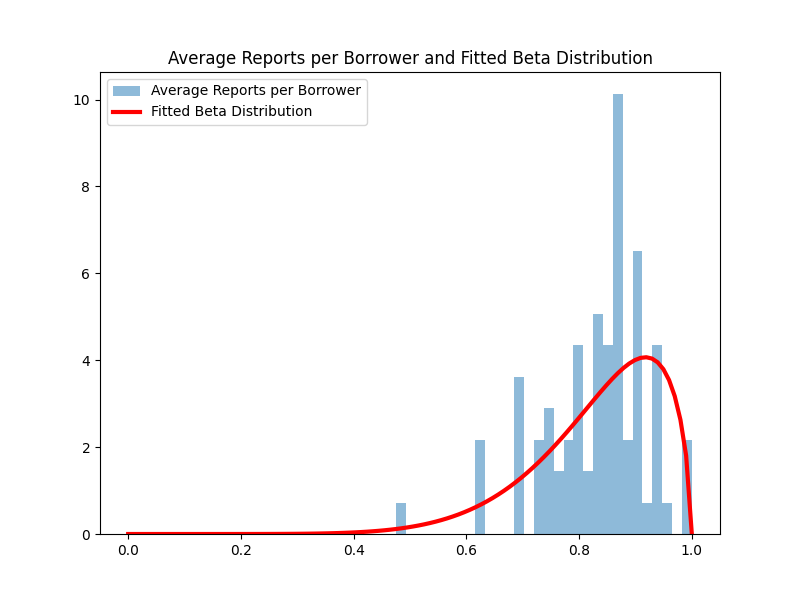}
\Description{Borrower-average reports with fitted beta distribution.}
\caption{Histogram of borrower-average recommender  reports with the estimated beta distribution.
\label{fig:beta_fitted}}
\end{figure}

\dcpadd{To set-up the simulation, we make use of} a  study  we conducted in collaboration with Makere University in Uganda, supported by the \textit{Global Challenges in Economics and Computation Prize}.\footnote{\dcpadd{The \textit{Global Challenges in Economics and Computation Prize} was funded partially by Meta, with \$20,000 US  awarded in 2020. The study did not make use of the adaptive-payment mechanism.  We conducted the study throughout 2021, with four different settings: lend to highest-rated borrowers while compensating recommenders for accuracy, lend to highest-rated and give constant compensation, lend to all borrowers and compensate for accuracy, and lend to all borrowers and give constant compensation. All reports and loan outcomes are incorporated into the data used to estimate the statistical model used in the  simulation.}} \dcpadd{We gathered loan applicants among farmers in four villages in Uganda, asked other members of each village to report the probability of each farmer repaying,} and made loans in the amount of approximately \$80 USD  to \dcpadd{each of} 79 smallholder farmers in Uganda. Of the borrowers, 85\% repaid their loan in full.
%
%
%
\dcpadd{For each borrower, we have an average report of repayment probability from others in the village, 
and we estimate the distribution of borrower repayment probabilities by fitting a beta distribution to these borrower-average recommender reports (see \Cref{fig:beta_fitted})}.
%
%
\dcpadd{With the mean fixed at $0.85$, the maximum likelihood estimator for the $\alpha$ and $\beta$ parameters of a beta 
distribution is
%
%
$\alpha = 0.85\cdot 9.9$ and $\beta=0.15\cdot 9.9$.}
We adopt i.i.d.~samples from this beta distribution
as  borrower $j$'s Bernoulli repayment probability, $\zeta_j = \mathbb{P}[O_j = 1]$.

\subsubsection{Recommender model}
\dcpadd{To model the repayment beliefs of recommenders for a given borrower, we 
estimate the variability in reports in
the data. For this,} \mjyadd{we adopt the borrower-average report for each borrower, and calculate the residual between each individual report for the borrower and the borrower-average report.
Pooling residuals across all borrower reports, the variance of the residuals, assuming 261 degrees of freedom after subtracting 79 from the 340 total reports, is $0.01312$. The mean of the borrower-average reports is $0.833$.
%
Considering a prototypical borrower with $\zeta_j=0.833$, and report variance  $0.01312$, this
corresponds to a beta distribution with parameters
$\alpha' = 7.99$ and $\beta' = 1.6$. The concentration 
of this beta distribution is  $\alpha' + \beta' = 9.59$. Concentration in the beta 
distribution is
inversely related to variance.\footnote{With concentration  $\kappa = \alpha + \beta$, and  mean  $\mu = \frac{\alpha}{\alpha + \beta}$, the variance is $\mu(1-\mu)/(\kappa+1)$.}
Given this, we model the belief of recommender $i$ for 
the repayment probability of borrower $j$, denoted $Q_{i,j}$, as coming from a beta distribution
with mean $\zeta_j$ and concentration $\alpha_j + \beta_j = 9.59$; i.e., we adopt
a constant concentration model for recommender beliefs.}

Formally, the generative model, for borrower $j$ and recommender $i$, is:
\begin{align}
    \zeta_j &\sim \mathcal{B}(\alpha, \beta), \quad \alpha = 0.85 \cdot 9.9,\  \beta = 0.15 \cdot 9.9, \quad \mbox{and}
\\
    Q_{i,j}& \sim \mathcal{B}(\alpha_j, \beta_j), \quad \alpha_j = 9.59\cdot  \zeta_j, \ \beta_j = 9.59(1-\zeta_j).
\end{align}

\dcpadd{In this way, the second-level beta distribution, $Q_{i,j}$,
has mean $\alpha_j/(\alpha_j+\beta_j)=\zeta_j$
and concentration $\alpha_j+\beta_j=9.59$.
The effect of this concentration
is that with  ten recommenders the standard error of the borrower-average report, with respect
to the true repayment probability, for a 
borrower $j$ with $\zeta_j=0.85$, is
just $0.035$.}

\subsubsection{Target decision function}

The \dcpadd{adaptive-payment} mechanism elicits reports \dcpadd{and sets the loan amount  according
to a decision rule that is equal to the target decision function.} 
\dcpadd{In this case, we adopt}
 a bounded-linear target decision function $f(Q)$, as in~\Cref{def:bldf}. 
 \dcpadd{Since the decision rule, $F(R)$, on reports $R$ is equal to the target decision function, going forward
 we write $F(R)$ as $f(R)$.}
 \mjyadd{We set the budget to ensure truthfulness as per~\Cref{thm:briberSufficient}, and we can therefore assume $R = Q$. In the sequel, we drop index $j$ for borrower $j$ for notational simplicity, and consider reports $R_i$
 (realized values $r_i$) for recommender $i$, and denote the loan outcome as $O$.}
 %
 The mechanism pays each recommender $i$ according to the adaptive-payment mechanism $\textsc{ada-pay}_i(R_i, O)$, as per~\Cref{def:adaPay}.

\dcpadd{We model the principal (lender) as receiving a payoff of a multiple 0.2 of the 
loan amount in the case of repayment $(O = 1)$ (representing interest), and a loss equal to the entire
loan amount otherwise $(O=0)$.
We also subtract the total payments it makes to each of $n$ recommenders. 
Altogether, we have  utility, $u_P$, to the principal of}
\begin{align} \label{eq:utility}
    u_P\dcpadd{(R,O)} &= 1.2 f(R) \cdot O - f(R) - \sum_{j \in [n]} \dcpadd{\textsc{ada-pay}_j(R_j,  O),}
\end{align}
where $O$ takes on 0 or 1. As desired, the non-payment term 
takes on $0.2f(R)$ in the case $O=1$ and $-f(R)$ in the case $O=0$.

We formulate $f(R)$ such that $f(R) \in [0,1]$, and interpret $f(R)$ as the 
\dcpadd{loan amount allocated} to the borrower. 
%
Following~\Cref{def:bldf},
we adopt a bounded-linear target decision function \dcpadd{on realized reports $r$}, and
\begin{align} \label{eq:bounded_linear_sim}
f(r)&
=
\bar f+
\operatorname{clip}_{[-\varepsilon,\varepsilon]}
\left(
\frac{\ell}{n}
\sum_{i=1}^n
\operatorname{clip}_{[-\alpha,\alpha]}(r_i-t)
\right),
\end{align}
where
$\operatorname{clip}_{[u,v]}(x)
:=
\min\{\max\{x,u\},v\}$,
for \(u\le v\).
All recommender reports are influential; i.e. $m = n$ in the context of~\Cref{def:bldf}). We assume that $n \geq 2$.

\begin{figure}[t]
\centering
\includegraphics[width=0.8\linewidth]{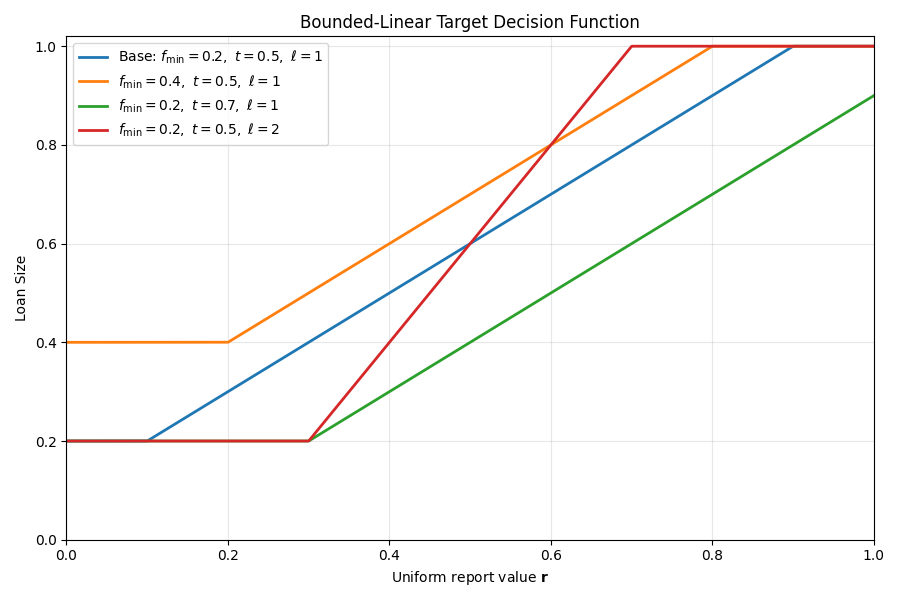}
\Description{Four versions of the bounded-linear target decision function.}
\caption{Target decision functions, with loan size (y-axis) against  report value $\mathbf{r}$,
assumed uniform with every recommender making the same report (x-axis).
%
$f_{\min}$ sets the minimum loan value. Clipping parameters, $\alpha$ and $\varepsilon$, also vary with $f_{\min}$, with $f_{\min} = 0.2 \Rightarrow \alpha = \varepsilon = 0.4$ and $f_{\min} = 0.4 \Rightarrow \alpha = \varepsilon = 0.3$.
Parameters $t$ and $\ell$ set the position of the center of the slope and the amount of slope of the target decision function, respectively. 
\label{fig:bounded_linear_target_decision_function}}
\end{figure}

We parameterize the target decision function by $f_{\min}\in (0,1]$ 
and centering parameter, $t\in [0.1,0.9]$.  We define
$\Bar{f}=\frac{1+f_{\min}}{2}$,
$\alpha = \varepsilon = \frac{1-f_{\min}}{2}$, and slope $\ell$, which we take as $\ell= 2.5$
for the simulation.
%
%
%
The decision, $f(r)$, varies from $\Bar{f}-\varepsilon=(1+f_{\min})/2-(1-f_{\min})/2=f_{\min}$
to $\Bar{f}+\varepsilon=(1+f_{\min})/2+(1-f_{\min})/2=1$,
and $f_{\min}$  determines
the minimum loan that is made (strictly positive, as
required by~\Cref{thm:briberSufficient}).
Also,  we have $\varepsilon$ and $\alpha\in[0,0.5]$, depending 
on the value of $f_{\min}$. \Cref{fig:bounded_linear_target_decision_function} shows how  $f_{\min}, t$, and $\ell$ change the shape of the bounded-linear target decision function.

\subsubsection{Adaptive-payment mechanism}\leavevmode\par\noindent

\dcpadd{Given the target decision function, we can derive the adaptive-payment mechanism.
Define
\[
z_i(r_i)
:=
\operatorname{clip}_{[-\alpha,\alpha]}(r_i-t).
\]

For  a dependence $\partial_{r_i}f(r)>0$ on recommender $i$'s report $r_i$,
we need each clipping function not to bind
on $r_i$. First, we need
\begin{align}
\label{eq:alphaneed}
-\alpha \leq r_i-t\leq \alpha.
\end{align}

Since $r_k \in [0,1]$, for any $k\in [n]$, then $z_k(r_k)$ can attain any value in the range
\[
-a\le z_k(r_k)\le b,
\ \mathrm{where} \qquad
a:=\min\{\alpha,t\}\geq 0,
\qquad
b:=\min\{\alpha,1-t\}\geq 0.
\]

Second, we need
\begin{align}
\label{eq:epsilonneed}
-\varepsilon&\leq \frac{\ell}{n}\sum_{k\in[n],k\neq i}z_k(r_k) + \frac{\ell}{n}z_i(r_i)\leq \varepsilon \quad \Leftrightarrow \quad -c \leq \sum_{k=1}^n z_k(r_k) \leq c,
\end{align}
where
\[
c:=\frac{n\cdot \varepsilon}{\ell}>0.
\]

The \(n-1\), $\alpha$-clipped reports other than $i$ can attain any sum in
\begin{align}
-(n-1)a
\le
\sum_{k\in [n],k\ne i}z_k(r_k)
\le
(n-1)b.
\end{align}

The function $\delta_i(r_i)$ in the adaptive-payment mechanism relies on knowing when there exist reports from other recommenders $k \neq i$ that cause $i$ to be influential. 

When $r_i \leq t$ and thus $z_i\leq 0$, the upper bound $(n-1)b$ on the sum of others' reports sets the upper bound on the most that other recommenders can offset $i$'s low report. Assume~\Cref{eq:alphaneed}, that the inner clip on recommender $i$ is inactive. Certainly, we can choose $\sum_{k\neq i}z_k(r_k)$ s.t. $r_i-t+\sum_{k\neq i}z_k(r_k)\leq c$ since we can set $z_k(r_k)=0$ for all $k\neq i$. 
The question is whether we can choose $z_k(r_k)$, $k\neq i$, s.t. $r_i-t+\sum_{k\neq i}z_k(r_k)\geq -c$. For this,
it is necessary and sufficient that
\begin{align}
\label{eq:need1}
-c \leq r_i - t + (n-1)b.
\end{align}

Similarly, consider  $r_i \geq t$,  and thus $z_i\geq 0$.  Assume~\Cref{eq:alphaneed}, that the inner clip on recommender $i$ is inactive. Certainly, we can choose $\sum_{k\neq i}z_k(r_k)$ s.t. $r_i-t+\sum_{k\neq i}z_k(r_k)\geq -c$ since we can set $z_k(r_k)=0$ for all $k\neq i$.
The question is whether we can choose $z_k(r_k)$, $k\neq i$, s.t. $r_i-t+\sum_{k\neq i}z_k(r_k)\leq c$. For this, it is necessary and sufficient that
\begin{align}
\label{eq:need2}
r_i -t - (n-1)a \leq c.
\end{align}

Putting~\Cref{eq:need1} and~\Cref{eq:need2} together, and rearranging, we have
\begin{align}
\label{eq:epsilonneednew}
t-c-(n-1)b
\leq
r_i
\leq
t+c+(n-1)a.
\end{align}

Condition~\Cref{eq:epsilonneednew} is necessary and sufficient for there to exist $r_{-i}$, s.t. there is no $\varepsilon$ clipping.
Together with condition~\Cref{eq:alphaneed}, this gives what is needed for $r_i$ to have influence for some $r_{-i}$.
Combining~\Cref{eq:epsilonneednew} and~\Cref{eq:alphaneed}, then an $r_i\in [0,1]$ has influence $\partial_{r_i}f(r)>0$ for some $r_{-i}$ if and only if
\begin{align}
\max(t-\alpha, t-c-(n-1)b)\leq r_i\leq \min(t+\alpha, t+c+(n-1)a).
\end{align}

We now return to the $\delta_i(r_i)$ term in the adaptive-payment mechanism expression,
\begin{align}
\label{eq:deltaj}
\delta_i(r_i) & =\max_{r_{-i}}\left|\partial_{r_i}f(r)\right|+\int_{0}^{1}\max_{r_{-i}}\left|\partial_{x}f(x,r_{-i})\right|dx.
\end{align}

Considering also that $r_i$ must be on $[0,1]$, we can define the range of values that $r_i$ can take such that  $\partial_{r_i}f(r)>0$ for some $r_{-i}$, as
\[
L
:=
\max\left(
0,
t-\alpha,\,
t-c-(n-1)b
\right),\quad \mbox{and}
\]
\[
U
:=
\min\left(
1,
t+\alpha,\,
t+c+(n-1)a
\right).
\]

By inspection, we have $U > L$ except when $f_{\min} = 1$, when $U = L$.
Considering the first term in~\Cref{eq:deltaj}, for $r_i\in[L,U]$, we have $\partial_{r_i} f(r_i,r_{-i})=\frac{\ell}{n}$, for some $r_{-i}$. Considering the second term in~\Cref{eq:deltaj}, by the same analysis we have
\[
\int_{0}^{1}\max_{r_{-i}}\left|\partial_{x}f(x,r_{-i})\right|dx=
\int_{L}^U\frac{\ell}{n}dx=\frac{\ell}{n}(U-L).
\]

Combining, we have 
\begin{align}
\label{eq:deltajnew}
\delta_i(r_i) & =\frac{\ell}{n}\mathbf{1}\{r_i\in[L,U]\}+\frac{\ell}{n}(U-L),
\end{align}
for indicator function $\mathbf{1}\{\cdot\}.$
The adaptive-payment mechanism is
\begin{align}
\textsc{ada-pay}_i(r_i,o)=\frac{\beta}{3n}\cdot\left(\frac{\int_{0}^{r_i}\delta_i(x)(o-x) dx}{\int_{0}^{1}\delta_i(x)dx}+1\right).\label{eq:adapaytwo}
\end{align}

For the denominator, we have
\begin{align}
\int_{0}^{1}\delta_i(x)dx=
\int_L^U\frac{\ell}{n}dx+\int_0^1\frac{\ell}{n}(U-L)dx=
\frac{\ell}{n}(U-L)+\frac{\ell}{n}(U-L)=\frac{2\ell}{n}(U-L).\label{eq:27}
\end{align}

For the numerator, first term, if $o=0$ then this is zero. If $o=1$, then this is
\begin{align*}
\int_0^{r_i}\delta_i(x)odx&=
\int_0^{r_i}\delta_i(x)\cdot 1dx
=
\int_L^{\min(r_i,U)}\frac{\ell}{n}dx +\int_0^{r_i}\frac{\ell}{n}(U-L)dx
\\
&=
\frac{\ell}{n}\cdot \max(0,\min(r_i,U)-L) +\frac{\ell}{n}(U-L)r_i.
\end{align*}

For the numerator, second term, we have
\begin{align*}
-\int_0^{r_i}\delta_i(x)xdx&=
-\int_L^{\min(r_i,U)}\frac{\ell}{n}xdx-\int_0^{r_i}\frac{\ell}{n}(U-L)xdx
\\
&=-\frac{\ell}{2n}\left[\max(0,\min(r_i,U)^2-L^2)\right]-\frac{\ell}{2n}(U-L)r_i^2.
\end{align*}

Thus
\begin{align}
&\int_0^{r_i}\delta_i(x)(o-x)\,dx \notag\\
&\quad =
\frac{\ell}{n}
\bigg[
(U-L) \left(o\cdot r_i-\frac{r_i^2}{2}\right) +
o\cdot \max(0,\min(r_i,U)-L)-\frac{\max(0,\min(r_i,U)^2-L^2)}{2}
\bigg].
\end{align}

Substituting this expression and~\Cref{eq:27}  into the
adaptive-payment  rule gives,
\begin{align}
&\textsc{ada-pay}_i(r_i,o)
=
\frac{\beta}{3n}
\left(
1+
\frac{
\int_0^{r_i}\delta_i(x)(o-x)\,dx
}{
\int_0^1\delta_i(x)\,dx
}
\right) =
\frac{\beta}{3n}
\left[
1
+\frac{1}{2}\left(o\cdot r_i-\frac{r_i^2}{2}\right)\right.
 \notag\\
\quad &\left.
+\frac{1}{2(U-L)}
\left(
o\cdot \max(0,\min(r_i,U)-L)-\frac{\max(0,\min(r_i,U)^2-L^2)}{2}
\right)
\right]\notag\\
&=
\frac{\beta}{3n}
\left[
1+\frac{o\cdot r_i}{2}-\frac{r_i^2}{4}
+\frac{o\cdot \max(0,\min(r_i,U)-L)}{2(U-L)}
-\frac{\max(0,\min(r_i,U)^2-L^2)}{4(U-L)}
\right].
\end{align}

Ignoring multipliers and terms that do not depend on $r_i$, this further simplifies to
\begin{align}
\textsc{ada-pay}_i(r_i,o)&
\propto
\left\{
\begin{array}{ll}
o\cdot r_i-\frac{r^2_i}{2} & \mbox{if $r_i\leq L, r_i\leq U$,}\\
o\cdot r_i-\frac{r^2_i}{2}+\frac{1}{U-L}\left(o\cdot r_i-\frac{r^2_i}{2}\right)+C_1 & \mbox{if $r_i> L, r_i<U$,}\\
o\cdot r_i-\frac{r^2_i}{2}+C_2& \mbox{if $r_i\geq L, r_i\geq U$.}
\end{array}
\right.
\end{align}

Here, $C_1=L(L-2o)/2(U-L)$ and $C_2=o-1/2(U+L)$. We see
in each component the structure of the strictly proper, quadratic scoring rule. 
This is an asymmetric, strictly-proper scoring rule (see~\Cref{fig:adapay_chart},
where the expected payment to a recommender is a strictly convex function). }
\begin{figure}[t]
\centering
\includegraphics[width=0.8\linewidth]{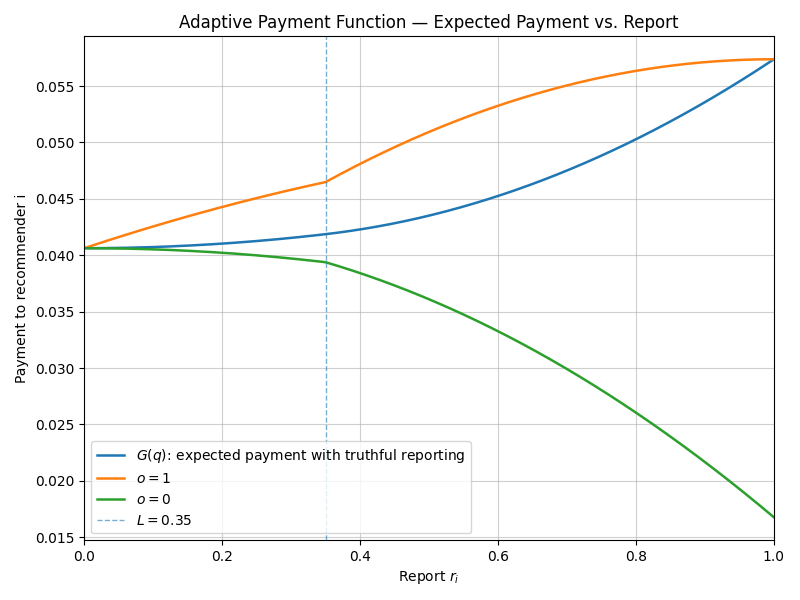}
\Description{Recommender payments from the adaptive payment mechanism.}
\caption{Conditional payments and expected payment
to a recommender $i$ in the adaptive-payment mechanism under truthful reporting and in the loan setting, for  $n=10$, $t=.75$, $f_{\min}=.2$, $\ell = 2.5$.
{\sc ada-pay} changes more steeply, and  concentrates more incentives, once the report $r_i$ becomes influential, which occurs for $r_i\geq L=0.35$. 
The expected payment, $G(q)$, for true belief $q$ of recommender $i$,
is strictly-convex, making the scoring rule strictly-proper.
\label{fig:adapay_chart} }
\end{figure}

\if false
    This allocation function gives us an adaptive-payment function of
    \[
    (f,\beta)\textsc{-ada-pay}_{i}(r_{i},o)=\begin{cases}
    \frac{\beta}{3n}\cdot\left(\frac{\int_{0}^{r_{i}}\delta_{i}(x)(o-x)\cdot dx}{\int_{0}^{1}\delta_{i}(x)\cdot dx}+1\right),
    \end{cases}
    \]
    where $\delta_i(r_i)$ becomes
    \begin{align*}
    \delta_{i}(r_{i}) & =\max_{r_{-i}}\left|\partial_{r_{i}}f(r)\right|+\int_{0}^{1}\max_{r_{-i}}\left|\partial_{r_{i}}f(r)\right|dr_{i} = \frac{2 \ell}{n},
    \end{align*}
    
    resulting in
    \[
    (f,\beta)\textsc{-ada-pay}_{i}(r_{i},o)=\begin{cases}
    \frac{\beta}{3n}\cdot\left(\frac{\int_{0}^{r_{i}}\delta_{i}(x)(o-x)\cdot dx}{\int_{0}^{1}\delta_{i}(x)\cdot dx}+1\right) = \frac{\beta}{3n} \left(or_i - \frac{r_i^2}{2} + 1 \right).
    \end{cases}
    \]

\fi


\subsection{Simulation Results}

We now present the results of the simulation.\footnote{\mjyadd{The simulation code can be found on \href{https://github.com/markjamesyork/elicitation_competing_incentives/tree/main}{Github (link here).}}}
For all results, we
 average across 100,000 trials. 

\subsubsection{Profit, ROI, and budget}

The profit to the lender (or principal) for a loan of size $f(R)$ and outcome $O$ is just 
utility, $u_P$~\Cref{eq:utility}.
The {\em return-on-investment} (ROI) to the lender for a loan of size $f(R)$ (recall, $f(R)>0$)
is
\begin{align} \label{eq:ROI}
    \mathrm{ROI}_P(R,O) &= \frac{u_P(R,O)}{f(R)}.
\end{align}
 
In this way, the ROI captures the profit per dollar that the lender lends to a borrower.

\begin{figure}[t]
\centering
\includegraphics[width=0.8\linewidth]{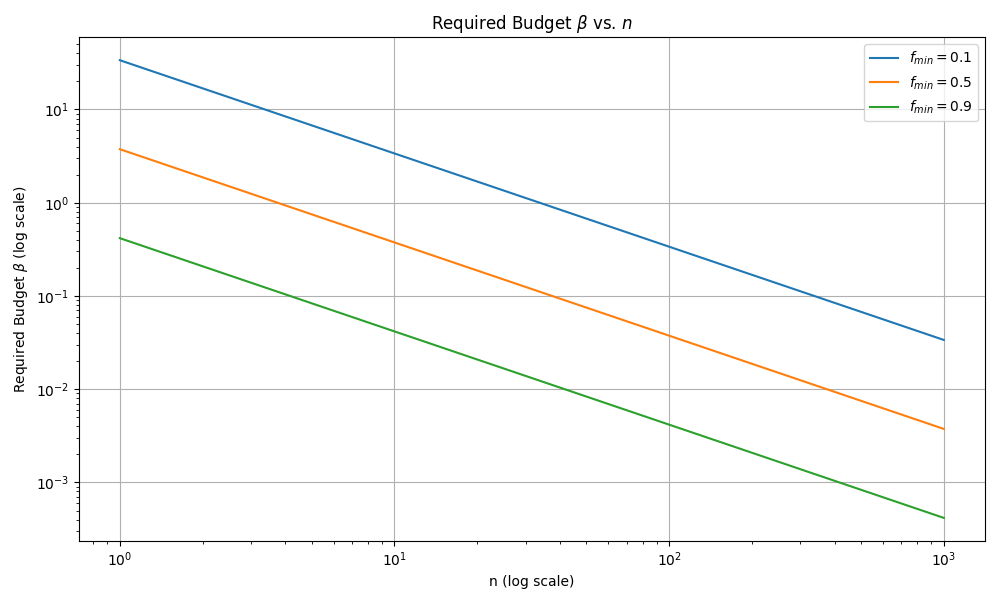}
\Description{Required budget to implement the bounded-linear target decision function.}
\caption{Required budget, $\beta$, to implement the target decision function in the loan setting versus the number of recommenders, $n$,
for three different values of $f_{\min}$.
For small $n$, a larger minimum loan allocation, $f_{\min}$, is required to
make the mechanism budget affordable.
\label{fig:budget_r_fmin}}
\end{figure}

\dcpadd{The required budget, $\beta$,
to make bribing  not useful for the briber decreases with $n$. 
The accuracy of the borrower-average recommendation  
also increases with the  number of recommenders, $n$.
Also, larger values of $f_{\min}$ make the decision function less sensitive, 
since there is more $\alpha$- and $\varepsilon$-clipping; 
see decision function~\Cref{eq:bounded_linear_sim}.
In turn, this reduces the required budget to implement the target decision function.
See~\Cref{fig:budget_r_fmin}, which shows  the required budget as a function
of the number of recommenders and the minimum loan size.  Here,
the briber's incentive $\upsilon = 0.1$, and 10\% of
the size of the loan; e.g., the briber may be a friend, 
receiving from the borrower this fraction of the loan amount.}

\subsubsection{Always-lend baseline}

\dcpadd{As a baseline, we consider the {\em always-lend policy}, which simply makes a loan of 1
to every borrower without eliciting any information from recommenders.
The expected,
always-lend profit is $0.85\times 0.2-0.15 \times 1=0.02$,
and this is also the amount of ROI, since the loan amount is just 1 all the time.
Another simple baseline is the {\em never-lend policy}, which has profit 0 (and with ROI technically undefined
since no loans are ever made).}

\subsubsection{Effect of varying $f_{\min}$ and varying centering parameter, $t$}
\begin{figure}[t]
\centering
\includegraphics[width=0.495\linewidth]{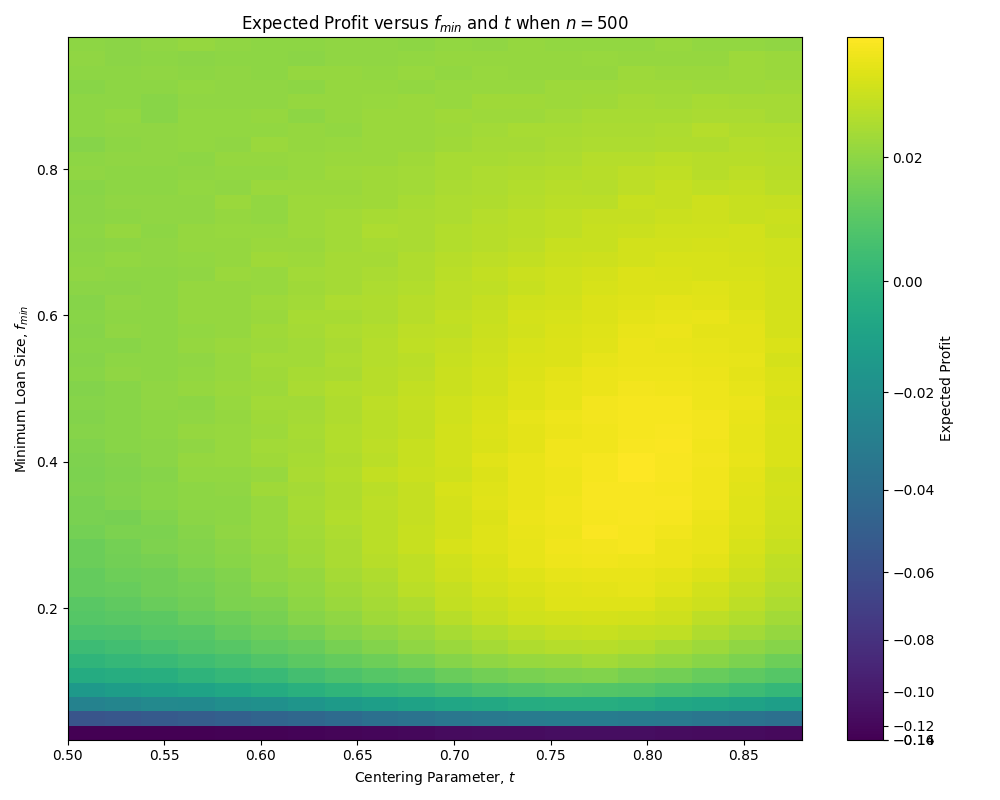}
\includegraphics[width=0.495\linewidth]{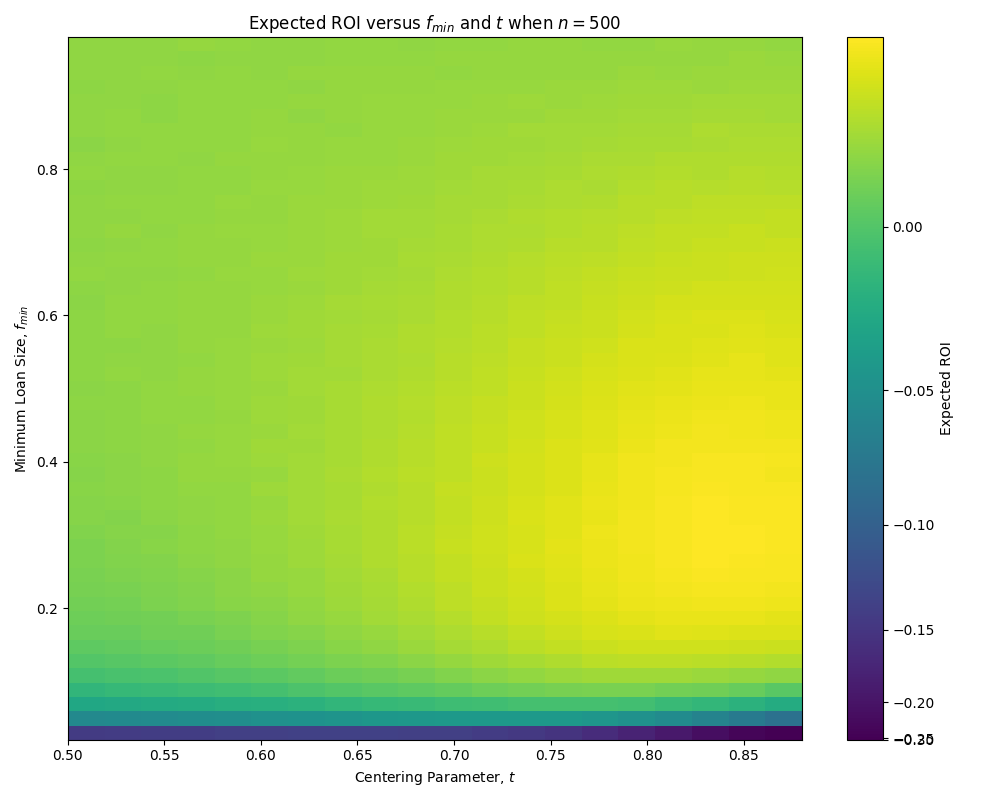}
\Description{Expected profit and ROI grid for combinations of t and $f_{\min}$ parameters.}
\caption{Expected profit and expected ROI with $n=500$ recommenders and $\ell=2.5$
in the loan setting, for different values of $f_{\min}$ and $t$.
For profit, optimal $t=0.8$ and optimal $f_{\min} = 0.4$. For ROI, optimal $t=0.84$ and optimal $f_{\min} = 0.28$. Maximizing ROI requires a higher choice of $t$ and a lower choice of $f_{\min}$. 
\label{fig:profit_roi_charts}}
\end{figure}

\dcpadd{For $n=500$ and $\ell=2.5$,~\Cref{fig:profit_roi_charts} shows the variation in
expected profit and expected ROI for different choices of $f_{\min}$ and $t$.} \mjyadd{For
expected profit, the optimal choice of centering parameter, $t$, 
and $f_{\min}$ is $0.8$ and $0.4$, 
respectively, yielding an expected profit of $0.0375$.
For expected ROI, the optimal choice of centering parameter, $t$, 
and $f_{\min}$ is  $0.84$ and $0.28$, 
respectively, leading to an expected ROI of $0.05$.}
\dcpadd{The optimal expected profit of $0.0375$ and optimal 
expected ROI of $0.05$ compares to an always-lend 
profit and ROI of $0.02$.  In particular,
the achievable ROI is more than double the ROI 
that can be achieved 
without using the adaptive-payment mechanism.} 

\dcpadd{We consistently see across different numbers of recommenders that the best choice
of $t$ and $f_{\min}$ for expected ROI, relative to the best choice for expected profit,
is to adopt a relatively higher value for $t$ 
and a relatively lower value for $f_{\min}$, so that we give smaller loans for the same mean report
together with
a baseline of giving out smaller loans.}
\begin{figure}[t]
\centering
\includegraphics[width=0.495\linewidth]{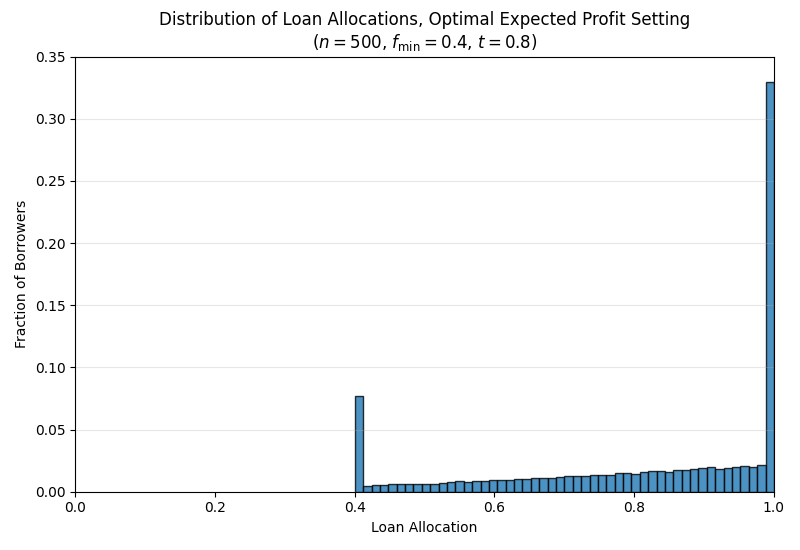}
\includegraphics[width=0.495\linewidth]{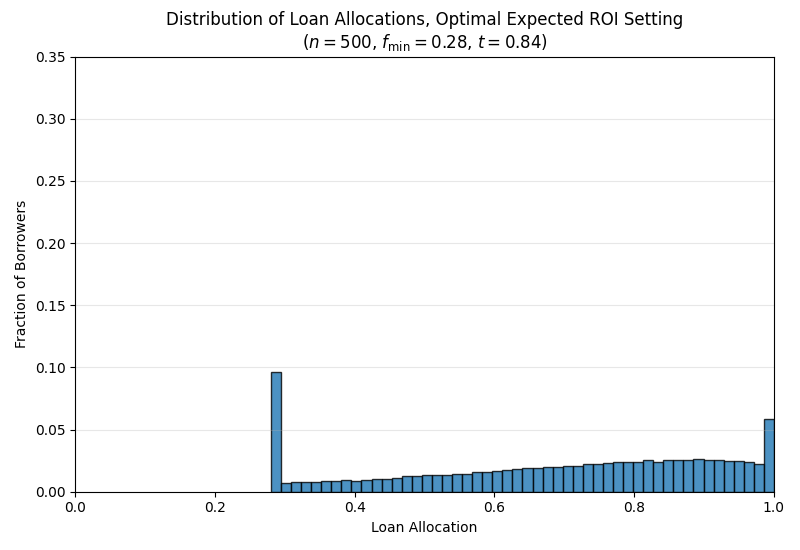}
\Description{Distributions of loan sizes.}
\caption{Distribution of loan sizes under the optimal expected profit (left) and optimal expected ROI (right) settings with 500 recommenders and $\ell = 2.5$. The optimal expected profit setting gives the full loan to 1/3 of borrowers with just 7.5\% of borrowers getting the minimum loan of 0.4. By contrast, the optimal expected ROI setting is  more cautious, giving just 6\% of borrowers the full loan, with smaller loans used and nearly 10\% receiving the minimum loan amount of 0.28. We  expect this from a lender who  seeks to make loans to maximize expected return. 
\label{fig:mean_report_vs_loan_size}}
\end{figure}
\begin{figure}[t]
\centering
\includegraphics[width=0.8\linewidth]{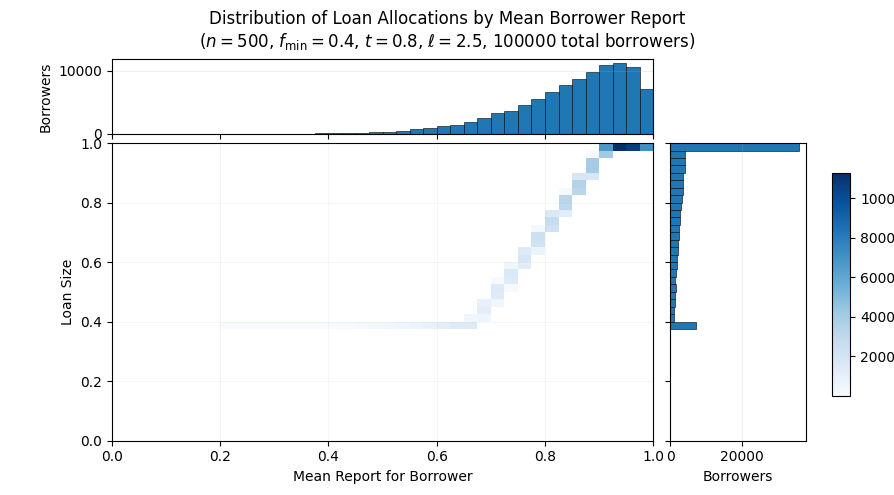}
\Description{Mean Report vs Loan Size, Profit.}
\caption{Borrower-average report versus assigned loan size: Optimal expected profit. $n=500,\ell=2.5$. Upper histogram: borrower-average reports. Right-hand histogram: distribution on loan sizes. Borrower-average reports linearly determine loan assignment until $\alpha$- or $\varepsilon$-clipping is active, here an average report of 0.69. Below that the minimum loan, $f_{\min}=0.4$ is adopted.
For borrowers with an average report  above $\sim$0.9, the mechanism provides the maximum loan size. \label{fig:sizevsreport-profit}}
\end{figure}
\begin{figure}[t]
\centering
\includegraphics[width=0.8\linewidth]{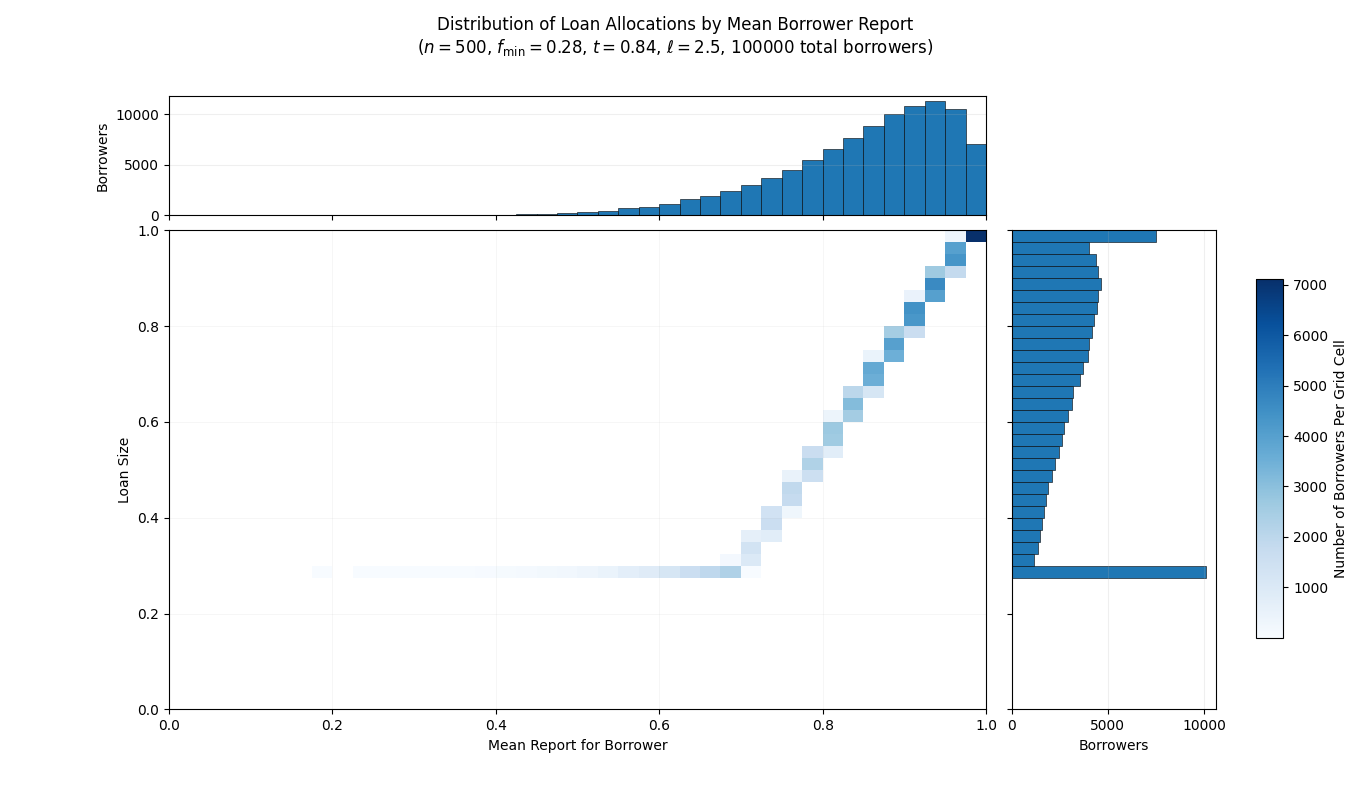}
\Description{Mean Report vs Loan Size, ROI.}
\caption{Borrower-average report versus assigned loan size: Optimal expected ROI. $n=500,\ell=2.5$. Upper histogram: borrower-average reports. Right-hand histogram: distribution on loan sizes.
Borrower-average reports linearly determine loan assignment until $\alpha$- or $\varepsilon$-clipping is active, here an average report of 0.70. Below that the minimum loan, $f_{\min}=0.28$ is adopted. In the optimal ROI setting, loan sizes only reach the maximum for the very highest-rated borrowers.
\label{fig:sizevsreport-roi}}
\end{figure}

\mjyadd{\Cref{fig:mean_report_vs_loan_size} shows how $f_{\min}$ sets the minimum loan size, 
while also  controlling the amount $t-\alpha = t-(1-f_{\min})/2$ such that 
when all reports are below this amount the  minimum loan is assigned.}
\dcpadd{See also \Cref{fig:sizevsreport-profit} and~\Cref{fig:sizevsreport-roi}, which show joint density plots
for borrower-average report and loan size as well as the marginal distributions on each (top and right, respectively). 
The profit-optimal setting gives the maximum loan to all borrowers with a mean report of over $\sim$0.9, while the ROI-optimal setting gives many borrowers 
the minimum loan and only the very highest-rated borrowers receive the maximum loan.}

%

\subsubsection{Effect of varying the number of recommenders, $n$}

\begin{figure}[t]
\centering
\includegraphics[width=0.495\linewidth]{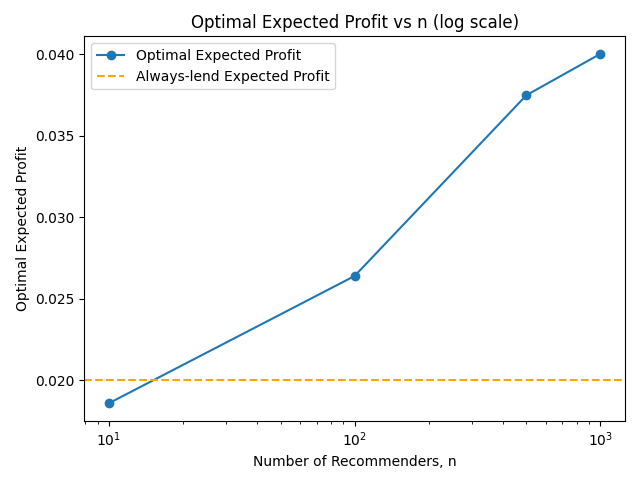}
\includegraphics[width=0.495\linewidth]{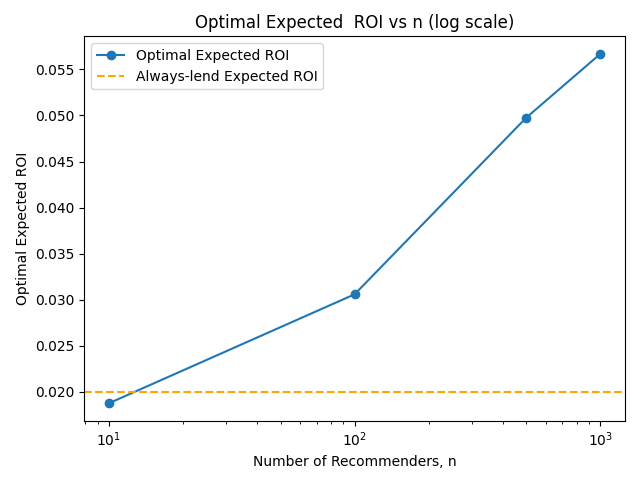}
\Description{Optimal expected profit and ROI under the optimal parameter settings for several numbers of recommenders.}
\caption{Optimal expected profit (left) and optimal expected ROI (right) in the loan setting with $\ell=2.5$,
varying  the number of recommenders $n$ and
choosing for each $n$, and each of profit and ROI, the optimal setting of $f_{\min}$ and $t$. 
For enough recommenders, the system adds  more profit and ROI than the always-lend policy.
\label{fig:max_profit_vs_n}}
\end{figure}

\mjyadd{We also vary $n$, with $n\in \{10, 100, 500,  1000\}$.
For each recommender size, $n$, we determine through simulation the optimal choice of 
centering parameter, $t$, and minimum loan, $f_{\min}$, for each of
expected profit and expected ROI. We fix $\ell=2.5$. As shown in~\Cref{fig:max_profit_vs_n}, 
each of the optimal expected profit and optimal expected ROI  is   monotonically-increasing with  $n$.
The optimal expected profit ranges  from $0.0186$ with 10 recommenders to $0.04$ with 1000 recommenders, 
compared to $0.02$ for always-lend.} \dcpadd{The optimal expected ROI ranges  from $0.0188$ with 10 recommenders to $0.0567$ with 1000 recommenders, compared to $0.02$ for always-lend}.  Further, by lowering the value of $\ell$ \dcpadd{to $\ell=1$,}
we find in simulation that  we can exceed the profit and ROI of the always-lend policy even with 10 recommenders.

\subsection{Discussion}

\mjyadd{The simulation, which is configured   on data on repayment and recommender reports from a study conducted in Uganda,  suggests  the adaptive-payment mechanism and the bounded-linear target decision function can create value for real-world decision makers. 
A practical tension is between the desire to use more recommenders, both for increased accuracy and 
in requiring a lower 
 budget  to overcome bribes, versus the desire to reduce the number of recommenders to a number that can easily be managed, would have expertise about the decision, and would each earn enough 
   after dividing the budget  to be motivated. 
We show that with as few as ten recommenders in our setting, 
each of these considerations can be met.

The adaptive-payment mechanism concentrates the recommender incentive where their report influences the decision.
While most decision scoring rules waste some of their budget compensating recommenders in report regions that 
are not relevant to the decision rule, 
the adaptive-payment mechanism  puts the incentive dollars where they count, 
providing a  practically-implementable system.

The principal has flexibility about how to use  recommender reports to make a decision.
The principal can choose a lower minimum loan allocation $f_{\min}$}  \dcpadd{and thus higher sensitivity to reports},
\mjyadd{they can save money by decreasing the slope  $\ell$ of the bounded-linear target decision function, and they can choose where to} 
\dcpadd{center the responsiveness of the decision rule to individual belief reports.}
\mjyadd{We believe  these simulation results pave the way for decision architects to cleverly craft how much information they want to pay for and how they want to use it to make decisions across various contexts, 
while handling the possibility of a briber who seeks to influence the decision.}
%

\section{Proofs}

\subsection{Proof of \Cref{thm:lower}}
We start by proving a few results that we will need to prove \Cref{thm:lower}.
\begin{lem}
\label{lem:averagePath}Suppose there is a list of functions, $f_{1},...,f_{n}:[0,1]^{n}\to\mathbb{R}_{\ge0}$ such that $\sum_{i=1}^{n}f_{i}(x)\le\beta~\forall x\in\mathbb{R}^{n}$, and $\mathcal{W}=(\mathcal{K},\mathcal{D}_{0},\mathcal{D}_{z})$ is some walk \Cref{def:walk}. Then, with $k=|\mathcal{K}|$, we have 
\[
\sum_{m=1}^{k}\mathbb{E}_{X\sim\mathcal{W}}\left[f_{[X^{m}\neq X^{m-1}]}\left(X^{m}\right)+f_{[X^{m}\neq X^{m-1}]}\left(X^{m-1}\right)\right]\le2\cdot\beta\cdot H(k),
\]
where $H$ is the harmonic number.
\end{lem}

\begin{proof}
 \mathauditadd{Relabel $\mathcal K=[k]$. }Dropping $X\sim\mathcal{W}$, and using the random variables as defined in \Cref{def:walk}, we have
\[
\mathbb{E}\left[f_{[X^{m}\neq X^{m-1}]}\left(X^{m}\right)+f_{[X^{m}\neq X^{m-1}]}\left(X^{m-1}\right)\right]=\mathbb{E}\left[f_{\Theta^{m}}\left(X^{m}\right)+f_{\Theta^{m}}\left(X^{m-1}\right)\right].
\]
For convenience, we define a few functions: For any set $T\subseteq\{1,..,k\}$, let's define the random variables 
\[
Y_{i}(T):=\begin{cases}
X_{i}^{0}+Z_{i} & \text{if }i\in T\\
X_{i}^{0} & \text{otherwise}.
\end{cases}
\]
Further, let $\Sigma$ be the set of all permutations on $\mathcal{K}$ and let for any $\sigma\in\Sigma$
\[
S^{\sigma}(m):=\{\sigma(1),..,\sigma(m)\}.
\]
We hence have 
\[
X^{j}=Y\left(S^{\Theta}(j)\right).
\]
Defining 
\begin{align*}
g_{\Theta^{m}}\left(S^{\Theta}(j)\right): & =f_{\Theta^{m}}\left(Y\left(S^{\Theta}(j)\right)\right)\\
 & =f_{\Theta^{m}}\left(X^{j}\right),
\end{align*}
we have 
\begin{align*}
\mathbb{E}\left[f_{\Theta^{m}}\left(X^{m}\right)+f_{\Theta^{m}}\left(X^{m-1}\right)\right] & =\mathbb{E}\left[g_{\Theta^{m}}\left(S^{\Theta}(m)\right)\right]+\mathbb{E}\left[g_{\Theta^{m}}\left(S^{\Theta}(m-1)\right)\right].
\end{align*}
We will now prove that 
\[
\sum_{m=1}^{k}\mathbb{E}\left[g_{\Theta^{m}}\left(S^{\Theta}(m-1)\right)\big|x^{0},z\right]\le H(k)\cdot\beta\quad\forall x^{0},z,
\]
which implies that 
\[
\sum_{m=1}^{k}\mathbb{E}\left[g_{\Theta^{m}}\left(S^{\Theta}(m-1)\right)\right]\le H(k)\cdot\beta.
\]

Let $\mathbb{I}[\cdot]$ denote the indicator function. We have
\begin{align*}
 & \sum_{m=1}^{k}\mathbb{E}\left[g_{\Theta^{m}}\left(S^{\Theta}(m-1)\right)\big|x^{0},z\right]\cdot k!\\
 & =\sum_{\sigma\in\Sigma}\sum_{m=1}^{k}g_{\sigma(m)}\left(S^{\sigma}(m-1)\right)\\
 & =\sum_{\sigma\in\Sigma}\sum_{m=1}^{k}\sum_{n=1}^{k}\sum_{Z}\mathbb{I}\left[\sigma(m)=n\text{ and }Z=S^{\sigma}\left(m-1\right)\right]g_{n}\left(Z\right)\\
 & =\sum_{n=1}^{k}\sum_{Z}g_{n}\left(Z\right)\sum_{m=1}^{k}\sum_{\sigma\in\Sigma}\mathbb{I}\left[\sigma(m)=n\text{ and }Z=S^{\sigma}\left(m-1\right)\right]\\
 & =\sum_{n=1}^{k}\sum_{Z}g_{n}\left(Z\right)\sum_{m=1}^{k}\sum_{\sigma\in\Sigma}\mathbb{I}\left[\sigma(m)=n\text{ and }Z=S^{\sigma}\left(m-1\right)\right]\mathbb{I}\left[|Z|=m-1\right]\mathbb{I}\left[n\notin Z\right].
\end{align*}

The first indicator function requires that the first $m-1$ elements of the permutation $\sigma$ are $Z$, the $m$-th element is $n$, and the last $k-m$ elements are $(\{1,..,k\}\setminus Z)\setminus\{n\}$. The degrees of freedom of such a permutation are the order of the first $m-1$ elements and the order of the last $k-m$ elements, we hence have 
\begin{align*}
 & =\sum_{n=1}^{k}\sum_{Z}g_{n}\left(Z\right)\sum_{m=1}^{k}(m-1)!(k-m)!\mathbb{I}\left[|Z|=m-1\right]\mathbb{I}\left[n\notin Z\right]\\
 & \le\sum_{Z}\sum_{m=1}^{k}(m-1)!(k-m)!\mathbb{I}\left[|Z|=m-1\right]\sum_{n=1}^{k}g_{n}\left(Z\right).
\end{align*}

Since $\sum_{i=1}^{n}f_{i}(x)\le\beta$, we have $\sum_{n=1}^{k}g_{n}\left(Z\right)\le\beta$ and hence
\begin{align*}
\text{\ensuremath{\sum_{\sigma\in\Sigma}\sum_{m=1}^{k}g_{\sigma(m)}\left(S^{\sigma}(m-1)\right)}} & \le\sum_{Z}\sum_{m=1}^{k}(m-1)!(k-m)!\mathbb{I}\left[|Z|=m-1\right]\cdot\beta\\
 & =\sum_{m=1}^{k}(m-1)!(k-m)!{k \choose m-1}\cdot\beta\\
 & =\sum_{m=1}^{k}\frac{k!}{(k-m+1)}\cdot\beta\\
 & =\sum_{m=1}^{k}\frac{k!}{m}\cdot\beta\\
 & =k!\cdot H(k)\cdot\beta.
\end{align*}
For the other term, we can make a similar argument: 
\begin{align*}
 & \sum_{m=1}^{k}\mathbb{E}\left[g_{\Theta^{m}}\left(S^{\Theta}(m)\right)\big|x^{0},z\right]\cdot k!\\
 & =\sum_{\sigma\in\Sigma}\sum_{m=1}^{k}g_{\sigma(m)}\left(S^{\sigma}(m)\right)\\
 & =\sum_{\sigma\in\Sigma}\sum_{m=1}^{k}\sum_{n=1}^{k}\sum_{Z}\mathbb{I}\left[\sigma(m)=n\text{ and }Z=S^{\sigma}\left(m\right)\right]g_{n}\left(Z\right)\\
 & =\sum_{n=1}^{k}\sum_{Z}g_{n}\left(Z\right)\sum_{m=1}^{k}\sum_{\sigma\in\Sigma}\mathbb{I}\left[\sigma(m)=n\text{ and }Z=S^{\sigma}\left(m\right)\right]\\
 & =\sum_{n=1}^{k}\sum_{Z}g_{n}\left(Z\right)\sum_{m=1}^{k}\sum_{\sigma\in\Sigma}\mathbb{I}\left[\sigma(m)=n\text{ and }Z=S^{\sigma}\left(m\right)\right]\mathbb{I}\left[|Z|=m\right]\mathbb{I}\left[n 
  \mathauditreplace{\in}{\notin} Z\right]\\
 & =\sum_{n=1}^{k}\sum_{Z}g_{n}\left(Z\right)\sum_{m=1}^{k}(m-1)!(k-m)!\mathbb{I}\left[|Z|=m\right]\mathbb{I}\left[n 
  \mathauditreplace{\in}{\notin} Z\right]\\
 & \le\sum_{Z}\sum_{m=1}^{k}(m-1)!(k-m)!\mathbb{I}\left[|Z|=m\right]\sum_{n=1}^{k}g_{n}\left(Z\right)\\
 & \le\sum_{Z}\sum_{m=1}^{k}(m-1)!(k-m)!\mathbb{I}\left[|Z|=m\right]\cdot\beta\\
 & =\sum_{m=1}^{k}(m-1)!(k-m)!{k \choose m}\cdot\beta\\
 & =\sum_{m=1}^{k}\frac{k!}{m}\cdot\beta\\
 & =k!\cdot H(k)\cdot\beta.
\end{align*}
The result follows straightforwardly. 
\end{proof}
\begin{lem}
\label{thm:sumOfIncentives} For any proper scoring rule $\textsc{sc}$, with $\beta\ge\textsc{sc}(x,y)\ge0,\ \forall x,y\in{[0,1]}$, and any monotone sequence $x^{(0)},..,x^{(K)}\in{[0,1]}$, such that $x^{(k+1)}-x^{(k)}=x^{(j+1)}-x^{(j)}\ \forall k,j\in\mathauditreplace{\{0,\ldots,K-1\}}{[K-1]}$, we have
\begin{align*}
 & \sum_{k=1}^{K}\textsc{d-sc}\left(x^{(k-1)},x^{(k)}\right)=\frac{1}{K}\cdot\textsc{d-sc}(x^{(0)},x^{(K)})
\end{align*}
with 
\[
\textsc{d-sc}(a,b):=\textsc{sc}(a,a)-\textsc{sc}(b,a)+\textsc{sc}(b,b)-\textsc{sc}(a,b).
\]
\end{lem}

\begin{proof}
Any scoring rule for dichotomous RVs can be written as 
\begin{align*}
\textsc{sc}(x,r) & =g(x)r+h(x)(1-r),
\end{align*}
for some $g(x)$ and $h(x)$. Using the identity 
\begin{align*}
\textsc{sc}(a,b)-\textsc{sc}(a,c) & =g(a)b+h(a)(1-b)-g(a)c-h(a)(1-c)\\
 & =\left(g(a)-h(a)\right)(b-c),
\end{align*}
we have 
\begin{align*}
 & \textsc{sc}(x,r)-\textsc{sc}(y,r)\\
 & =\left(\textsc{sc}(x,r)-\textsc{sc}(x,x)\right)+\left(\textsc{sc}(y,x)-\textsc{sc}(y,r)\right)+\textsc{sc}(x,x)-\textsc{sc}(y,x)\\
 & =\left(g(x)-h(x)\right)(r-x)+\left(g(y)-h(y)\right)(x-r)+\textsc{sc}(x,x)-\textsc{sc}(y,x)\\
 & =\left(\left(g(x)-h(x)\right)(x-y)+\left(g(y)-h(y)\right)(y-x)\right)\frac{x-r}{y-x}+\textsc{sc}(x,x)-\textsc{sc}(y,x)\\
 & =\left(\textsc{sc}(x,x)-\textsc{sc}(x,y)+\textsc{sc}(y,y)-\textsc{sc}(y,x)\right)\frac{x-r}{y-x}+\textsc{sc}(x,x)-\textsc{sc}(y,x)\\
 & =\left(\textsc{sc}(x,x)-\textsc{sc}(y,x)\right)\frac{y-r}{y-x}+\left(\textsc{sc}(y,y)-\textsc{sc}(x,y)\right)\frac{x-r}{y-x}.
\end{align*}
With $r=0$, we have 
\[
\textsc{sc}(x,0)-\textsc{sc}(y,0)=\left(\textsc{sc}(x,x)-\textsc{sc}(y,x)\right)\frac{y}{y-x}+\left(\textsc{sc}(y,y)-\textsc{sc}(x,y)\right)\frac{x}{y-x},
\]
and with $r=1$, we have 
\[
\textsc{sc}(x,1)-\textsc{sc}(y,1)=\left(\textsc{sc}(x,x)-\textsc{sc}(y,x)\right)\frac{y-1}{y-x}+\left(\textsc{sc}(y,y)-\textsc{sc}(x,y)\right)\frac{x-1}{y-x}.
\]
Subtracting the second equality from the first, we obtain 
\begin{align}
\frac{\textsc{sc}(x,x)-\textsc{sc}(y,x)+\textsc{sc}(y,y)-\textsc{sc}(x,y)}{y-x} & =\textsc{sc}(x,0)-\textsc{sc}(y,0)-\left(\textsc{sc}(x,1)-\textsc{sc}(y,1)\right)\nonumber \\
\textsc{sc}(x,x)-\textsc{sc}(y,x)+\textsc{sc}(y,y)-\textsc{sc}(x,y) & =(y-x)\left(\textsc{sc}(x,0)-\textsc{sc}(y,0)-\textsc{sc}(x,1)+\textsc{sc}(y,1)\right).\label{eq:scoring-rule-ID-1}
\end{align}
Now, for an increasing sequence $x^{(0)},..,x^{(K)}$ with $|x^{(k+1)}-x^{(k)}|=\frac{\left|x^{(K)}-x^{(0)}\right|}{K}\quad\forall k$, we have 
\begin{align*}
 & \sum_{k=0}^{K-1}\textsc{d-sc}(x^{(k)},x^{(k+1)})\\
 & =\sum_{k=0}^{K-1}\left(\textsc{sc}(x^{(k)},x^{(k)})-\textsc{sc}(x^{(k+1)},x^{(k)})+\textsc{sc}(x^{(k+1)},x^{(k+1)})-\textsc{sc}(x^{(k)},x^{(k+1)})\right)\\
 & =\sum_{k=0}^{K-1}(x^{(k+1)}-x^{(k)})\overbrace{\left(\textsc{sc}(x^{(k)},0)-\textsc{sc}(x^{(k+1)},0)-\textsc{sc}(x^{(k)},1)+\textsc{sc}(x^{(k+1)},1)\right)}^{\ge0}\ (\mathrm{From\eqref{eq:scoring-rule-ID-1}})\\
 & =\frac{\left|x^{(K)}-x^{(0)}\right|}{K}\cdot\sum_{k=\mathauditreplace{0}{1}}^{K-1}\left(\textsc{sc}(x^{(k)},0)-\textsc{sc}(x^{(k+1)},0)-\textsc{sc}(x^{(k)},1)+\textsc{sc}(x^{(k+1)},1)\right)\\
 & =\frac{\left|x^{(K)}-x^{(0)}\right|}{K}\cdot\left(\textsc{sc}(x^{(0)},0)-\textsc{sc}(x^{(K)},0)-\textsc{sc}(x^{(0)},1)+\textsc{sc}(x^{(K)},1)\right)\\
 & =\frac{1}{K}\cdot\left(\textsc{sc}(x^{(0)},x^{(0)})-\textsc{sc}(x^{(K)},x^{(0)})-\textsc{sc}(x^{(0)},x^{(K)})+\textsc{sc}(x^{(K)},x^{(K)})\right)\\
 & =\frac{1}{K}\cdot\textsc{d-sc}(x^{(0)},x^{(K)}),
\end{align*}
and similarly for a decreasing sequence. Hence, the result follows.
\end{proof}
\begin{lem}
\label{lem:worstcaseBels}For any mechanism $\textsc{mech}=(\textsc{F},\textsc{pay})$ with budget $\beta\ge0$, any $\gamma\mathauditreplace{>}{\ge}0$, $k\in[n]$, \mathauditadd{and }$\delta\in\mathauditreplace{(}{[}0,1]$, there exists a $q\in\mathcal{Q}$, $t\in[n]$, $x\in[0,1]$, $\revisionredreplace{c_t=\gamma}{c_{t}\in\{-\gamma,\gamma\}}$, with $|q_{t}-x|\le\delta$, such that
\begin{align*}
\frac{\gamma\cdot\Lambda_{\textsc{F}}(k,\delta)^{2}}{\revisionredreplace{16}{8}\cdot k\cdot\delta\cdot\frac{\beta}{\gamma}\cdot H(k)+4\cdot k\cdot\Lambda_{\textsc{F}}(k,\delta)}\le & \mathbb{E}_{O\sim q_{t}}\left[\textsc{pay}_{t}\left((x,q_{-t}),O\right)-\textsc{pay}_{t}(q,O)\right]+c_{t}\left(\textsc{F}\left(x,q_{-t}\right)-\textsc{F}\left(q\right)\right)
\end{align*}
where $H$ is the harmonic number.
\revisionredadd{If signed competing incentives are allowed, the same conclusion holds with $c_t\in\{-\gamma,\gamma\}$ and with $8$ in place of $16$.}
\end{lem}

\begin{proof}
For convenience, we will sometimes use the notation 
\begin{align*}
\textsc{pay}_{i}(r_{i},q_{i}|r_{-i}): & =\mathbb{E}_{O\sim q_{i}}\left[\textsc{pay}_{i}(r,O)\right]\\
 & =\underbrace{\textsc{pay}_{i}(r,1)}_{=:g_{i}(r)}\cdot q_{i}+\underbrace{\textsc{pay}_{i}(r,0)}_{=:h_{i}(r)}\cdot(1-q_{i}).
\end{align*}
We have $\textsc{pay}_{i}(r_{i},o|r_{-i})=\textsc{pay}_{i}(r,o)~\forall o\in\mathcal{O}$. By definition of our mechanism (\Cref{def:mechanism}), $\textsc{pay}_{i}(r_{i},q_{i}|r_{-i})$ is a strictly proper scoring rule in $r_{i},q_{i}$. For any $a,b\in\mathcal{R}$ that differ only at index $i$, let 
\[
\textsc{d-pay}(a,b):=\textsc{pay}_{i}(a_{i},a{}_{i}|a_{-i})-\textsc{pay}_{i}(b_{i},a_{i}|a_{-i})+\textsc{pay}_{i}(b_{i},b_{i}|a_{-i})-\textsc{pay}_{i}(a_{i},b_{i}|a_{-i}),
\]
which we can interpret as a notion of the cost of misreporting. We will see below how to 
use this quantity. We have
\begin{align*}
&\textsc{d-pay}(a,b) \\
& =g(a)a_{i}+h(a)(1-a_{i})-g(b)a_{i}-h(b)(1-a_{i})+g(b)b_{i}+h(b)(1-b_{i})-g(a)b_{i}-h(a)(1-b_{i})\\
 & =\left(g(a)-h(a)-g(b)+h(b)\right)a_{i}+h(a)-h(b)+\left(g(b)-h(b)-g(a)+h(a)\right)b_{i}+h(b)-h(a)\\
 & =\left(g(a)-h(a)-g(b)+h(b)\right)a_{i}+\left(g(b)-h(b)-g(a)+h(a)\right)b_{i}\\
 & =\left(b_{i}-a_{i}\right)\left(g_{i}\left(b\right)-g_{i}\left(a\right)+h_{i}\left(a\right)-h_{i}\left(b\right)\right).
\end{align*}
For any $\mathcal{W}\in\textsc{walks}(k,\delta)$, we have
\begin{align*}
\textsc{cost}: & =\mathbb{E}_{x\sim\mathcal{W}}\left[\sum_{m=1}^{k}\textsc{d-pay}(x^{m-1},x^{m})\right]\\
 & \revisionredreplace{\le}{=}\sum_{m=1}^{k}\mathbb{E}_{x\sim\mathcal{W}}\left[\left\Vert x^{m}-x^{m-1}\right\Vert _{1}\revisionredadd{\lvert}\left(g_{[x^{m}\neq x^{m-1}]}\left(x^{m}\right)-g_{[x^{m}\neq x^{m-1}]}\left(x^{m-1}\right)\right)\revisionredadd{\rvert}\right]\\
 & +\sum_{m=1}^{k}\mathbb{E}_{x\sim\mathcal{W}}\left[\left\Vert x^{m}-x^{m-1}\right\Vert _{1}\revisionredadd{\lvert}\left(h_{[x^{m}\neq x^{m-1}]}\left(x^{m-1}\right)-h_{[x^{m}\neq x^{m-1}]}\left(x^{m}\right)\right)\revisionredadd{\rvert}\right]\\
 & \le\delta\cdot\sum_{m=1}^{k}\mathbb{E}_{x\sim\mathcal{W}}\left[\revisionredadd{\lvert}g_{[x^{m}\neq x^{m-1}]}\left(x^{m}\right)-g_{[x^{m}\neq x^{m-1}]}\left(x^{m-1}\right)\revisionredadd{\rvert}+\revisionredadd{\lvert}h_{[x^{m}\neq x^{m-1}]}\left(x^{m-1}\right)-h_{[x^{m}\neq x^{m-1}]}\left(x^{m}\right)\revisionredadd{\rvert}\right]\\
 & \le\delta\cdot\sum_{m=1}^{k}\mathbb{E}_{x\sim\mathcal{W}}\left[g_{[x^{m}\neq x^{m-1}]}\left(x^{m}\right)+g_{[x^{m}\neq x^{m-1}]}\left(x^{m-1}\right)\right]\\
 & +\delta\cdot\sum_{m=1}^{k}\mathbb{E}_{x\sim\mathcal{W}}\left[h_{[x^{m}\neq x^{m-1}]}\left(x^{m-1}\right)+h_{[x^{m}\neq x^{m-1}]}\left(x^{m}\right)\right].
\end{align*}

Since $g_{i}\left(x\right)=\textsc{pay}_{i}\left(x,1\right)$ we have 
\[
\sum_{i=1}^{k}g_{i}\left(x\right)\le\beta\quad\forall x,
\]
and similarly for $h$, hence we can apply \Cref{lem:averagePath} twice and we have
\[
\textsc{cost}\le4\cdot\delta\cdot\beta\cdot H(k).
\]
Let $\Lambda(k,\delta)$ be the sensitivity function from \Cref{def:sensitivityLower} and $\mathcal{W}$ its walk. We will sometimes simply write $\Lambda$ without its arguments, for simplicity.  \mathauditadd{If $\Lambda=0$, the claim is immediate, so assume $\Lambda>0$. } We hence have for any $\omega\ge0$ 
\begin{align*}
\omega\cdot\Lambda-\revisionredreplace{4}{2}\cdot\delta\cdot\beta\cdot H(k) & \le\omega\cdot\mathbb{E}_{X\sim\mathcal{W}}\left[\sum_{i=1}^{k}\left|\textsc{F}(x^{i})-\textsc{F}(x^{i-1})\right|\right]-\mathauditdelete{\frac{1}{2}}\textsc{cost}\\
 & =\omega\cdot\mathbb{E}_{X\sim\mathcal{W}}\left[\sum_{i=1}^{k}\left|\textsc{F}(x^{i})-\textsc{F}(x^{i-1})\right|\right]-\mathauditdelete{\frac{1}{2}}\mathbb{E}_{x\sim\mathcal{W}}\left[\sum_{m=1}^{k}\textsc{d-pay}(x^{m-1},x^{m})\right].
\end{align*}

By the pigeonhole principle, there must be a path $x$ and a step $m$ such that
\begin{equation}
\omega\cdot\Lambda-\revisionredreplace{4}{2}\cdot\delta\cdot\beta\cdot H(k)\le k\left(\omega\left|\textsc{F}(x^{m})-\textsc{F}(x^{m-1})\right|-\mathauditdelete{\frac{1}{2}}\textsc{d-pay}(x^{m-1},x^{m})\right).
\label{eq:lemma7-walk-step}
\end{equation}
Let now $z^{0},...,z^{J}$ be the linear interpolation between $x^{m-1}$ and $x^{m}$. We have
\begin{align*}
\textsc{d-pay}(x^{m-1},x^{m}) & =\textsc{d-pay}(z^{0},z^{J})\\
 & =J\cdot\sum_{j=1}^{J}\textsc{d-pay}(z^{j-1},z^{j}),
\end{align*}
where the last equality follows from \Cref{thm:sumOfIncentives}. Plugging this in, we have 
\begin{align*}
\frac{1}{k}\left(\Lambda\cdot\omega-\revisionredreplace{4}{2}\cdot\delta\cdot\beta\cdot H(k)\right)\le & \omega\left|\sum_{j=0}^{J-1}\left(\textsc{F}\left(z^{j+1}\right)-\textsc{F}\left(z^{j}\right)\right)\right|-\mathauditdelete{\frac{1}{2}}J\cdot\sum_{j=1}^{J}\textsc{d-pay}(z^{j-1},z^{j})\\
\frac{1}{kJ^{2}}\left(\Lambda\cdot\omega-\revisionredreplace{4}{2}\cdot\delta\cdot\beta\cdot H(k)\right)\le\frac{1}{J} & \sum_{j=0}^{J-1}\left(\frac{\omega}{J}\left|\left(\textsc{F}\left(z^{j+1}\right)-\textsc{F}\left(z^{j}\right)\right)\right|-\mathauditdelete{\frac{1}{2}}\textsc{d-pay}(z^{j},z^{j+1}\right).
\end{align*}
Applying the pigeonhole principle again, it follows that there are report profiles $z,z'$ such that (i) they differ only in one dimension (by at most $\delta$); \revisionredadd{relabel them if necessary so that $\textsc F(z')\ge\textsc F(z)$;} and (ii)
\begin{align}
\frac{1}{kJ^{2}}\left(\Lambda\cdot\omega-\revisionredreplace{4}{2}\cdot\delta\cdot\beta\cdot H(k)\right) & \le\frac{\omega}{J}\cdot\mathauditdelete{\text{sign}\left(\textsc{F}\left(z'\right)-\textsc{F}\left(z\right)\right)\cdot}\left(\textsc{F}\left(z'\right)-\textsc{F}\left(z\right)\right)-\mathauditdelete{\frac{1}{2}}\textsc{d-pay}(z,z').
\label{eq:lemma7-substep}
\end{align}
With $c:=\frac{\omega}{J}\mathauditdelete{\cdot\text{sign}\left(\textsc{F}\left(z'\right)-\textsc{F}\left(z\right)\right)}\revisionredadd{>0}$, we have

\begin{align*}
\frac{1}{kJ^{2}}\left(\Lambda\cdot\mathauditdelete{|}c\mathauditdelete{|}\cdot J-\revisionredreplace{4}{2}\cdot\delta\cdot\beta\cdot H(k)\right) & \le c\cdot\left(\textsc{F}\left(z'\right)-\textsc{F}\left(z\right)\right)-\mathauditdelete{\frac{1}{2}}\textsc{d-pay}(z,z').
\end{align*}

\begin{align*}
\frac{1}{kJ}\left(\Lambda\cdot\mathauditdelete{|}c\mathauditdelete{|}-\frac{1}{J}\cdot\revisionredreplace{4}{2}\cdot\delta\cdot\beta\cdot H(k)\right) & \le c\left(\textsc{F}\left(z'\right)-\textsc{F}\left(z\right)\right)-\mathauditdelete{\frac{1}{2}}\textsc{d-pay}(z,z')
\end{align*}
This holds for any $J\in\mathbb{N}_{>0}$, let's pick
\[
J\ \mathauditdelete{=}\mathauditadd{:=\max\bigl\{1,}\left\lceil\frac{\revisionredreplace{8}{4}\cdot\delta\cdot\beta\cdot H(k)}{\revisionredreplace{\gamma}{c}\cdot\Lambda}\right\rceil\mathauditadd{\bigr\}},
\qquad
\revisionredadd{\epsilon:=J-\frac{8\cdot\delta\cdot\beta\cdot H(k)}{\gamma\cdot\Lambda}\in[0,1].}
\]
\mathauditdeleteddisplay{J=\frac{4\cdot\delta\cdot\beta\cdot H(k)}{c\cdot\Lambda}+\epsilon}
where $\left\lceil \cdot\right\rceil$ means rounding up to the next integer. With this $J$, \mathauditadd{set $\omega:=\gamma J$, so $c=\gamma$.} We have
\begin{align*}
\frac{1}{kJ}\left(\Lambda\cdot c-\frac{1}{J}\cdot\revisionredreplace{4}{2}\cdot\delta\cdot\beta\cdot H(k)\right) & =\frac{c\mathauditadd{^{2}}\cdot\Lambda^{2}\cdot\left(\revisionredreplace{4}{2}\cdot\delta\cdot\beta\cdot H(k)+\epsilon\cdot c\cdot\Lambda\right)}{k\cdot\left(\revisionredreplace{8}{4}\cdot\delta\cdot\beta\cdot H(k)+\epsilon\cdot c\cdot\Lambda\right)^{2}}\\
 & \ge\frac{c\mathauditadd{^{2}}\cdot\Lambda^{2}\cdot\left(\revisionredreplace{4}{2}\cdot\delta\cdot\beta\cdot H(k)+\epsilon\cdot c\cdot\Lambda\right)}{k\cdot\left(\revisionredreplace{8}{4}\cdot\delta\cdot\beta\cdot H(k)+2\cdot\epsilon\cdot c\cdot\Lambda\right)^{2}}\\
 & =\frac{c\mathauditadd{^{2}}\cdot\Lambda^{2}}{k\cdot4\cdot\left(\revisionredreplace{4}{2}\cdot\delta\cdot\beta\cdot H(k)+\epsilon\cdot c\cdot\Lambda\right)}\\
 & \ge\frac{c\mathauditadd{^{2}}\cdot\Lambda^{2}}{k\cdot4\cdot\left(\revisionredreplace{4}{2}\cdot\delta\cdot\beta\cdot H(k)+c\cdot\Lambda\right)}.
\end{align*}
Hence, for any $\gamma\mathauditreplace{>}{\ge}0$, there is a $q$, an $i$, an $r_i$, and $\revisionredreplace{c_i=\gamma}{c_i\in\{-\gamma,\gamma\}}$ such that
\begin{align*}
\frac{\gamma\cdot\Lambda^{2}}{k\cdot4\cdot\left(\revisionredreplace{4}{2}\cdot\delta\cdot\frac{\beta}{\gamma}\cdot H(k)+\Lambda\right)} & \le-\mathauditdelete{\frac{1}{2}}\textsc{d-pay}((q_{-i},r_i),q)+c_i\left(\textsc F((q_{-i},r_i))-\textsc F(q)\right)\\
 & \le\textsc{pay}_{i}(r_i,q_i|q_{\neg i})-\textsc{pay}_{i}(q_i,q_i|q_{\neg i})+c_i\left(\textsc F((q_{-i},r_i))-\textsc F(q)\right).
\end{align*}
\revisionredadd{The last inequality uses that each directional scoring loss is at most $\textsc{d-pay}$. For the signed refinement, use $-\frac{1}{2}\textsc{cost}$ and $-\frac{1}{2}\textsc{d-pay}$ above, choose the direction with the smaller scoring loss, and choose the sign of $c_i$ to reward the resulting change in $\textsc F$. Then $4$ replaces $8$ in $J$, yielding $c_i\in\{-\gamma,\gamma\}$ and coefficient $8$.}
\end{proof}
We are now ready to prove \Cref{thm:lower}: 
\begin{proof}
Fix some $k\in[n]$, $\delta\in\mathauditreplace{(}{[}0,1]$, and let $q$, $t$, $x$, $c_{t}$  \mathauditadd{be } as in \mathauditadd{the signed refinement of }\Cref{lem:worstcaseBels}. Suppose that the prior $D$ only has support at
\begin{align*}
P_{i}(o|y_{i}) & =p_{i}(o|y_{i}):=\begin{cases}
q_{i} & \text{if }o=1\\
1-q_{i} & \text{if }o=0
\end{cases}\quad\forall y_{i}\\
C_{i} & =c_{i}:=\begin{cases}
c_{t} & i=t\\
0 & i\neq t
\end{cases}\\
Y & =y:=0.
\end{align*}
Then 
\begin{align*}
\textsc{util}_{i}\left(p,y,c_{i},r\right) & =\mathbb{E}_{O\sim p_{i}(\cdot|y)}\left[\textsc{pay}_{i}(r,O)+c_{i}\cdot\textsc{F}(r)\right]\\
 & =\textsc{pay}_{i}(r_{i},q_{i}|r_{-i})+c_{i}\cdot\textsc{F}(r).
\end{align*}
Hence, we have
\begin{align*}
\textsc{eutil}_{i}\left(s,p_{i},y_{i},c_{i}\right) & :=\mathbb{E}\left[\textsc{util}_{i}\left(P_{i},Y,C_{i},R\right)\big|s,p_{i},y_{i},c_{i}\right].
\end{align*}

Since all priors have only point-support, this is not an incomplete-information game anymore, and we can identify strategies with distributions over reports, we hence write, with slight abuse of notation, 
\begin{equation}
\textsc{eutil}_{i}\left(s,p_{i},y_{i},c_{i}\right)=\mathbb{E}_{R\sim s}\left[\textsc{pay}_{i}\left(R_{i},q_{i}\bigg|R_{-i}\right)+c_{i}\cdot\textsc{F}\left(R\right)\right].\label{eq:utilProofLower}
\end{equation}
For any $i\neq t$ we have 
\[
\textsc{eutil}_{i}\left(s,p_{i},y_{i},c_{i}\right)=\mathbb{E}_{R\sim s}\left[\textsc{pay}_{i}\left(R_{i},q_{i}\bigg|R_{-i}\right)\right].
\]
Since  $\textsc{pay}_{i}\left(r_{i},q_{i}\bigg|r_{-i}\right)$ is a strictly proper scoring rule, we have for any $i\neq t$ that $R_{i}=q_{i}$with probability 1. For $t$ we hence have 
\begin{align*}
\textsc{eutil}_{t}\left(s,p_{t},y_{t},c_{t}\right) & =\mathbb{E}_{R_{t}\sim s_{t}}\left[\textsc{pay}_{t}\left(R_{t},q_{t}\bigg|q_{-t}\right)+c_{t}\cdot\textsc{F}\left(R_{t},q_{-t}\right)\right],
\end{align*}
and hence 
\[
R_{t}\in\arg\max_{r_{t}\in\mathcal{R}_{t}}\left(\textsc{pay}_{t}\left(r_{t},q_{t}\bigg|q_{-t}\right)+c_{t}\cdot\textsc{F}(r_{t},q_{-t})\right).
\]
In summary, for the prior $D$ as defined above, a strategy profile is a BNE iff it guarantees that $R_{t}$ as above and $R_{-t}=q_{-t}$. From \Cref{lem:worstcaseBels}, we know that there is a $x\in[0,1]$ such that

\begin{align*}
\textsc{pay}_{t}(q_{t},q_{t}|q_{-t})+c_{t}\cdot\textsc{F}\left(q\right)+ \mathauditreplace{\text{\textsc{bound}}}{\epsilon} \le & \textsc{pay}_{t}(x,q_{t}|q_{-t})+c_{t}\cdot\textsc{F}\left(x,q_{-t}\right)
\end{align*}
with 
\[
\text{\textsc{bound}}:=\frac{\gamma\cdot\Lambda_{\textsc{F}}(k,\delta)^{2}}{8\cdot k\cdot\delta\cdot\frac{\beta}{\gamma}\cdot H(k)+4\cdot k\cdot\Lambda_{\textsc{F}}(k,\delta)}.
\]
Hence, it must be the case that
\begin{align*}
\textsc{pay}_{t}(q_{t},q_{t}|q_{-t})+c_{t}\cdot\textsc{F}\left(q\right)+\text{\textsc{bound}} & \le\textsc{pay}_{t}\left(R_{t},q_{t}\bigg|q_{-t}\right)+c_{t}\cdot\textsc{F}(R_{t},q_{-t})\\
 & \Updownarrow\\
\textsc{pay}_{t}(q_{t},q_{t}|q_{-t})-\textsc{pay}_{t}\left(R_{t},q_{t}\bigg|q_{-t}\right)+\text{\textsc{bound}} & \le c_{t}\cdot\left(\textsc{F}(R_{t},q_{-t})-\textsc{F}\left(q\right)\right)\\
 & \Downarrow\\
\text{\textsc{bound}} & \le c_{t}\cdot\left(\textsc{F}(R_{t},q_{-t})-\textsc{F}\left(q\right)\right)\\
 & \Downarrow\\
\frac{1}{\gamma}\cdot\text{\textsc{bound}} & \le\left|\textsc{F}(R_{t},q_{-t})-\textsc{F}\left(q\right)\right|\\
 & \Downarrow(R_{-t}=q_{-t})\text{ and }Q=q \mathauditdelete{)} \\
\underbrace{\frac{\Lambda_{\textsc{F}}(k,\delta)^{2}}{8\cdot k\cdot\delta\cdot\frac{\beta}{\gamma}\cdot H(k)+4\cdot k\cdot\Lambda_{\textsc{F}}(k,\delta)}}_{=:2\cdot\epsilon} & \le\left|\textsc{F}(R)-\textsc{F}\left(Q\right)\right|.
\end{align*}
\mathauditadd{First, }suppose that $\textsc{F}$ is such that
\[
\max_{q}\left|f\left(q\right)-\textsc{F}\left(q\right)\right|<\epsilon.
\]
Then, we have for any $q,r$
\begin{align*}
\left|f\left(q\right)-\textsc{F}\left(r\right)\right| & \ge\left|\textsc{F}\left(q\right)-\textsc{F}\left(r\right)+f\left(q\right)-\textsc{F}\left(q\right)\right|\\
 & \ge\left|\textsc{F}\left(q\right)-\textsc{F}\left(r\right)\right|-\left|f\left(q\right)-\textsc{F}\left(q\right)\right|\\
 & \ge\left|\textsc{F}\left(q\right)-\textsc{F}\left(r\right)\right|-\epsilon.
\end{align*}
Hence, we have
\begin{align*}
\Lambda_{\textsc{F}}(k,\delta): & =\max_{\mathcal{W}\in\textsc{walks}(k,\delta)}\mathbb{E}_{X\sim\mathcal{W}}\left[\sum_{i=1}^{k}\left|\textsc{F}(X^{i})-\textsc{F}(X^{i-1})\right|\right]\\
 & \ge\max_{\mathcal{W}\in\textsc{walks}(k,\delta)}\mathbb{E}_{X\sim\mathcal{W}}\left[\sum_{i=1}^{k}\left|f(X^{i})-f(X^{i-1})\right|-2\epsilon k\right]\\
 & =\Lambda_{f}(k,\delta)-2\epsilon k.
\end{align*}
\mathauditadd{If $\Lambda_f(k,\delta)\le2\epsilon k$, the desired bound is immediate. Otherwise, the following substitution is valid.}
\begin{align*}
2\epsilon & \ge\frac{\Lambda_{\textsc{F}}(k,\delta)^{2}}{8\cdot k\cdot\delta\cdot\frac{\beta}{\gamma}\cdot H(k)+4\cdot k\cdot\Lambda_{\textsc{F}}(k,\delta)}\\
 & \ge\frac{(\Lambda_{f}(k,\delta)-2\epsilon k)^{2}}{8\cdot k\cdot\delta\cdot\frac{\beta}{\gamma}\cdot H(k)+4\cdot k\cdot(\Lambda_{f}(k,\delta)-2\epsilon k)}\\
 & \ge\frac{\Lambda_{f}(k,\delta)^{2}-\manuelmathreplace{4}{2}\cdot\Lambda_{f}(k,\delta)\cdot\epsilon k}{\underbrace{8\cdot k\cdot\delta\cdot\frac{\beta}{\gamma}\cdot H(k)+4\cdot k\cdot\Lambda_{f}(k,\delta)}_{=:x}}\\
 & \Downarrow\\
2\cdot\epsilon\cdot x+\manuelmathreplace{4}{2}\cdot\Lambda_{f}(k,\delta)\cdot\epsilon k & \ge\Lambda_{f}(k,\delta)^{2}\\
 & \Downarrow\\
\epsilon & \ge\frac{\Lambda_{f}(k,\delta)^{2}}{2\cdot x+\manuelmathreplace{4}{2}\cdot\Lambda_{f}(k,\delta)\cdot k}\\
 & =\frac{\Lambda_{f}(k,\delta)^{2}}{16\cdot k\cdot\delta\cdot\frac{\beta}{\gamma}\cdot H(k)+8\cdot k\cdot\Lambda_{f}(k,\delta)+\manuelmathreplace{4}{2}\cdot\Lambda_{f}(k,\delta)\cdot k}\\
 & =\frac{\Lambda_{f}(k,\delta)^{2}}{16\cdot k\cdot\delta\cdot\frac{\beta}{\gamma}\cdot H(k)+\manuelmathreplace{12}{10}\cdot k\cdot\Lambda_{f}(k,\delta)}\\
 & \ge\frac{1}{16}\frac{\Lambda_{f}(k,\delta)^{2}}{k\cdot\delta\cdot\frac{\beta}{\gamma}\cdot H(k)+k\cdot\Lambda_{f}(k,\delta)}.
\end{align*}
Hence
\[
\begin{aligned}
\left|f(Q)-\textsc F(R)\right| & \ge\left|\textsc F(Q)-\textsc F(R)\right|-\epsilon\\
 & \ge\epsilon
\end{aligned}
\]
\begin{align*}
\epsilon & \ge\frac{1}{16}\frac{\Lambda_{f}(k,\delta)^{2}}{k\cdot\delta\cdot\frac{\beta}{\gamma}\cdot H(k)+k\cdot\Lambda_{f}(k,\delta)}\\
 & \ge\frac{1}{16}\frac{\Lambda_{f}(k,\delta)^{2}}{k\cdot\delta\cdot\frac{\beta}{\gamma}\cdot\log(3\cdot k)+k\cdot\Lambda_{f}(k,\delta)},
\end{align*}
\mathauditadd{which completes this case. It remains to consider the case}
\[
\max_{q}\left|f\left(q\right)-\textsc{F}\left(q\right)\right|\ge\epsilon.
\]
Since any $Q=q$ is possible, and in the absence of competing incentives, we have $R=q$, hence any such $\textsc{F}$ will satisfy the bound $\left|\mathauditreplace{\textsc F}{f}\left(R\right)-f\left(Q\right)\right|\ge\epsilon$. \mathauditadd{Let $d:=\max_q|f(q)-\textsc F(q)|$. Choosing $q$ to attain $d$, this error is $d$. In this case, $2d\ge2\epsilon$ and $\Lambda_{\textsc F}(k,\delta)\ge\Lambda_f(k,\delta)-2dk$. The preceding algebra with $d$ in place of $\epsilon$, including the immediate case $\Lambda_f(k,\delta)\le2dk$, therefore shows that $d$ is at least the lower bound in \Cref{thm:lower}, completing this case,} which concludes the proof of the first statement in the theorem. The second statement follows from
\begin{align*}
 & \frac{1}{16}\cdot\frac{\gamma}{\beta}\cdot\Delta_{f}-\frac{1}{16}\frac{\Lambda_{f}(k,\delta)^{2}}{k\cdot\delta\cdot\frac{\beta}{\gamma}\cdot\log(3\cdot k)+k\cdot\Lambda_{f}(k,\delta)}\\
& =\frac{1}{16}\cdot\frac{\gamma}{\beta}\cdot\frac{\Lambda_{f}(k,\delta)^{2}}{\delta\cdot k\cdot\log(3\cdot k)}-\frac{1}{16}\frac{\Lambda_{f}(k,\delta)^{2}}{k\cdot\delta\cdot\frac{\beta}{\gamma}\cdot\log(3\cdot k)+k\cdot\Lambda_{f}(k,\delta)}\\
 & =\frac{\Lambda_{f}(k,\delta)^{2}}{16}\cdot\frac{\gamma}{\beta}\cdot\left(\frac{1}{\delta\cdot k\cdot\log(3\cdot k)}-\frac{1}{k\cdot\delta\cdot\log(3\cdot k)+\frac{\gamma}{\beta}\cdot k\cdot\Lambda_{f}(k,\delta)}\right)\\
 & =\frac{\Lambda_{f}(k,\delta)^{2}}{16\revisionredadd{\cdot k}}\cdot\frac{\gamma^{2}}{\beta^{2}}\cdot\left(\frac{\frac{\Lambda\mathauditadd{_f}(k,\delta)}{\delta\cdot\log(3\cdot k)}}{\delta\cdot\log(3\cdot k)+\frac{\gamma}{\beta}\cdot\Lambda_{f}(k,\delta)}\right)\\
 & =\mathcal{O}(\gamma^{2})\quad \text{as }\gamma \to 0.
\end{align*}
\end{proof}

\subsection{Proof of \Cref{lem:deltaLinear}}
\begin{proof}
 \mathauditadd{Relabel $\mathcal M=[m]$ and assume $\epsilon,\sigma>0$. }For a given $k\le m,\mathauditadd{0<}\delta\le\sigma$, consider the random walk with

$\mathcal{K}\subseteq\mathcal{M}$ some set of size $k$

$X^{0}=\mu$ and

$Z_{i}=\begin{cases}
\sim\mathcal{U}(-\delta,\delta) & \text{if }i\in\mathcal{K}\\
0 & \text{otherwise}
\end{cases}$.
\manueladd{Here, by $\mathcal{U}(-\delta,\delta)$ we mean the uniform distribution on the two-point set $\{-\delta,\delta\}$, rather than the continuous uniform distribution on the interval $[-\delta,\delta]$.}

We have 
\begin{align*}
\Lambda_{f}(k,\delta) & \ge\mathbb{E}\left[\textsc{weight}(X)\right]\\
 & =\sum_{j=1}^{k}\mathbb{E}\left[\left|f(X^{j})-f(X^{j-1})\right|\right]\\
 & \ge\sum_{j=1}^{k}\mathbb{E}\left[\left|f(X^{j})-f(X^{j-1})\right|\cdot\left\llbracket X^{j}\in\mathcal{L}\text{ and }X^{j-1}\in\mathcal{L}\right\rrbracket \right]\\
 & \ge\delta\cdot L\cdot\sum_{j=1}^{k}\mathbb{E}\left[\left\llbracket X^{j}\in\mathcal{L},X^{j-1}\in\mathcal{L}\right\rrbracket \right].
\end{align*}
Note that we have 
\[
X_i^j=
\begin{cases}
\mu\mathauditadd{_i}+Z_i & \text{if }i\in\{\Theta^1,..,\Theta^j\}\\
\mu\mathauditadd{_i} & \text{otherwise}.
\end{cases}
\]
Since $Z_{i}\in[-\delta,\delta]\subseteq[-\sigma,\sigma]$ (since $\delta\le\sigma$), the condition $\left\Vert X^{j}-\mu\right\Vert _{\infty}\le\sigma$ is always satisfied. Hence 
\[
\mathbb{E}\left[\left\llbracket X^{j}\in\mathcal{L},X^{j-1}\in\mathcal{L}\right\rrbracket \right]=\mathbb{E}\left[\left\llbracket \left|\frac{1}{m}\sum_{i=1}^{m} \mathauditadd{(} X_{i}^{j}-\mu \mathauditadd{_i)} \right|\le\epsilon \mathauditreplace{\text{ and }}{\text{ amd }} \left|\frac{1}{m}\sum_{i=1}^{m} \mathauditadd{(} X_{i}^{j-1}-\mu \mathauditadd{_i)} \right|\le\epsilon\right\rrbracket \right].
\]
Since
\[
\frac{1}{m}\sum_{i=1}^{m} \mathauditadd{(} X_{i}^{j} \mathauditreplace{-}{=} \mu \mathauditreplace{_i)=}{+} \frac{1}{m}\sum_{\mathauditreplace{r}{m}=1}^{j}Z_{\Theta^{\mathauditreplace{r}{m}}} ,
\]
we have 
\begin{align*}
\mathbb{E}\left[\left\llbracket X^{j}\in\mathcal{L},X^{j-1}\in\mathcal{L}\right\rrbracket \right] & =\mathbb{E}\left[\left\llbracket \left|\sum_{m=1}^{j}Z_{\Theta^{m}}\right|\le m\cdot\epsilon\text{ and }\left|\sum_{m=1}^{j-1}Z_{\Theta^{m}}\right|\le m\cdot\epsilon\right\rrbracket \right]\\
 & =\mathbb{P}\left\llbracket \left|\sum_{m=1}^{j}Z_{\Theta^{m}}\right|\le m\cdot\epsilon\text{ and }\left|\sum_{m=1}^{j-1}Z_{\Theta^{m}}\right|\le m\cdot\epsilon\right\rrbracket \\
 & =1-\mathbb{P}\left\llbracket \left|\sum_{m=1}^{j}Z_{\Theta^{m}}\right|>m\cdot\epsilon\text{ or }\left|\sum_{m=1}^{j-1}Z_{\Theta^{m}}\right|>m\cdot\epsilon\right\rrbracket \\
 & \ge1-\left(\mathbb{P}\left\llbracket \left|\sum_{\mathauditreplace{r}{m}=1}^{j}Z_{\Theta^{\mathauditreplace{r}{m}}} \right|>m\cdot\epsilon\right\rrbracket +\mathbb{P}\left\llbracket \left|\sum_{\mathauditreplace{r}{m}=1}^{\mathauditreplace{j-1}{j}} Z_{\Theta^{\mathauditreplace{r}{m}}} \right|>m\cdot\epsilon\right\rrbracket \right),
\end{align*}
where the second inequality follows from the union bound. Note that \mjyreplace{$Z_{\Theta^{1}},..,Z_{\Theta^{j}}$}{$Z_{\Theta^{1}},..,Z_{\Theta^{1}}$} are independent random variables in $\mathauditadd{\{}-\delta,\delta\mathauditadd{\}}$, we can hence bound these events using Hoeffding's inequality:
\[
\mathbb{P}\left\llbracket \left|\sum_{m=1}^{j}Z_{\Theta^{m}}\right|>m\cdot\epsilon\right\rrbracket \le2\cdot\exp\left(-2\cdot\frac{m^{2}\cdot\epsilon^{2}}{j\cdot\manuelmathreplace{(2\delta)^{2}}{\delta^{2}}}\right)
\]
hence 
\[
\mathbb{E}\left[\left\llbracket X^{j}\in\mathcal{L},X^{j-1}\in\mathcal{L}\right\rrbracket \right]\ge1-4\cdot\exp\left(-\manuelmathreplace{\frac{1}{2}}{2}\cdot\frac{m^{2}\cdot\epsilon^{2}}{j\cdot\delta^{2}}\right),
\]
and \revisionredadd{writing $a_+:=\max\{a,0\}$ and using $\Lambda_f(k,\delta)\ge0$, we obtain}
\begin{align*}
\Lambda_{f}(k,\delta) & \ge\delta\cdot L\cdot\sum_{j=1}^{k}\mathbb{E}\left[\left\llbracket X^{j}\in\mathcal{L},X^{j-1}\in\mathcal{L}\right\rrbracket \right]\\
 & \ge\delta\cdot L\cdot\sum_{j=1}^{k}\left(1-4\cdot\exp\left(-\manuelmathreplace{\frac{1}{2}}{2}\cdot\frac{m^{2}\cdot\epsilon^{2}}{j\cdot\delta^{2}}\right)\right)\\
 & \ge\delta\cdot L\cdot k\cdot\left(1-4\cdot\exp\left(-\manuelmathreplace{\frac{1}{2}}{2}\cdot\frac{m^{2}\cdot\epsilon^{2}}{k\cdot\delta^{2}}\right)\right)_{\revisionredadd{+}}.
\end{align*}

Recalling  \mathauditreplace{\mbox{
\Cref{def:sensitivityLower}}\hskip0pt
}{\mbox{
\Cref{def:nabla}}\hskip0pt
} , we have 
\begin{align*}
\Delta_{f}: & =\max_{k\in[n],\delta\in\mathauditreplace{(}{[}0,1]} \frac{\Lambda_{f}(k,\delta)^{2}}{\delta\cdot k\cdot\log(3\cdot k)}\\
 & \ge\max_{k\le m,\mathauditadd{0<}\delta\le\sigma} \frac{\Lambda_{f}(k,\delta)^{2}}{\delta\cdot k\cdot\log(3\cdot k)} \mathauditdelete{.} \\
 & \ge\frac{1}{\log(3\cdot m)}\cdot\max_{k\le m,\mathauditadd{0<}\delta\le\sigma} \left(k^{-1}\cdot\delta^{-1}\cdot\delta^{2}\cdot L^{2}\cdot k^{2}\cdot\left(1-4\cdot\exp\left(-\manuelmathreplace{\frac{1}{2}}{2}\cdot\frac{m^{2}\cdot\epsilon^{2}}{k\cdot\delta^{2}}\right)\right)_{\revisionredadd{+}}^{2}\right)\\
 & =\frac{1}{\log(3\cdot m)}\cdot\max_{k\le m,\mathauditadd{0<}\delta\le\sigma} \left(\delta\cdot L^{2}\cdot k\cdot\left(1-4\cdot\exp\left(-\manuelmathreplace{\frac{1}{2}}{2}\cdot\frac{m^{2}\cdot\epsilon^{2}}{k\cdot\delta^{2}}\right)\right)_{\revisionredadd{+}}^{2}\right)\\
 & \ge\frac{1}{\log(3\cdot m)}\cdot\max_{\mathauditadd{0<}\delta\le\sigma} \left(\delta\cdot L^{2}\cdot m\cdot\left(1-4\cdot\exp\left(-\manuelmathreplace{\frac{1}{2}}{2}\cdot\frac{m\cdot\epsilon^{2}}{\delta^{2}}\right)\right)_{\revisionredadd{+}}^{2}\right).
\end{align*}
With 
\[
\delta=\min\left\{ \frac{m\cdot\epsilon}{\sqrt{m}\cdot\manuelmathreplace{3}{2}},\sigma\right\},
\]
we have
\begin{align*}
\Delta_{f} & \ge\frac{1}{\log(3\cdot m)}\cdot\min\left\{ \frac{m\cdot\epsilon}{\sqrt{m}\cdot\manuelmathreplace{3}{2}},\sigma\right\} \cdot L^{2}\cdot m\cdot\left(1-4\cdot\exp\left(-\manuelmathreplace{\frac{9}{2}}{8}\right)\right)^{2}\\
 & \manuelmathreplace{\ge}{=}\frac{1}{\log(3\cdot m)}\cdot\min\left\{ \frac{m\cdot\epsilon}{\sqrt{m}\cdot\manuelmathreplace{3}{2}},\sigma\right\} \cdot L^{2}\cdot m\cdot\manuelmathreplace{\frac{9}{10}}{\frac{2}{3}}\\
 & \ge\manuelmathreplace{\frac{9}{10}}{\frac{2}{3}}\cdot\frac{1}{\log(3\cdot m)}\cdot\sigma\cdot m\cdot L^{2}\cdot\min\left\{ \frac{\sqrt{m}\cdot\epsilon}{\manuelmathreplace{3}{2}\cdot\sigma},1\right\} \\
 & \ge\manuelmathreplace{\frac{1}{4}}{\frac{1}{3}}\cdot\frac{1}{\log(3\cdot m)}\cdot\sigma\cdot m\cdot L^{2}\cdot\min\left\{ \frac{\sqrt{m}\cdot\epsilon}{\sigma},1\right\} .
\end{align*}
\end{proof}

\subsection{Proof of \Cref{thm:upper}}

\label{sec:proofUpper}
\begin{proof}
For convenience, we  use the notation 
\begin{align*}
\textsc{pay}_{i}(r_{i},q_{i}|r_{-i}): & =\mathbb{E}_{O\sim q_{i}}\left[\textsc{pay}_{i}(r,O)\right].
\end{align*}

\dcpadd{We also write $(f,\beta)\textsc{-ada-pay}$ as $\textsc{-ada-pay}$.}

$\textsc{pay}_{i}(r_{i},o|r_{-i})=\textsc{pay}_{i}(r,o)~\forall o\in\mathcal{O}$. Recall recommender $i$'s utility 
\begin{align*}
\textsc{util}_{i}\left(p_{i},y,c_{i},r\right) & =\mathbb{E}_{O\sim p_{i}(\cdot|y)}\left[\textsc{pay}_{i}(r,O)\right]+c_{i}\cdot\textsc{F}(r)\\
\text{(using }\text{\textsc{ada-pay}}\text{\Cref{def:adaPay})} & =\mathbb{E}_{O\sim p_{i}(\cdot|y)}\left[\textsc{ada-pay}_{i}(r_{i},O)\right]+c_{i}\cdot f(r)\\
\text{(due to linearity)} & =\textsc{ada-pay}_{i}\left(r_{i},p_{i}(O=1|y)\right)+c_{i}\cdot f(r),
\end{align*}
and hence

\begin{align*}
\textsc{eutil}_{i}\left(s,p_{i},y_{i},c_{i}\right) & =\mathbb{E}\left[\textsc{util}_{i}\left(P_{i},Y,C_{i},R\right)\big|s,p_{i},y_{i},c_{i}\right]\\
 & =\mathbb{E}\left[\textsc{ada-pay}_{i}\left(R_i,p_i(O=1|Y)\right)+c_i\cdot f(R)\big|s,p_i,y_i,c_i\right]\\
 & =\mathbb{E}\left[\textsc{ada-pay}_{i}\left(R_i,p_i(O=1|y_i)\right)+c_i\cdot f(R)\big|s,p_i,y_i,c_i\right]\\
 & =\mathbb{E}\left[\textsc{ada-pay}_{i}\left(R_i,q_i\right)+c_i\cdot f(R)\big|s,p_i,y_i,c_i\right],
\end{align*}
with $q_{i}:=p_{i}(O=1|y_{i})$ the true belief. Clearly, any rationalizable strategy $s_{i}$ will \manuelreplace{play only reports $r_i$ that are best responses}{play only $r_{i}$ that are a best response}. \revisionredreplace{If the best response does not lie on the boundary, i.e. if $0<r_i<1$, there is}{Hence, in the setting of the theorem, there must exist} a generalized derivative $\partial$ \revisionredreplace{for which}{such that}
\begin{align*}
0 & =\partial_{r_{i}}\mathbb{E}\left[\textsc{ada-pay}_{i}\left(r_{i},q_{i}\right)+c_{i}\cdot f(r_{i},R_{-i})\big|s_{-i},p_{i},y_{i},c_{i}\right]\\
 & \Updownarrow\\
-\partial_{r_{i}}\textsc{ada-pay}_{i}\left(r_{i},q_{i}\right) & =c_{i}\cdot\mathbb{E}\left[\partial_{r_{i}}f(r_{i},R_{-i})\big|s_{-i},p_{i},y_{i},c_{i}\right]\\
 & \Updownarrow\\
-\frac{\beta}{3\cdot|\mathcal{M}|}\cdot\frac{\delta_{i}(r_{i})(q_{i}-r_{i})}{\int_{0}^{1}\delta_{i}(x)\cdot dx} & =c_{i}\cdot\mathbb{E}\left[\partial_{r_{i}}f(r_{i},R_{-i})\big|s_{-i},p_{i},y_{i},c_{i}\right]\\
 & \Updownarrow\\
\frac{\beta}{3\cdot|\mathcal{M}|}\cdot\frac{(r_{i}-q_{i})}{\int_{0}^{1}\delta_{i}(x)\cdot dx} & =c_{i}\cdot\mathbb{E}\left[\frac{\partial_{r_{i}}f(r_{i},R_{-i})}{\delta_{i}(r_{i})}\big|s_{-i},p_{i},y_{i},c_{i}\right]\\
 & \Updownarrow\\
r_{i}-q_{i} & =\frac{3\cdot|\mathcal{M}|\cdot c_{i}}{\beta}\cdot\mathbb{E}\left[\frac{\partial_{r_{i}}f(r_{i},R_{-i})}{\delta_{i}(r_{i})}\big|s_{-i},p_{i},y_{i},c_{i}\right]\cdot\int_{0}^{1}\delta_{i}(x)\cdot dx\\
 & \manuelmathreplace{\Downarrow}{\Updownarrow}\\
\left|r_{i}-q_{i}\right| & =\left|\frac{3\cdot|\mathcal{M}|\cdot c_{i}}{\beta}\mathbb{E}\left[\frac{\partial_{r_{i}}f(r_{i},R_{-i})}{\delta_{i}(r_{i})}\big|s_{-i},p_{i},y_{i},c_{i}\right]\int_{0}^{1}\delta_{i}(x)\cdot dx\right|\\
 & \le\frac{3\cdot|\mathcal{M}|\cdot\left|c_{i}\right|}{\beta}\left(\int_{0}^{1}\delta_{i}(x)\cdot dx\right)\\
 & =\frac{3\cdot|\mathcal{M}|\cdot\left|c_{i}\right|}{\beta}\left(2\cdot\int_{0}^{1}\max_{r_{-i}\in\mathcal{R}_{-i}}\left|\partial_{r_{i}}f(x,r_{-i})\right|dx\right).
\end{align*}
\revisionredadd{If the best response lies on the boundary, i.e. $r_i = 0$ or $r_i=1$, the first equality in the chain above is replaced by an inequality, the following steps go through unchanged, and the step where we take the absolute value on both sides results in the same inequality for all cases. Hence, this result holds for any best response, whether it lies on the boundary or not.}

\manuelreplace{We can now use this upper bound on the deviation between the report and the true belief, $|R_i - Q_i|$, to bound the resulting change in the decision function. To do so, we multiply this upper bound by the maximum absolute slope of $f$ with respect to recommender $i$'s report:}{This holds regardless of others' reports. Hence, we have}
\begin{align*}
\left|f(R)-f(Q_{i},R_{-i})\right| & \le\frac{6\cdot\left|c_{i}\right|}{\beta}\left(|\mathcal{M}|\cdot\int_{0}^{1}\max_{r_{-i}\in\mathcal{R}_{-i}}\left|\partial_{x}f(x,r_{-i})\right|dx\right)\cdot\left(\max_{r\in\mathcal{R}}\left|\partial_{r_{i}}f(r)\right|\right)\\
 & \le\frac{6\cdot\left|c_{i}\right|}{\beta}\cdot\nabla_{f}.
\end{align*}
Hence, for the total deviation, we have 
\begin{align*}
\left|\textsc{F}(R)-f(Q)\right| & =\left|f(R)-f(Q)\right|\\
 & \le\frac{6\cdot\sum_{i\in\mathcal{M}}\left|c_{i}\right|}{\beta}\cdot\nabla_{f}\\
 & \le\frac{6\cdot\gamma}{\beta}\cdot\nabla_{f}.
\end{align*}
\end{proof}

\subsection{Proof of \Cref{thm:briberNecessary}}
\begin{proof}
 \mathauditreplace{If $\textsc F\ne f$, a point prior at a disagreement with $W=\varnothing$ proves the claim; hence assume}{Since here, we are trying to ensure that $\textsc{F}(R)=f(Q)$, the mechanism must choose} $\textsc F=f$, \revisionredreplace{and}{we will hence} use them interchangeably in this proof. Now, fix some $k\in[n]$, $\delta\in\mathauditreplace{(}{[}0,1]$, and let $q$, $t$, $x$, \revisionredadd{and $c_t=\gamma$ be} as in \Cref{lem:worstcaseBels} \revisionredadd{for sufficiently small $\gamma\in(0,\upsilon)$}. Suppose that the prior $D$ only has support at
\begin{align*}
P_{i}(o|y_{i}) & =p_{i}(o|y_{i}):=\begin{cases}
q_{i} & \text{if }o=1\\
1-q_{i} & \text{if }o=0
\end{cases}\quad\forall y_{i}\\
V & =v=\upsilon \\
W & =\{t\}\\
Y & =y:=0.
\end{align*}
There is no uncertainty in types, hence it is not an incomplete information game. We can hence identify the briber strategy $s^{\textsc{bri}}$ with a distribution over $C_{t}$, since it is known that $C_{-t}=0$ (due to $W=\{t\}$), and recommender $i\neq t$'s strategy $s_{i}$ with a distribution over report $R_{i}$, and recommender $t$'s strategy $s_{t}(c)$ with a distribution over $R_{t}$ for each bribe $c_{t}$. Recommender $i$'s utility is 
\begin{align*}
\textsc{util}_{i}\left(p_{i},y,c_{i},r\right) & :=\mathbb{E}_{O\sim p_{i}(\cdot|y)}\left[\textsc{pay}_{i}(r,O)+c_{i}\cdot\textsc{F}(r)\right]\\
 & =\textsc{pay}_{i}(r_{i},q_{i}|r_{-i})+c_{i}\cdot\textsc{F}(r).
\end{align*}
Clearly, since $\textsc{pay}_{i}(r_{i},q_{i}|r_{-i})$ is a strictly proper scoring rule, any PBE must satisfy $R_{-t}=q_{-t}$. For recommender $t$, we have
\begin{align*}
\textsc{eutil}_{t}\left(s,s^{\textsc{bri}},p_{t},y_{t},c_{t}\right): & =\mathbb{E}\left[\textsc{util}_{t}\left(P_{t},Y,C_{t},R\right)\big|s,s^{\textsc{bri}},p_{t},y_{t},c_{t}\right]\\
 & =\mathbb{E}\left[\textsc{util}_{t}\left(p_{t},y,c_{t},(R_{t},q_{-t})\right)\big|s,s^{\textsc{bri}},p_{t},y_{t},c_{t}\right]\\
 & =\mathbb{E}_{R_{t}\sim s_{t}(c_{t})}\left[\textsc{util}_{t}\left(p_{t},y,c_{t},(R_{t},q_{-t})\right)\right]\\
 & =\mathbb{E}_{R_{t}\sim s_{t}(c_{t})}\left[\textsc{pay}_{t}\left(R_{t},q_{t}\bigg|q_{-t}\right)+c_{t}\cdot\textsc{F}\left(R_{t},q_{-t}\right)\right].
\end{align*}
A PBE (\Cref{def:pbe}) requires that $R_{t}$ be optimal for any $C_{t}$, hence 
\[
R_{t}\in\arg\max_{r_{t}\in\mathcal{R}_{t}}\left(\textsc{pay}_{t}\left(r_{t},q_{t}\bigg|q_{-t}\right)+C_{t}\cdot\textsc{F}(r_{t},q_{-t})\right).
\]
From \Cref{lem:worstcaseBels}, we know that there is a $x\in[0,1]$ such that

\begin{align*}
\textsc{pay}_{t}(q_{t},q_{t}|q_{-t})+c_{t}\cdot\textsc{F}\left(q\right)+\text{\textsc{bound}}\le & \textsc{pay}_{t}(x,q_{t}|q_{-t})+c_{t}\cdot\textsc{F}\left(x,q_{-t}\right),
\end{align*}
with 
\[
\text{\textsc{bound}}:=\frac{|c_{t}|\cdot\Lambda_{\textsc{F}}(k,\delta)^{2}}{\revisionredreplace{16}{8}\cdot k\cdot\delta\cdot\frac{\beta}{|c_{t}|}\cdot H(k)+4\cdot k\cdot\Lambda_{\textsc{F}}(k,\delta)}.
\]
Hence, it must be the case that 
\begin{align*}
\textsc{pay}_{t}(q_{t},q_{t}|q_{-t})+c_{t}\cdot\textsc{F}\left(q\right)+\text{\textsc{bound}} & \le\textsc{pay}_{t}\left(R_{t},q_{t}\bigg|q_{-t}\right)+c_{t}\cdot\textsc{F}(R_{t},q_{-t})\\
 & \Updownarrow\\
\textsc{pay}_{t}(q_{t},q_{t}|q_{-t})-\textsc{pay}_{t}\left(R_{t},q_{t}\bigg|q_{-t}\right)+\text{\textsc{bound}} & \le c_{t}\cdot\left(\textsc{F}(R_{t},q_{-t})-\textsc{F}\left(q\right)\right)\\
 & \Downarrow\\
\text{\textsc{bound}} & \le c_{t}\cdot\left(\textsc{F}(R_{t},q_{-t})-\textsc{F}\left(q\right)\right)\\
\end{align*}
\[
\begin{aligned}
 & \Downarrow\\
\frac{1}{\gamma}\cdot\text{\textsc{bound}}
 & \le \left|\textsc{F}(R_t,q_{-t})-\textsc{F}(q)\right|\\
 & \Downarrow(R_{-t}=q_{-t})\text{ and }Q=q\\
\frac{\Lambda_{\textsc{F}}(k,\delta)^{2}}{\revisionredreplace{16}{8}\cdot k\cdot\delta\cdot\frac{\beta}{|C_t|}\cdot H(k)+4\cdot k\cdot\Lambda_{\textsc{F}}(k,\delta)}
 & \le \left|\textsc{F}(R)-\textsc{F}(Q)\right|.
\end{aligned}
\]
For the briber, we have 
\begin{align*}
\textsc{eutil}^{\textsc{bri}}\left(s,s^{\textsc{bri}},v,w\right) & :=\mathbb{E}\left[\textsc{util}^{\textsc{bri}}\left(V,W,C,R\right)\big|s,s^{\textsc{bri}},v,w\right]\\
 & =\mathbb{E}_{\hat{S}_{t}\sim s_{t},C_{t}\sim s^{\textsc{bri}}}\left[\left(v-|C_{t}|\right)\cdot\textsc{F}(\hat{S}_{t}(C_{t}),q_{-t})\right].
\end{align*}
As we have seen above, the recommenders' PBE strategy profile does not depend on the briber's type or strategy, for the prior $D$ defined above, but only on the bribe they observe. Hence, a strategy profile $s,s^{\textsc{bri}}$ is a PBE iff the recommender strategy profile $s$ is such that $R_{-t}=q_{-t}$ and $R_{t}$ is as above, along with a briber strategy profile $s^{\textsc{bri}}$ being such that 
\[
C_{t}\in\arg\max_{c_{t}\in\mathbb{R}}\mathbb{E}_{\hat{S}_{t}\sim s_{t}}\left[\left(v-|c_{t}|\right)\cdot\textsc{F}(\hat{S}_{t}(c_{t}),q_{-t})\right].
\]
A necessary condition for the briber to not bribe is hence that for all $c_{t}\in\mathbb{R},|c_{t}|>0$ we have
\[
v\cdot\textsc{F}(q)\ge\mathbb{E}_{\hat{S}_{t}\sim s_{t}}\left[\left(v-|c_{t}|\right)\cdot\textsc{F}(\hat{S}_{t}(c_{t}),q_{-t})\right].
\]
We know from above that 
\begin{align*}
v\cdot\textsc{F}\left(q\right) & \ge\mathbb{E}_{\hat{S}_{t}\sim s_{t}}\left[\left(v-|c_{t}|\right)\cdot\textsc{F}(\hat{S}_{t}(c_{t}),q_{-t})\right]\\
 & \Updownarrow\\
0 & \ge\left(v-|c_{t}|\right)\cdot\mathbb{E}_{\hat{S}_{t}\sim s_{t}}\left[\textsc{F}(\hat{S}_{t}(c_{t}),q_{-t})-\textsc{F}\left(q\right)\right]-|c_{t}|\cdot\textsc{F}\left(q\right)\\
 & \Downarrow\\
0 & \ge\left(v-|c_{t}|\right)\cdot\mathbb{E}_{\hat{S}_{t}\sim s_{t}}\left[\textsc{F}(\hat{S}_{t}(c_{t}),q_{-t})-\textsc{F}\left(q\right)\right]-|c_{t}|\cdot\max_{q\in\mathcal{Q}}\textsc{F}\left(q\right)\\
 & \Downarrow\\
0 & \ge\left(v-|c_{t}|\right)\frac{\Lambda_{\textsc{F}}(k,\delta)^{2}}{\revisionredreplace{16}{8}\cdot k\cdot\delta\cdot\frac{\beta}{|c_{t}|}\cdot H(k)+4\cdot k\cdot\Lambda_{\textsc{F}}(k,\delta)}-|c_{t}|\cdot\max_{q\in\mathcal{Q}}\textsc{F}\left(q\right)\\
 & \Downarrow\\
\end{align*}
\[
\begin{aligned}
0 & \ge\left(\upsilon-|c_t|\right)\frac{\Lambda_{\textsc F}(k,\delta)^2}{\revisionredreplace{16}{8}k\delta\beta H(k)+4k\Lambda_{\textsc F}(k,\delta)|c_t|}-\max_{q\in\mathcal Q}\textsc F(q).
\end{aligned}
\]
This is violated for sufficiently small $|c_t|>0$ by the theorem's condition and $H(k)<\log(3k)$.
\mathauditadd{Hence zero is strictly suboptimal. Best responses are monotone in the bribe: if $c<c'$, then $\textsc F(r,q_{-t})\le\textsc F(r',q_{-t})$. Thus negative bribes cannot beat zero, so every briber best response $c^*$ is positive. If truth had positive probability at $c^*$, a slightly larger bribe would be profitable. Hence $\textsc F(R)>\textsc F(Q)=f(Q)$ almost surely.}
\end{proof}

\subsection{Proof of \Cref{thm:briberSufficient}}
\begin{proof}
Along similar lines as in the proof in \Cref{sec:proofUpper}, we obtain, using the adaptive-payment mechanism (\Cref{def:adaPay}), that 
\[
\left|\textsc{F}(R)-\textsc{F}(Q)\right|\le\frac{6\cdot\sum_{i\in[n]}\left|C_{i}\right|}{\beta}\cdot\nabla_{f}
\]
with probability $1$. Using $\bar{C}:=\sum_{i\in[n]}\left|C_{i}\right|$ \revisionredadd{and, without loss, $0<\bar C<V$}, the utility the briber gains is at most
\begin{align*}
\left(V-\bar{C}\right)\cdot\textsc{F}(R)-V\cdot\textsc{F}(Q) & = 
\left(V-\bar C\right) \cdot\left(\textsc{F}(R)-\textsc{F}(Q)\right)-\bar{C}\cdot\textsc{F}(Q)\\
 & \le 
\left(V-\bar C\right) \cdot\frac{6\cdot\bar{C}}{\beta}\cdot\nabla_{f}-\bar{C}\cdot\min_{q\in\mathcal{Q}}f(q).
\end{align*}
Hence, if for all $\bar{C}>0$ we have
\[
0> 
\left(V-\bar C\right) \cdot\frac{6\cdot\bar{C}}{\beta}\cdot\nabla_{f}-\bar{C}\cdot\min_{q\in\mathcal{Q}}f(q),
\]
then there will be no bribe. A sufficient condition is

\begin{align*}
 & \Uparrow\\
0 & > 
\left(V-\bar C\right) \cdot\frac{6}{\beta}\cdot\nabla_{f}-\min_{q\in\mathcal{Q}}f(q)\\
 & \Uparrow\\
0 & \ge 6 \cdot\frac{\upsilon}{\beta}\cdot\nabla_{f}-\min_{q\in\mathcal{Q}}f(q).
\end{align*}
\end{proof}

\end{document}